\documentclass[aps,prb, twocolumn,groupedaddress]{revtex4-2}
\usepackage[colorlinks=true,linkcolor=blue,urlcolor=blue,citecolor=blue]{hyperref}
\usepackage{graphicx}
\usepackage[caption=false]{subfig}
\usepackage{tikz}
\usepackage{times}
\usepackage{lipsum}
\usepackage{mathrsfs,amsmath}
\usepackage{tabularx}
\usepackage{booktabs}
\usepackage{natbib}

\graphicspath{{figures/}}

\newcommand{\Kay}[1]{\color{red}}

\newcommand{\be}{\begin{equation}}
\newcommand{\ee}{\end{equation}}
\newcommand{\bea}{\begin{eqnarray}}
\newcommand{\eea}{\end{eqnarray}}
\newcommand{\nn}{\nonumber}
\newcommand{\ca}[1]{\mathcal{#1}}
\newcommand{\eq}[1]{(\ref{#1})}
\newcommand{\Eq}[1]{Eq.~(\ref{#1})}
\newcommand{\Eqs}[1]{Eqs.~(\ref{#1})}
\newcommand{\rmd}{\mathrm{d}}
\newcommand{\rme}{\mathrm{e}}
\newcommand{\sign}{\text{sign}}

\begin{document}

\title{Depinning and flow of a vortex line in an uniaxial random medium}

\author{Federico El\'ias}
\email{federico.elias@cab.cnea.gov.ar}
\affiliation{Centro At\'omico Bariloche and Instituto Balseiro,
CNEA, CONICET and Universidad Nacional de Cuyo, 8400 Bariloche, Argentina}

\author{Kay J\"org Wiese}
\email{wiese@lpt.ens.fr}
\affiliation{Laboratoire de physique, Departement de physique de l’ENS, Ecole normale superieure,
UPMC Univ. Paris 06, CNRS, PSL Research University, 75005 Paris, France}

\author{Alejandro B. Kolton}
\email{alejandro.kolton@cab.cnea.gov.ar}
\affiliation{Centro At\'omico Bariloche and Instituto Balseiro,
CNEA, CONICET and Universidad Nacional de Cuyo, 8400 Bariloche, Argentina}

\date{\today}

\begin{abstract}
We study numerically and analytically the dynamics of a single directed elastic string driven through a 3-dimensional disordered medium. 
In the quasistatic limit   the string is super-rough in the direction of the driving force, with roughness exponent $\zeta_{\parallel} = 1.25\pm 0.01$, dynamic exponent   $z_{\parallel}=  1.43  \pm 0.01$, correlation-length exponent $\nu= 1.33 \pm 0.02$, 
depinning exponent $\beta = 0.24\pm 0.01$, and avalanche-size   exponent $\tau_{\parallel} = 1.09 \pm 0.03$. In the transverse direction we find 
$\zeta_{\perp} = 0.5 \pm 0.01$, 
$z_{\perp} = 2.27  \pm 0.05$,  and $\tau_{\perp} =1.17\pm 0.06$. 
Our results show that transverse fluctuations do not alter the  critical exponents in the driving direction, as predicted by the  planar   approximation (PA) proposed in 1996 by Ertas and Kardar (EK).
We check the PA  
for the measured force-force correlator, comparing to  the functional renormalization group and   numerical simulations.
Both  Random-Bond (RB) and Random-Field (RF)  disorder yield   a single   universality class,
indistinguishable from the one of an elastic string in a two-dimensional random medium.
While   relations $z_{\perp}=z_{\parallel}+1/\nu$ and $\nu=1/(2-\zeta_{\parallel})$ of EK  are   satisfied, 
 the transversal movement is that of a  Brownian, with a  clock set locally by the forward movement. 
This implies  $\zeta_\perp = (2-d)/2$,  distinct from EK.  
Finally   at small driving velocities the distribution of local parallel displacements has a negative skewness, while in the transverse direction it is a  Gaussian.  
For  large scales,  the system can be described by anisotropic effective temperatures defined from generalized fluctuation-dissipation relations. 
In the fast-flow regime   the local displacement distributions become Gaussian in both directions and the effective temperatures vanish   as $T^{\perp}_{\tt eff}\sim 1/v$ and $T^{\parallel}_{\tt eff}\sim 1/v^3$ for RB disorder and   as $T^{\perp}_{\tt eff} \approx T^{\parallel}_{\tt eff} \sim 1/v$ for RF disorder. 
\end{abstract}

\maketitle

\tableofcontents


\section{Introduction}
\label{sc:intro}

Many driven  systems display a depinning transition from a static to a sliding state, at a finite    value of an applied force or stress. 
Examples are field-driven domain walls in ferromagnetic 
~\cite{DurinZapperi2006b,FerreMetaxasMouginJametGorchonJeudy2013,DurinBohnCorreaSommerLeDoussalWiese2016}   
or ferroelectric
materials \cite{Kleemann2007,ParuchGuyonnet2013}, 
cracks under stress in heterogeneous materials
\cite{BonamySantucciPonson2008,Ponson2009,LePriolChopinLeDoussalPonsonRosso2020}, 
contact lines of liquids on a rough substrate~\cite{MoulinetRossoKrauthRolley2004,LeDoussal2009}, 
imbibition of fluids in porous and fractured media~\cite{PlanetSantucciOrtin2009}, 
reaction fronts in porous media~\cite{AtisDubeySalinTalonLeDoussalWiese2015},
solid-solid friction~\cite{BayartSvetlizkyFineberg2015}, sheared 
amorphous solids or yield-stress fluids~\cite{NicolasFerreroMartensBarrat2017}, 
dislocation arrays in 
sheared crystals~\cite{SethnaBierbaumDahmenGoodrichGreerHaydenKentDobiasLeeLiarteNiQuinnRajuRocklinShekhawatZapperi2017}, current-driven 
vortex lattices in superconductors~\cite{NattermannScheidl2000,GiamarchiBhattacharya2002,ledoussal2010,  KwokWelpGlatzKoshelevKihlstromCrabtree2016,ThomannGeshkenbeinBlatter2017, SadovskyyKoshelevKwokWelpGlatz2019,EleyGlatzWilla2021}, skyrmion lattices in ferromagnets~\cite{SchulzRitzBauerHalderWagnerFranzPfleidererEverschorGarst2012},  or earthquakes models~\cite{JaglaKolton2010,JaglaLandesRosso2014}. Among   these systems, the family of {\em directed elastic manifolds in random media}, where interactions between constituents  are purely elastic,  and topological defects   absent, have become the framework of choice for understanding quantitatively   universal properties   of the depinning transition~\cite{Kardar1998,Fisher1998,NattermannScheidl2000,Wiese2021}.

Directed elastic manifolds embedded in a space of dimension $D$ have an internal dimension $d$ and an $N$-dimensional displacement field, 
such that $D=N+d$~\footnote{See ~\cite{NattermannScheidl2000} for a general description.}. 
 We   focus   on the overdamped zero-temperature dynamics   when the manifold is driven by a force $f$ in one of the $N$ directions in a quenched random potential. In such a case,  the competition of elasticity and disorder yields a critical force $f_c$, such that for $f<f_c$ the steady-state velocity vanishes, $v=0$, while for $f>f_c$ the interfaces slides according to a velocity-field characteristics $v(f)>0$.
While one can impose the force $f$ and then measure the velocity $v(f)$, in most experiments the driving velocity $v$ is imposed, and the pinning force $f$   measured. This is achieved by confining the elastic manifold within  a quadratic potential, given by the demagnetization field in magnets, gravity in contact-line depinning, or the bulk elasticity in earthquakes.

The case $(d,N)=(d,1)$ of a directed elastic interface has been studied in   detail~\cite{NattermannStepanowTangLeschhorn1992,NarayanFisher1993,LeschhornNattermannStepanowTang1997,ChauveGiamarchiLeDoussal2000}. 
It was found that the depinning transition at $f_c$
is continuous, reversible,
and occurs at a well-defined
characteristic threshold force $f_c$ ~\cite{KoltonBustingorryFerreroRosso2013}, below which the interface remains pinned in a metastable state. At $f_c$ the interface is marginally blocked and the instability is described by a localized soft-spot or eigenvector~\cite{CaoBouzatKoltonRosso2018}.  
Just above the threshold, the mean velocity $v$ is   
given by the depinning law 
$v\sim (f-f_c)^{\beta}$, with $\beta$ a non-trivial 
critical exponent. A divergent correlation 
length $\xi \sim (f-f_c)^{-\nu}$ and a divergent 
correlation time $\tau \sim \xi^{z}$ characterize the 
jerky motion as we approach $f_c$ from above. 
Below the length scale $\xi$ the  interface is self-affine, with the displacement field growing 
as $u \sim x^{\zeta}$. 
Similarly $v\sim \xi^{\zeta-z}$ and $\beta=\nu(z-\zeta)$. 
The critical exponents have been estimated 
analytically \cite{ChauveLeDoussalWiese2000a,LeDoussalWieseChauve2002,FedorenkoStepanow2003} and numerically 
\cite{Leschhorn1993,LeschhornNattermannStepanowTang1997,RotersHuchtLubeckNowakUsadel1999,RossoHartmannKrauth2003,RossoLeDoussalWiese2007,FerreroBustingorryKolton2013}. 
The different universality classes are distinguished by  $d$, the 
range~\cite{RamanathanFisher1998,ZapperiCizeauDurinStanley1998,RossoKrauth2002,DuemmerKrauth2007,LaursonIllaSantucciToreTallakstadMaloyAlava2013} 
or nature~\cite{BoltzKierfeld2014}
of the elastic interactions, the anisotropic~\cite{TangKardarDhar1995} 
or isotropic correlations of the 
pinning force~\cite{FedorenkoLeDoussalWiese2006b,BustingorryKoltonGiamarchi2010}, 
and by the presence of  non-linear 
terms in addition to the pinning force~\cite{AmaralBarabsiStanley1994,TangKardarDhar1995,RossoKrauth2001b,GoodmanTeitel2004,LedoussalWiese2003,ChenZapperiSethna2015}. 
Boundary \cite{AragonKoltonLeDoussalWieseJagla2016} or AC-driven \cite{GlatzNattermannPokrovsky2003} 
  elastic interfaces have been   studied as well.
If the so-called statistical tilt symmetry (STS) holds, each  depinning universality class has exactly two independent exponents, $\zeta$ and $z$. 
At large velocities, the   disorder mimics thermal fluctuations with a velocity-dependent effective temperature, vanishing in the infinite-velocity limit.   
The particular case $(d,N)=(1,1)$, which   represents an elastic line in a random medium, is relevant for bulk magnets \cite{DurinBohnCorreaSommerLeDoussalWiese2016,terBurgBohnDurinSommerWiese2021},
and magnetic domain walls in ultra-thin   ferrimagnetic materials at very low temperatures~\cite{AlbornozFerreroKoltonJeudyBustingorryCuriale2021}. 
Variants are systems with long-ranged elasticity, as 
  contact lines of liquid menisci~\cite{MoulinetRossoKrauthRolley2004,LeDoussal2009} or 
planar crack propagation~\cite{LePriolChopinLeDoussalPonsonRosso2020}. When STS is broken, an additional KPZ term is generated, as in  
  reaction fronts in porous media~\cite{AtisDubeySalinTalonLeDoussalWiese2015}.

\begin{figure}[t]
\centering
\includegraphics[width=.89\columnwidth]{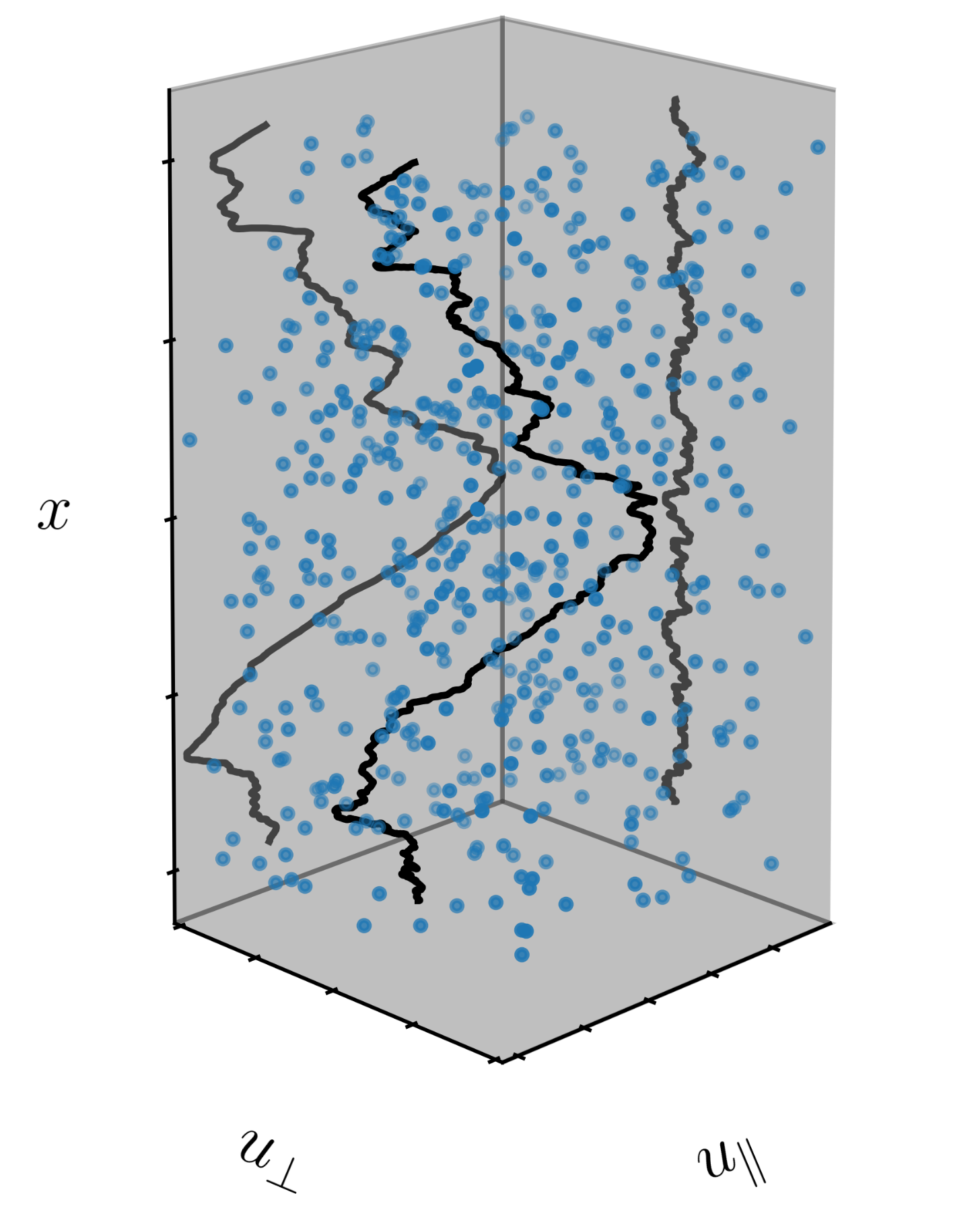}
\caption{Snapshot of an elastic line moving in an isotropic three-dimensional random environment. The projected profiles display the $x$-dependent displacements  $u_{\parallel}$ (parallel) and $u_{\perp}$    (perpendicular) to the  driving.}
\label{fg:esquema}
\end{figure}

The depinning transition of a one-dimensional directed elastic line   with a two-dimensional scalar displacement field $(d,N)=(1,2)$ 
in a $D=d+N=3$-dimensional medium   is realized  for an isolated flux line (FL) in a superconductor. The 2-dimensional 
displacement field has  one component $u_{\parallel}$ parallel to the driving direction, 
and a perpendicular component $u_{\perp}$, while the string is directed in the $x$ direction as shown in Fig \ref{fg:esquema}.
For an overdamped dynamics the study of such a system has been pioneered by Ertas and Kardar more than twenty five years ago~\cite{ErtasKardar1996}. 
Compared with the $(d,N)=(1,1)$ case of interface depinning considerable less progress has been made since then, with the exception of a precise study of the critical force in the case of   isotropic disorder~\cite{koshelevkolton2011}, relevant for superconductor applications~\cite{KwokWelpGlatzKoshelevKihlstromCrabtree2016,Civale2019}. 
This is     due to the difficulties the flux-line problem presents in addition to an elastic interface (and   fewer applications). Indeed, even for harmonic elasticity and   uncorrelated disorder one has to deal with a 2-component (vector) displacement field instead of the scalar   displacement field of the interface. One of the important consequences of this difference is that Middleton's theorems~\cite{Middleton1992}, which lie at the heart of   FRG calculations for depinning \cite{LeDoussalWieseChauve2002,ChauveLeDoussalWiese2000a,DobrinevskiLeDoussalWiese2011b,DobrinevskiLeDoussalWiese2014a}, and which greatly accelerate numerical simulations  \cite{RossoKrauth2001b,RossoKrauth2001a}
are not valid for the FL, making  a precise study of its large-scale properties difficult. In addition, renormalization-group calculations rely on an expansion in $\epsilon=4-d$, with  $d=1$  far away from the upper critical dimension. 
In Ref.~\onlinecite{ErtasKardar1996}  Ertas and Kardar postulated anisotropic scaling forms for the directions parallel and perpendicular to the driving, with  exponents $\zeta_{\perp}$, $\zeta_{\parallel}$, $z_{\perp}$, $z_{\parallel}$, $\nu$, $\beta$. The latter  were calculated, both via FRG  and via direct numerical simulations. 
They describe the   anisotropic self-affine steady-state   of the line below the depinning correlation length $\xi \sim (f-f_c)^{-\nu}$,
\begin{eqnarray}
\langle [u_{\parallel}(x,t)-u_{\parallel}(0,0)]^2  \rangle &=& x^{2\zeta_{\parallel}} g_{\parallel}(t/x^{z_{\parallel}}), \\
\langle [u_{\perp}(x,t)-u_{\perp}(0,0)]^2  \rangle &=& x^{2\zeta_{\perp}} g_{\perp}(t/x^{z_{\perp}}),
\end{eqnarray}
with $\beta=\nu(z_\parallel-\zeta_\parallel)$ and $g_\parallel(y)$ and $g_\perp(y)$ universal functions.

The central idea of  the 1996   Ref.~\onlinecite{ErtasKardar1996} by Ertazs and Kardar (EK) is to propose a  ``planar approximation'',  postulating that the exponents corresponding to the direction parallel to the driving 
are equal to those for $N=1$, i.e.\ without a perpendicular direction. In a second step, exponents in the perpendicular direction are obtained, using the results in the parallel direction.  
EK then  obtained the exponents $\zeta_{\parallel}=1$, and $\zeta_{\perp}=\zeta_{\parallel}-d/2=1/2$, using a FRG calculation at 1-loop order (followed by a numerical check). 

They then claimed that this result holds to all orders in perturbation theory, using an argument advanced in 1993 for $N=1$ by Narayan and Fisher 
\cite{NarayanFisher1993}, and in agreement with numerical simulations at the time. 
In the meantime it has been established   that for $N=1$ the exponent $\zeta>1$:  The 2-loop FRG  \cite{LeDoussalWieseChauve2002,ChauveLeDoussalWiese2000a} calculation of 2000 points out a mechanism absent from \cite{NarayanFisher1993}, which invalidates the argument for $\zeta\equiv (4-d)/3$ at depinning, while the latter  remains valid in equilibrium. On the numerical side, it was shown 
that $\zeta>1$~\cite{LeschhornTang1993,RossoKrauth2001b},  though the 2-point function behaves as ($\ca A$ is a number)
\be
\left< \left[ u(x,t)-u(0,t)\right]^2\right> = \ca A |x|^2 L^{2\zeta-2},
\ee
which can easily be misinterpreted as $\zeta=1$. 
 The most precise current value is $\zeta=1.25$ \cite{FerreroBustingorryKolton2013,GrassbergerDharMohanty2016}, and probably exactly $\zeta=5/4$    \cite{Wiese2021,ShapiraWieseUnpublished}. 
Assuming   Ertasz and Kardar to keep $\zeta_\parallel|_{N>1}= \zeta\big|_{N=1}$, one may wonder whether they would keep the relation of  $\zeta_{\perp}=\zeta_{\parallel}-d/2$, or replace it as well.

In this paper we revisit the problem of depinning and flow of an elastic string in three dimensions using numerical simulations and analytical arguments to address     the above issues. Both for RB and RF  disorder we confirm the validity of the anisotropic scaling forms and of the  planar approximation for an extended set of exponents, together with the scaling relation $z_{\perp}=z_{\parallel}+1/\nu$ between dynamic exponents, as predicted in Ref.~\onlinecite{ErtasKardar1996}. Our results agree with the ``improved'' EK relation $\zeta_{\parallel}= \zeta|_{N=1}=1.25$, while our perpendicular roughness is $\zeta_\perp=0.5  \pm 0.001$. We shall argue below that this is simply  the thermal exponent of a $d$-dimensional elastic system, 
\be
\zeta_\perp= \frac{2-d}2.
\ee
We also calculate the force-force correlator, the avalanche-size and waiting-time distributions, as well as  the joint distribution of the two components of the center-of-mass jumps in the quasistatic regime. 
We  finally  unveil a skewed distribution of local parallel displacements at low velocities, and 
 show that the dynamical structure of the string at large scales has non-trivial features.

The large-velocity or fast-flow regime of elastic manifolds, usually considered trivial, present nevertheless interesting open issues. Using a perturbative analysis, Koshelev and Vinokur \cite{KoshelevVinokur1994} introduced the concept of shaking temperatures for moving vortex lattices by considering the effect of disorder as an effective thermal noise in the co-moving frame. Later on, for the same problem, it was numerically shown than an effective velocity-dependent temperature can be defined from generalized fluctuation-dissipation theorems~\cite{KoltonExartierCugliandoloDominguezGronbechJensen2002}.
A similar analysis can be performed for the fast-flow regime of elastic manifolds in general. For interfaces with harmonic elasticity in isotropic media, Functional Renormalization Group (FRG) calculations show that the large-scale structure is well described by the Edwards-Wilkinson equation with an effective temperature $T$ which,
at large velocities, vanishes as $T\sim 1/v$ \cite{ChauveGiamarchiLeDoussal2000}. In contrast,   a perturbative analysis for a single vortex line finds that the shaking temperature is zero in the longitudinal direction and vanishes as $1/v$ in the transverse direction~\cite{NattermannScheidl2000},   contradicting the FRG prediction \cite{ChauveGiamarchiLeDoussal2000} for interfaces
in combination with the  planar approximation. 
Moreover, a single monomer driven in a two-dimensional disordered medium which may be thought of as  the limiting case of very short correlation lengths along $x$, shows an effective temperature vanishing as $1/v^3$ in the parallel direction and $1/v$ in the transverse direction~\cite{kolton2006}, hence adding a third different prediction for the same property. 
Understanding these aparent discrepancies is an open issue we address here.

We find that in the comoving frame
the system can accurately be  described by   two effective temperatures, defined from a generalized fluctuation-dissipation relation. 
In particular, at large velocities $v$, for microscopic disorder of the random-bond (RB) type, this effective temperature vanishes as $T^{\perp}_{\tt eff}\sim 1/v$ in the transversal direction, and as $T^{\parallel}_{\tt eff}\sim 1/v^3$ in the  driving direction, while 
for microscopic disorder of the  random-field (RF) type  we find $T^{\parallel}_{\tt eff}\sim T^{\perp}_{\tt eff}\sim 1/v$.
Interestingly, applying the EK planar approximation in the fast-flow regime, the dependency on $v$ of the effective temperatures in the RB case is incompatible with the longitudinal ``shaking-temperature'' obtained from a perturbative analysis \cite{NattermannScheidl2000}, and the effective ``Edwards-Wilkinson'' temperature  $\sim 1/v$  predicted in Ref.~\onlinecite{ChauveGiamarchiLeDoussal2000}. Nevertheless, the latter prediction agrees with our result for  microscopic RF disorder. 
Finally, for RB  disorder we confirm that with increasing $v$ the aspect ratio of the string width changes from being elongated in the driving direction, to being elongated in the perpendicular direction, as qualitatively predicted by Scheidl and Nattermann ~\cite{NattermannScheidl2000}. 
In contrast,  the 
change of aspect ratio dissapears for RF disorder which crosses over from anisotropic fluctuations at intermediate velocities to isotropic ones at large velocities.

This paper is organized as follows: In Section \ref{sc:model} we describe the model, the protocol and  its numerical implementation. In section \ref{sc:properties} we define the observables of interest. 
Section \ref{sc:ana-result} summarizes the theoretical results confirmed in our simulations, and proposes a new  mechanism for the perpendicular direction. Due to its simplicity, we are able to  predict many observables analytically. 
In section \ref{sc:result} the main results are presented, followed by conclusions in  section \ref{sc:conclusion}.


\section{Model and observables}
\label{s:Model+Observables}

\subsection{Model and implementations}
\label{sc:model}
We study a driven one-dimensional elastic line (string), directed along the $x$-direction, in a three-dimensional space at zero temperature. It is parameterized by the time-dependent 2-dimensional vector field ${\bf u}(x, t) = \{u_{\parallel}(x, t), u_{\perp}(x, t) \}$ which measures its continuum displacements along the parallel and perpendicular directions with respect to the driving direction (Fig.~\ref{fg:esquema}).
The overdamped dynamics of the system is  
\begin{equation}
    \eta \partial_t {\bf u}(x, t) =  c \partial^2_x {\bf u}(x, t)+{\bf F}_p\big({\bf u}(x,t), x\big)+ {\bf f\big(\bf u(x,t)\big)}.
    \label{eq:langevin}
\end{equation}
Here $\eta$ and $c$ are the friction and elastic constants, ${\bf f}(\bf u)$ is the driving force and ${\bf F}_p({\bf u}, x)$ is a statistically isotropic random pinning force in the $\{u_{\parallel}, u_{\perp}\}$ plane. 
Eq.~(\ref{eq:langevin}) is a minimal model for a single vortex line in a type-II superconductor induced by a magnetic field pointing in the $x$-direction~\cite{Blatter1994,Tinkham2004}. The left-hand side then represent Bardeen-Stephen friction. The first term on the right-hand side is a harmonic approximation for the elastic tension of a single vortex, the second   the coupling to the  defects in the material, and the third     a force due to a uniform applied current or some other way to impose a uniform mean velocity. In this work, to impose a steady-state mean velocity along the $u_\parallel$ direction ${\bf v}=\{v,0\}$ we use
a moving parabolic trap, 
\begin{equation}
    {\bf f}[{\bf u}] = m^2 \left[ {\bf w}(t)-{\bf u} \right].
    \label{eq:drive_force}
\end{equation}
with ${\bf w}(t) = \{ w, 0\} = \{ v t, 0\}$ and 
$m\sim 1/L$, with $L$ the size of the string. With this driving protocol, both the instantaneous center-of-mass velocity and the driving force fluctuate, as they do in most eperiments. 
We   checked that the fixed-force ensemble (i.e.\ the one obtained by using a fixed force ${\bf f}=\{f,0\}$, and where the center-of-mass velocity fluctuates) yields equivalent results for large enough systems (see Appendix \ref{Force-controlled driving versus velocity-controlled driving}). 
Moreover,  driving with a confining potential allows us  to  measure the effective force-correlator, the  central object of  functional-renormalization-group calculations \cite{LeDoussalWiese2007,LeDoussalWieseMoulinetRolley2009,WieseBercyMelkonyanBizebard2019,terBurgWiese2020,terBurgBohnDurinSommerWiese2021,Wiese2021}.
 
For RB disorder, 
we consider a pinning force ${\bf F}_p=\{F^{\parallel}_p,F^{\perp}_p\}=\{-\partial_{u_\parallel}V,-\partial_{u_\perp}V\}$ derived from a pinning potential 
\be
\overline{ V({\bf u},x ) V({\bf u'},x')} = \delta(x-x') R(\bf u-\bf u'), 
\label{RB-disorder}
\ee
where $\overline{(...)}$ denotes average over disorder realizations and 
$R(u)$ is assumed to be a short-ranged function.

For RF disorder we use  
\begin{equation}
 \overline{ F^{\alpha}_p({\bf u}, x)F^{\beta}_p({\bf u}', x') } = 
 \delta_{\alpha \beta}\Delta( |{\bf u}-{\bf u'}|) \delta( x-x'),
 \label{eq:dis_correlator-RF}
\end{equation}
 and $\Delta(u)$ a short-ranged correlated function. 
Note that this type of disorder is not conservative, thus   may allow    a momomer $x$  to perform periodic orbits even in the presence of a finite viscosity. As an alternative, we could use \Eq{RB-disorder}, with $R(u)\sim |u|$. This could be done by generating Fourier modes for the disorder, multiply it with an appropriate kernel, and then   Fourier transforming. We will not do this since it is (i) prohibitively costly to implement, and (ii) not necessary close to the depinning transition. Indeed we   show below that as long as $m$ is sufficiently small, if a momomer position $\bf u(x)$ is updated, $u_\parallel(x)$ necessarily increases, forbidding periodic orbits.
Both potentials represent  isotropic disorder in the $\bf u$ plane (``model A'' in Ref.~\cite{ErtasKardar1996}).

We   discretize the system along the $x$-direction, such that $\partial^2_x {\bf u} \to  {\bf u}(x+1)+{\bf u}(x-1)-2 {\bf u}(x) $, and implement different uncorrelated pinning potentials on the resulting layers $x=1,\dots,L$, with periodic boundary conditions. We use dimensionless units with $\eta=c=1$, and $R(0)=\Delta(0)= \ca O(1)$.

For computational convenience two different implementations for RB disorder   were used, one for   finite velocities, and the other for   the quasi-static $v\to 0^+$ limit. 

(i) At finite velocities we directly solve Eq.~(\ref{eq:langevin}) using standard finite-difference techniques, and consider for each monomer $x$ the $V({\bf u})$ to be the sum of Gaussian wells randomly distributed according to a   Poissonian distribution. 

(ii) In the quasi-static limit the dynamics of the system becomes very slow,  and the direct integration of Eq.~(\ref{eq:langevin}) is computationally inefficient. We therefore use a cellular automaton. This algorithm is a generalization of the algorithm used in  Ref.~\cite{LeDoussalWiese2009} for a particle dragged in a 2-dimensional disordered landscape. 
Here we generate a random pinning-energy landscape in a cubic lattice, described by the integer indices $\{x,n_\parallel, n_\perp\}$, using a uniform $[0, 1)$ box distribution on each site, with uncorrelated energies at different sites. The 2-dimensional position of the monomers at each layer $x$ is discretized on this grid, and thus given by two integers, ${\bf u}(x,t)= \{n_\parallel(x),n_\perp(x)\}$. 
For a given configuration we compute the total energy of each monomer in  all nine nearest neighboring positions, keeping the monomers of neighboring layers fixed. This energy for layer $x$ is the sum of the random pinning energy, the quadratic potential well with curvature $m^2$ centered at $\{w, 0\}$, and the elastic   energy involving     layers $x+1$ and $x-1$. In an iteration, each monomer moves to the neighboring site with the lowest energy. This elementary step is performed synchronously for all $x$ and repeated $M$ times for a fixed $w$. In the quasistatic limit, $M$ is chosen $\infty$, i.e.\ $\bf w$ is increased only after the string gets   stuck. For $v>0$,   $M>1$ is  fixed (i.e.\ we let the string make $M$ steps for each step $\delta w$ in $w$). This protocol simulates a velocity that fluctuates around the mean velocity 
$v \sim \delta w/M$.
We  checked that this algorithm yields, in the low-velocity limit, the same universal results as the direct integration of the continuous displacements in the smooth pinning potential. 
Although most of our results are for RB disorder,  
we give some results, both for low and high velocities, for a RF type of disorder 
in order to detect possible dependencies on the microscopic disorder.

\subsection{Observables of interest}
\label{sc:properties} 
We now define some observables of interest. The steady-state force-velocity characteristics $f(v)$ is computed from the mean force in the steady state,
\begin{equation}
    f(v) = 
     m^2[v t-u_{\parallel}(t)].
    \label{eq:forcevelocity}
\end{equation}
Here $u_{\parallel}(t)$ is the parallel component of the center-of-mass position 
\begin{equation}
 {\bf u}(t) =  {\langle {\bf u}(x,t)\rangle}.
 \label{eq:cm}
\end{equation}
with $  \langle (...) \rangle:= L^{-1}\int \rmd x  (...)$. 
The velocity-force characteristics, and in particular the critical force $f_c := \lim_{v\to 0} f(v)$, have a well-defined scaling behavior when $m\to 0$~\cite{LeDoussalWiese2007}. For small $v$, we expect $v\sim [f(v)-f_{\rm c}]^\beta$.

To characterize the geometry of the string we define the mean quadratic widths in both directions
\begin{equation}
{W^2_\alpha}(t) = \overline{\langle ({u_\alpha}(x, t)-{ u_\alpha}(t))^2\rangle},
 \label{eq:w2}
\end{equation}
with $\alpha = \{ \parallel, \perp \}$.
We also define the local displacement components with respect to the center of mass 
$\delta u_\alpha(x,t):=u_{\alpha}(x,t)-u_{\alpha}(t)$
and consider their distributions 
\begin{eqnarray}
P^u_{\alpha}(s)=\overline{\langle \delta(s-\delta u_\alpha(x,t)\rangle},
\label{eq:localdisppdf}
\end{eqnarray}
as well as the joint distribution
\be
P^u(s_\parallel,s_\perp)=
\overline{\langle \delta(s_\parallel-\delta u_\parallel(x,t))
\delta(s_\perp-\delta u_\perp(z,t)) 
\rangle}.
\label{eq:localdispjointpdf}
\ee
We define the anisotropic structure factor as
\begin{equation}
{\cal S}_\alpha(q, t) = \overline{\left| \langle u_\alpha(x, t) e^{- i q x}\rangle\right|^2},
 \label{eq:sq}
\end{equation}
where $\alpha = \{ \parallel, \perp \}$ and $q = 2 \pi n / L$ with $n = 0, \dots , L/2+1$.

In the quasistatic $v\to 0^+$ regime,  we consider the center of mass 
\be
{\bf u}(t) \to  {\bf u}(w)
\ee
 as a function of $w = v t$ instead of $t$. (Which object is referred to will be clear from the context.) 
  For this we let the string   relax completely before shifting  $w \to w+\delta w$. To study avalanches, we compute the PDF 
\begin{eqnarray}
P_{\alpha}(S)=\overline{\langle \delta(S- L u_{\alpha}(w+\delta w)+L u_{\alpha}(w)) \rangle}.
\label{eq:pdfavalanches}
\end{eqnarray}
Discontinuous jumps of $u_\alpha(w)$ versus $w$ are identified as shocks or avalanches. They occur at a discrete set of points, $w=w_i$, 
producing a finite set of jumps of sizes $S^{\alpha}_i=L(u_\alpha(w_i+\delta w)-u_\alpha(w_i))$, each one measuring the effective number of sites involved in the avalanche.
The ``waiting time'' between consecutive jumps is defined as $\delta w_i = w_i - w_{i-1}$. Although $\delta w_i$ is not   a time, it yields the   true waiting times $\delta w_i/v$ between   successive avalanches in the $v\to 0^+$ limit. 
We are interested in its distribution 
\begin{equation}
    P^w(s)= \frac{1}{N}\sum_{i=1}^{N} \delta(s-\delta w_i)
\label{eq:waitingtimedist}
\end{equation}
for a large sequence of $N\gg 1$    avalanches.

Finally, we   consider the steady-state force-force correlation function in the driving direction, defined as~\cite{LeDoussalWiese2007,RossoLeDoussalWiese2007},
\begin{equation}
    \Delta_{\alpha \beta}(w-w') = L^{d}\overline{f_\alpha (w)f_\beta(w')}^c,
\label{eq:correlator}
\end{equation}
where 
$f_\parallel(w) = m^2 [w-u_\parallel(w)]$ is the (spatially averaged) parallel force for each metastable state at fixed $w$, and 
$f_\perp(w) =  - m^2 u_\perp(w)$ is the perpendicular force. 
Due to symmetry, the cross correlator $\Delta_{\parallel \perp}(w)$ vanishes, and we are left with two (a priori) independent correlation functions. 
From the correlation theorem these functions can be computed using the Fourier transform as $\Delta = \mathscr{F}^{-1}\{ |\hat{f}|^2\}$, with 
$f \xrightarrow{\mathscr{F}} {\hat f}$.
This quantity is a central object of the theory and can be compared quantitatively  with FRG calculations and experiments \cite{RossoLeDoussalWiese2007,LeDoussalWiese2007,LeDoussalWiese2009,LeDoussalWieseMoulinetRolley2009,WieseBercyMelkonyanBizebard2019,terBurgWiese2020,terBurgBohnDurinSommerWiese2021,Wiese2021}.

\section{Analytical Results}
\label{sc:ana-result}

\subsection{Analytical results for the planar approximation}
\label{Analytical results for the planar approximation: Improved EK model}

\begin{figure}
\centering
\includegraphics[width=.95\columnwidth]{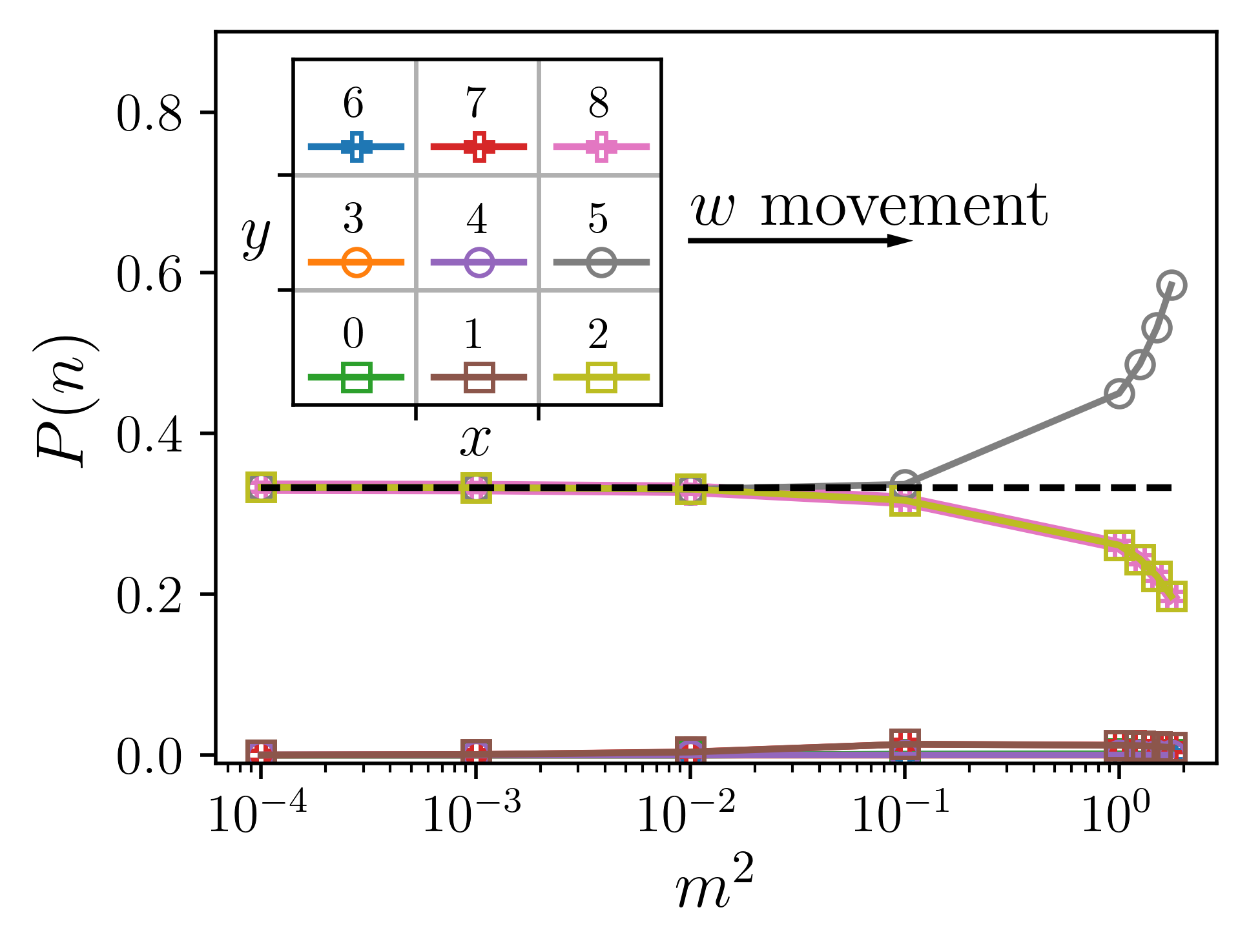}
\caption{Single monomer jump probability for the eight possible directions from position $4$, in the cellular automata model.}
\label{fig:singlemonomerjumps}
\end{figure}
EK consider a range of models, of which we focus on the simplest one, their {\em model A}.
Omitting some additional terms, and changing to our notations, the essence of 
EK's  Eqs.~(2.10a)-(2.10b) is
\bea
\label{2.10a}
\eta \partial_t u_\parallel(x,t) &=& c \nabla_x^2 u_\parallel(x,t) + F + f_\parallel\big(x,{\bf u}(x,t)\big), \\ 
\label{2.10b}
\eta \partial_t u_\perp(x,t) &=& c \nabla_x^2 u_\perp (x,t)+ F + f_\perp\big(x,{\bf u}(x,t)\big) .~~~
\eea
EK then say in Eq.~(3.2) that \Eq{2.10a} can be reduced to 
\be
\eta \partial_t u_\parallel(x,t) = c \nabla_x^2 u_\parallel(x,t) + F + f'\big(x,u_\parallel(x,t)\big). 
\ee
EK argue that as long as $f_\parallel(x,{\bf u})$ is short-ranged correlated, so will be 
 $f'(x,u_\parallel)$.
 We agree with their analysis, and would like to justify it as follows: If we assume the pinning energies to be bounded, so will be the possible forces. The critical force $f_{\rm c}$ becomes large for $m\to 0$, thus (in the cellular automaton model) each monomer can choose among the three forward neighbors, while the three backward neighbors as well as the two sideway neighbors can be neglected. The locally chosen force (including the elastic forces)   is the maximum force among the three possible choices, i.e.\ the maximally possible descent in energy. 
 That this image is correct, and each monomer only moves forward, is shown in Fig.~\ref{fig:singlemonomerjumps}.

 As the minimum of three (correlated) random variables,  we expect it to have (roughly) the same statistics as one of them (say the middle one). Thus it  is   short-ranged, with a minimal history dependence.
 As an immediate consequence, 
\be\label{theory:zeta-parallel}
\zeta_\parallel\Big|_{N\ge 2} = \zeta\Big|_{N=1}.
\ee
For the line ($d=1$) this yields \cite{GrassbergerDharMohanty2016,Wiese2021,ShapiraWieseUnpublished}
\be\label{theory:zeta-parallel:d=1}
\zeta_\parallel\Big|_{d=1} = \frac54.
\ee
As for $N=1$, the time-integrated response function is protected by STS 
\cite{NarayanFisher1993,ErtasKardar1996}
\be
\zeta_\parallel + \frac1\nu = 2.
\ee
EK then argue that for the mean forces
\be
{\bf F}({\bf v}) = F(v) \hat v \, \qquad \hat v := \frac {\bf v}  v  
\ee
one has 
\bea
\frac{\partial}{\partial v_\parallel} {\bf F}({\bf v}) &=& \frac {\partial F(v)}{\partial v}  \left({ 1 \atop 0 }\right),  \\
\frac{\partial}{\partial v_\perp} {\bf F}({\bf v}) &=&  \frac {F(v)}{v} \left({ 0  \atop 1 }\right) .
\eea
The only non-vanishing components are 
\bea
\frac{\partial}{\partial v_\parallel} {\bf F}_\parallel ({\bf v}) &=& \frac {\rmd F(v)}{\rmd v},    \\
\frac{\partial}{\partial v_\perp} {\bf F}_\perp ({\bf v}) &=&  \frac {F(v)}{v}  . 
\eea
Using that  $v\sim (F-F_c)^\beta$, 
$\beta = \nu (z_\parallel-\zeta_\parallel)$, 
this leads to the scaling of time scales in the two directions 
\bea
\tau_\parallel &\sim& \frac {\xi^2} { \frac {\rmd F(v)}{\rmd v} } \sim \xi ^{2+ (\beta-1)/\nu} = \xi ^{z_\parallel  },\\
\tau_\perp &\sim &\frac {\xi^2} { \frac {  F(v)}{   v} } \sim \xi ^{2+ \beta/\nu} = \xi ^{z_\perp  }.
\eea
As a consequence 
\be\label{rel:zperpfromzparallel}
z_\perp = z_\parallel + \frac1\nu \equiv  z_\parallel + 2-\zeta_\parallel.
\ee
While EK follow NF \cite{NarayanFisher1993} in introducing a mean-field theory, and then expanding around it, this was not necessary in the field-theoretic work of Refs.~\cite{LeDoussalWieseChauve2003,LeDoussalWieseChauve2002,ChauveLeDoussalWiese2000a}. Here we follow the latter approach. 
What is yet missing is a   treatment of the transversal directions. We write this in differential form as 
\be
\eta \rmd u_\perp(x,t) = \left[ c \nabla^2 u_\perp(x,t) - m^2 u_\perp(x,t)\right] \rmd t + \rmd f_\perp (x, u_\parallel).
\label{33}
\ee
The last term can be rewritten in different ways:
\bea
\rmd f_\perp (x, u_\parallel) &=& \partial_{u_\parallel}  f_\perp (x, u_\parallel) \rm d u_\parallel \nn\\
&=& \partial_{u_\parallel}  f_\perp (x, u_\parallel) \partial_t u_\parallel(x,t) \rmd t .
\eea
We now suppose that $\partial_{u_\parallel} f_\perp (x, u_\parallel) $ is a white noise in $u_\parallel$, s.t.  
\bea\label{35}
\rmd f_\perp (x, u_\parallel) &=& \sqrt \sigma \kappa(x,u_\parallel) \rmd u_\parallel ,\\
\left< \kappa(x,u) \kappa(x',u') \right> &=& \delta^d(x-x')\delta(u-u').
\eea
Alternatively, we can use a white noise in time, 
\bea\label{EOM-trans}
\rmd f_\perp (x, u_\parallel) &=& \sqrt {\sigma  \dot u_\parallel(x,t)}\zeta(x,t) \rmd t ,\\
\left< \zeta(x,t) \zeta(x',t') \right> &=& \delta^d(x-x')\delta(t-t').
\label{EOM-trans2}
\eea
Let us derive some consequences of these equations. First of all, consider the motion of the center of mass of $u_\perp(x,t)$. We claim it satisfies the stochastic differential equation
\bea
\partial_t   u_\perp(t) &:=& 
\partial_t \frac 1 {L^d}\int_x u_\perp(x,t) \nn\\
& =& - m^2 u_\perp(t) + {L^{-d/2} }\sqrt{\sigma \dot u_\parallel(t)} \zeta(t), \\
\left< \zeta(t) \zeta(t') \right> &=& \delta(t-t').
\eea
To prove the equivalence to \Eqs{EOM-trans}-\eq{EOM-trans2}, one first checks the second moment of the driving term, 
\bea
&&\left<   \frac {1 } {L^d}\int_x  \sqrt{\sigma \dot u_\parallel(x,t)} \zeta(x,t) \; \frac {1} {L^d}\int_{y}  \sqrt{\sigma \dot u_\parallel(y,t')} \zeta(y,t')\right>  \nn\\
&&\qquad  = \frac 1 {L^{2d}} \int_x \sigma \dot u(x,t) \delta(t-t') \equiv \frac{\sigma}{L^d} \dot u(t) \delta(t-t').\qquad 
\eea
Second, one uses that the process is Gaussian, implying that this is the only cumulant one has to check. 
In the same way, one derives that $u_\perp(t)$ is a Gaussian process with variance (in the limit of $m\to 0$), 
\be\label{37}
\frac12 \left< \left[ u_\perp(t)-u_\perp(t') \right]^2 \right> = \frac{\sigma}{L^d} |u_\parallel(t) - u_\parallel(t')|.
\ee
At $m>0$,  the center of mass   performs an Ornstein-Uhlenbeck process as a function of $w$. As a result  we obtain, see appendix \ref{Correlations of an Ornstein-Uhlenbeck process}
\bea
\Delta_{\perp\perp}(w) &:=& m^4 L^d \overline{u_\perp(w'+w) u_\perp(w')}^c\nn\\
& = &\sigma  m^2 \rme^{-m^2 w}.
\label{eq:Dperpperp}
\eea
\Eq{37} immediately implies relations \eq{eq:smperp} and $s_\perp \sim \sqrt{ s_\parallel}$ observed in Fig.~\ref{fig:jointavalanche}.
More precisely
\be\label{S-perp-from-S-parallel}
  \left < S_\perp^2 \right>\Big|_{S_\parallel} = 2 \sigma  S_\parallel.
\ee 
Let us finally address the question of the roughness exponent $\zeta_\perp$. Differential equations \eq{33} and \eq{35}  yield
\bea
\frac 12 \left< \left[ u_\perp(x,u_\parallel) - u_\perp(x',u_\parallel) \right]^2 \right> &=& \sigma |x-x'|^{ 2\zeta_\perp},\\
\zeta_\perp &=& \frac{2-d}2. \label{theory:zeta-perp}
\eea
The parallel coordinate $u_\parallel$ acts as a local time. The tricky point is whether this local time can be used as a global time. Our simulations show that this is indeed the case in $d=1$ \footnote{We learned from L.~Ponson that the perpendicular roughness for fracture in $d=1$ is $\zeta_\perp\simeq 0$, consistent with the ``thermal'' exponent for LR elasticity $\zeta_\perp = (1-d)/2$.}.
Even if this cannot be rigorously asserted, we  can show that the 2-point function is indeed equivalent to the thermal one:   
The thermal equilibrium is characterized by a probability distribution of a monomer with coordinate $u$, given positions $u_{i}$, $i=1,...,n$ of the $n$ neighbors, 
\be\label{P(u|...)}
P(u|  u_1,...,u_n) = \exp\left( - \frac{\sum_{i=1}^n (u-  u_i)^2 + m^2 u^2}{2 T_{\rm eff}} \right). 
\ee
It is achieved by a Langevin equation for a selected monomer with coordinate $u$, here written in a discretized form and a time step $\delta t=1$ 
\bea\label{Langevin-eq-4-eq}
 u(t+1) -u(t) &=&  \sum_{i=1}^n [  u_i-u (t)] - m^2 u(t)+ \sqrt {2T_{\rm eff}} \eta_t   \ ,\qquad  \\
\left < \eta_t \eta_{t'}\right>& =& \delta_{t,t'}.
\eea
(This is a Kronecker-$\delta$.)
If the string (or manifold) is in thermal equilibrium, then running \Eq{Langevin-eq-4-eq} for the selected  monomer  ensures that it remains in equilibrium. If it is not in equilibrium, then running \Eq{Langevin-eq-4-eq} for  the selected  monomer   ensures that it will get   into equilibrium with its neighbors according to the measure \eq{P(u|...)}. Running the same equation for each monomer in turn, and repeating the procedure for all monomers, one ensures that   thermal equilibrium is reached for the whole string.

Let us finally comment on EK and their result that $\zeta_\perp = \zeta_\parallel -d/2$. There are several assumptions in their calculation which need to be questioned: The first and strongest is that $\Delta(u_\parallel,u_\perp)$ only depends on $u_\parallel$. The simplest interpretation is that $\Delta(u_\parallel,u_\perp)$ is constant in the perpendicular direction, or at least  extremely long-ranged correlated, violating basic assumptions of the system, and the particle simulation in Ref.~\cite{LeDoussalWiese2009}. The next-to-simplest assumption is that $\Delta(u_\parallel,u_\perp)= \Delta(u_\parallel)\delta(u_\perp)$, i.e.\ extremely short-ranged correlated. 
In this case, however, the transversal dependence needs to appear in the FRG equations, which it does not. In this context let us remind that when one starts with short-ranged correlated disorder for $N=1$ in a simulation, one can clearly sees that $\Delta(w)$ acquires a finite range as $m$ is decreased. This is correctly described by the FRG. 

Next, we are doubtful about the appearance of $z_\parallel$ in EK's Eq.~(6.16): We believe that since $\Delta(u_\parallel,u_\perp)$ is a static quantity, its renormalization cannot contain information about the dynamical exponents. Rather, the rescaling term in Eq.~(6.16) should reduce to $(\epsilon - 2\zeta_\perp)C_\perp(u)$, similar to what happens in  Eq.~(6.15), where the rescaling term simplifies to   $(\epsilon - 2\zeta_\perp)C_\perp(u)$. With this in mind, we can rewrite Eq.~(6.16) as 
\bea
\zeta_\perp &=& \frac {4-d}2 + z_\parallel - z_\perp  \nn\\
&=& \left\{ 
\begin{array}{lcc}
 \frac {4-d}2 &\mbox{~if~}& z_\perp \to z_\parallel
\\
\rule{0mm}{3ex} \zeta_\parallel - \frac {d}2   &\mbox{~if~}& z_\perp \to z_\parallel + 
\frac1\nu
\end{array}
\right .  
\eea
The first is the Larkin (dimensional-reduction) result, expected if the disorder is constant in space. 
The second is the one given by EK, which contradicts our    result \eq{theory:zeta-perp}, and our numerical simulations, see \Eq{eq:roughnessexponentsperp}, while \Eqs{theory:zeta-perp}  and   \eq{eq:roughnessexponentsperp} agree.

\subsection{Analytical results in the fast-flow regime}
The standard argument for the amplitude of the 2-point function in the fast-flow regime is constructed as follows:
First, one considers the 2-point function at equal times set to 0, 
\be\label{60}
 \left<  u(q,0) u(-q,0) \right> = \int_{t>0}\int_{t'>0} \Delta\big(v(t -t')\big) \rme^{-(q^2+m^2)(t+t')}.
\ee
For RF disorder, since $\Delta(w)$ monotonically and faster than exponentially decays to 0 for increasing $w$, one can approximate for large $v$
\bea
\label{52}
\Delta(v t ) &\simeq& 2 \ca A \delta (vt ) \equiv \frac{2\ca A}{v}\delta(t), \\
\ca A &:=&  \int_0^\infty \Delta(w) \,\rmd w .
\label{53}
\eea
With this, \Eq{60} reduces to 
\bea
 \left<  u(q,0) u(-q,0) \right> &\simeq&  \frac {2 \ca A} v \int_{t>0}  \nn  \rme^{-2 (q^2+m^2)t }\\
  &=&  
  \frac{T^{\rm RF}_{\rm eff}}{q^2+m^2 }, \qquad T^{\rm RF}_{\rm eff}=\frac {\ca A} v. 
\eea
For RB disorder, the situation is different as  $\Delta(w) = - R''(w)$, where now $R(w)$ is  fast  decaying.
As a result 
\bea
\Delta\big(v (t-t') \big) &=& - R''\big(v (t-t')\big) \nn\\
&=&  \frac1{v^2} \partial_t \partial_{t'} R \big(v (t-t')\big)  \nn\\
&\simeq&   - \frac{2 \ca B}{v^3}\delta ''(t), \\
\ca B &=&\int  _0^\infty R(w) \rmd w.
\eea
A numerical simulation   shows that thermal noise correlated as $ \delta''(t)$ leads to a non-vanishing variance for each site, uncorrelated between neighboring sites. It does not contribute to the structure factor. 
Thus the leading contribution should come from \Eqs{52}-\eq{53}, where the RG-flow  from RB to RF  is cut such that below, in section \ref{Crossover to the fast-flow regime}, we observe $\ca A\simeq 1/v^2$. A proper theoretical explanation remains outstanding. 

For a single over-damped particle in a one-dimensional force field with a finite correlation length, we solve the problem  analytically in section \ref{Single monomer diffusion}, yielding in agreement with the above scalings  $T^{\rm RB}_{\rm eff}\sim 1/v^3$ and $T^{\rm RF}_{\rm eff}\sim 1/v$ at large velocities.

\section{Numerical Results}
\label{sc:result}

\begin{figure}
    \centering
    \includegraphics[width=.95\columnwidth]{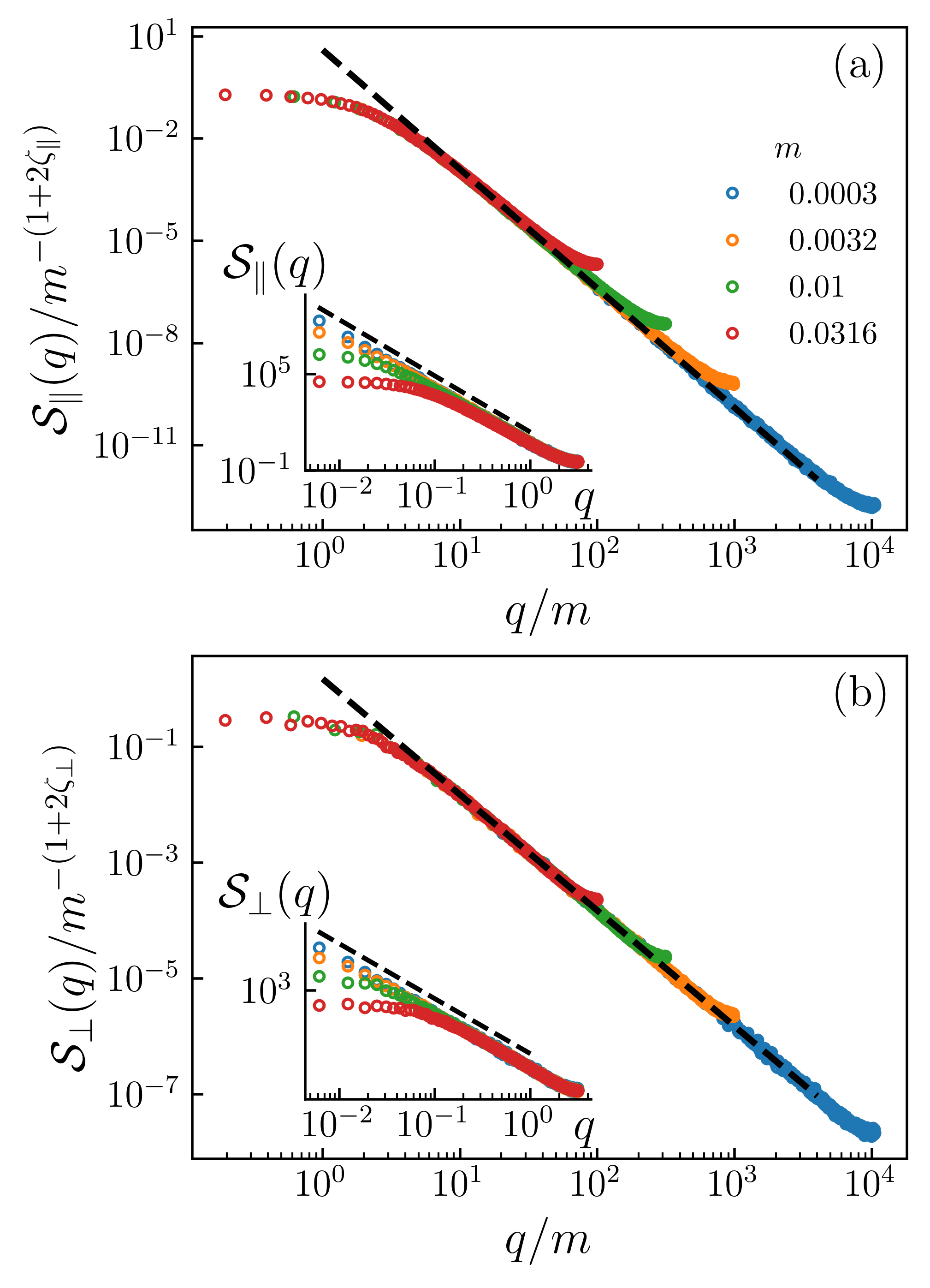}
    \caption{
    Scaled structure factors for quasi-static motion as a function of confinement strength $m$ for the parallel (a) and transverse (b) directions. Both directions display a self-affine structure up to the confinement length $L_m\sim m^{-1}$. 
    Dashed lines in (a) and (b) correspond to power-laws ${\cal S}_\parallel(q)\sim q^{-(1+2\zeta_\parallel)}$ and ${\cal S}_\perp(q) \sim q^{-(1+2\zeta_\perp)}$, yielding the depinning roughness $\zeta_\parallel \approx 1.25$ and $\zeta_\perp \approx 0.5$ respectively. 
    Insets show raw data for each case.
    }
    \label{fig:Sq_para_vs_m}
\end{figure}

\subsection{Steady state in the quasistatic regime}
\label{sc:Steady-state in the quasistatic regime}

We start analyzing the geometry of the elastic string in the quasistatic steady-state regime, using the cellular automaton. 
In Fig.~\ref{fig:Sq_para_vs_m} we show the structure factors for the parallel (a) and perpendicular (b) directions, as a function of the parameter $m$  in Eq.~(\ref{eq:drive_force}). 
Since $v\to 0^+$, the depinning correlation length $\xi$  is not set by the distance to $f_c$, but by the confining potential, $\xi_m \approx \frac 1 m$.
Indeed, in both cases we   observe that the string becomes flat beyond  $\xi$, while below (but above the lattice constant here set to 1) a self-affine random-manifold regime is observed. 
The inset of the two figures validates the scalings  
\begin{eqnarray}
{\cal S}_{\alpha}(q)\sim q^{-(1+2\zeta_\alpha)}
G_\alpha(q \xi_m),
\end{eqnarray}
with $G_{\alpha}(y)=\text{const}$ 
for $y \ll 1$ ($\xi_m \gg 1/q$) and $G_{\alpha}(y)=y^{1+2\zeta_\alpha}$ for $y\gg 1$ ($\xi_m \ll 1/q$). 
These results also imply that $W^2_\alpha \sim \xi_m^{2\zeta_\alpha}\approx L^{2\zeta_\alpha}$ if $mL$ is kept constant when increasing $L$. 
The roughness exponents in the two directions (see dashed lines) are different 
\begin{eqnarray}
\zeta_\parallel &=& 1.25 \pm 0.01,
\label{eq:roughnessexponentsparallel}
\\
\zeta_\perp &=& 0.50 \pm 0.01.
\label{eq:roughnessexponentsperp}
\end{eqnarray}
These results are compatible with the earlier result of Ref.~\cite{KoshelevVinokur1994}.
We find a value of $\zeta_\parallel\approx 5/4$, indistinguishable from the roughness exponent of the driven one-dimensional quenched Edwards-Wilkinson interface obtained from numerical simulations \cite{FerreroBustingorryKolton2013,GrassbergerDharMohanty2016}, and consistent with 2-loop functional renormalization group calculations \cite{LeDoussalWieseChauve2002,ChauveLeDoussalWiese2000a}.
In the perpendicular direction the exponent is the same as for a moving line in presence of thermal noise.
Both results are in agreement with our theoretical predictions in \Eqs{theory:zeta-parallel:d=1} and \eq{theory:zeta-perp}.

 The statistical tilt symmetry applied to the parallel direction   implies that
\begin{equation}
\nu=\frac{1}{2-\zeta_\parallel}=1.33\pm 0.02.
\label{eq:nufromsts}
\end{equation}
These results are consistent with the planar approximation of Ertas and Kardar \cite{ErtasKardar1996}, and with their 1-loop analysis. 
As we discussed in section \ref{Analytical results for the planar approximation: Improved EK model}, they are inconsistent with the higher-order results. In particular, they contradict EK's   $\zeta_\perp=5\zeta_\parallel/2-2$ \cite{ErtasKardar1996}.

As the model of \cite{ErtasKardar1996} and its numerical implementation are equivalent to ours, the numerical discrepancy can be explained by noting that the $x$ and $L$ dependence of the correlation function 
\be
B_\alpha(x,L) := \overline{\langle [u_\alpha(x,t)-u_\alpha(0,t)]^2 \rangle}\Big|_L
\ee
used in Ref~\cite{ErtasKardar1996}  cannot detect roughness exponents larger than one for a fixed sample size $L$.
As first observed in Ref.~\cite{LeschhornTang1993} for $N=1$ ($\ca A$ is a number)
\begin{eqnarray}
B (x,L) &\simeq & \ca A L^{2\zeta-2} x^2 \mbox{~~if~} \zeta >1 ,\\
B(x,L) &\simeq& \ca A x^{2\zeta} \mbox{~~if~} \zeta <1 .\end{eqnarray}
Taking into account that  $\zeta_\parallel>1$ and $\zeta_\perp<1$, the first line applies to $\zeta_\parallel$, and the second to $\zeta_\perp$. 
Therefore, if we use the scaling $B(x)\sim x^{2\zeta}$ to determine $\zeta$, its value is only correct when $\zeta<1$, but saturates at $1$ whenever $\zeta>1$ (In practice, it is even difficult to see the exponent 1, and one tends to measure something slightly smaller \cite{Wiese2021}). We believe that this is what Ertas and Kardar~\cite{ErtasKardar1996} saw in their simulation.

Besides validating the planar approximation, the results of Eq.~(\ref{eq:roughnessexponentsparallel}) imply that whenever the harmonic elasticity is an approximation for a more complicated elasticity, the model becomes physically unrealistic for large enough sizes $L$ because local slopes $\overline{\langle (\rmd u_\parallel/\rmd x)^2 \rangle} = B(1,L) \simeq L^{2\zeta_\parallel-2}$ diverge with $L$ ~\cite{LeschhornTang1993}. This   motivates one to either  include anharmonic corrections to the elasticity, or other effects such as overhangs and pinch-off loops to the model. 
This not withstanding, the predictions of Eq.~(\ref{eq:roughnessexponentsparallel}) may describe   the geometry at intermediate scales, below a putative crossover to a different regime. 
This scenario is present in recent experiments on creep \cite{GrassiKoltonJeudyMouginBustingorryCuriale2018} and depinning \cite{AlbornozFerreroKoltonJeudyBustingorryCuriale2021} displaying super-rough magnetic domain walls in ultrathin ferromagnetic films.

\begin{figure}
    \centering
    \includegraphics[width=.95\columnwidth]{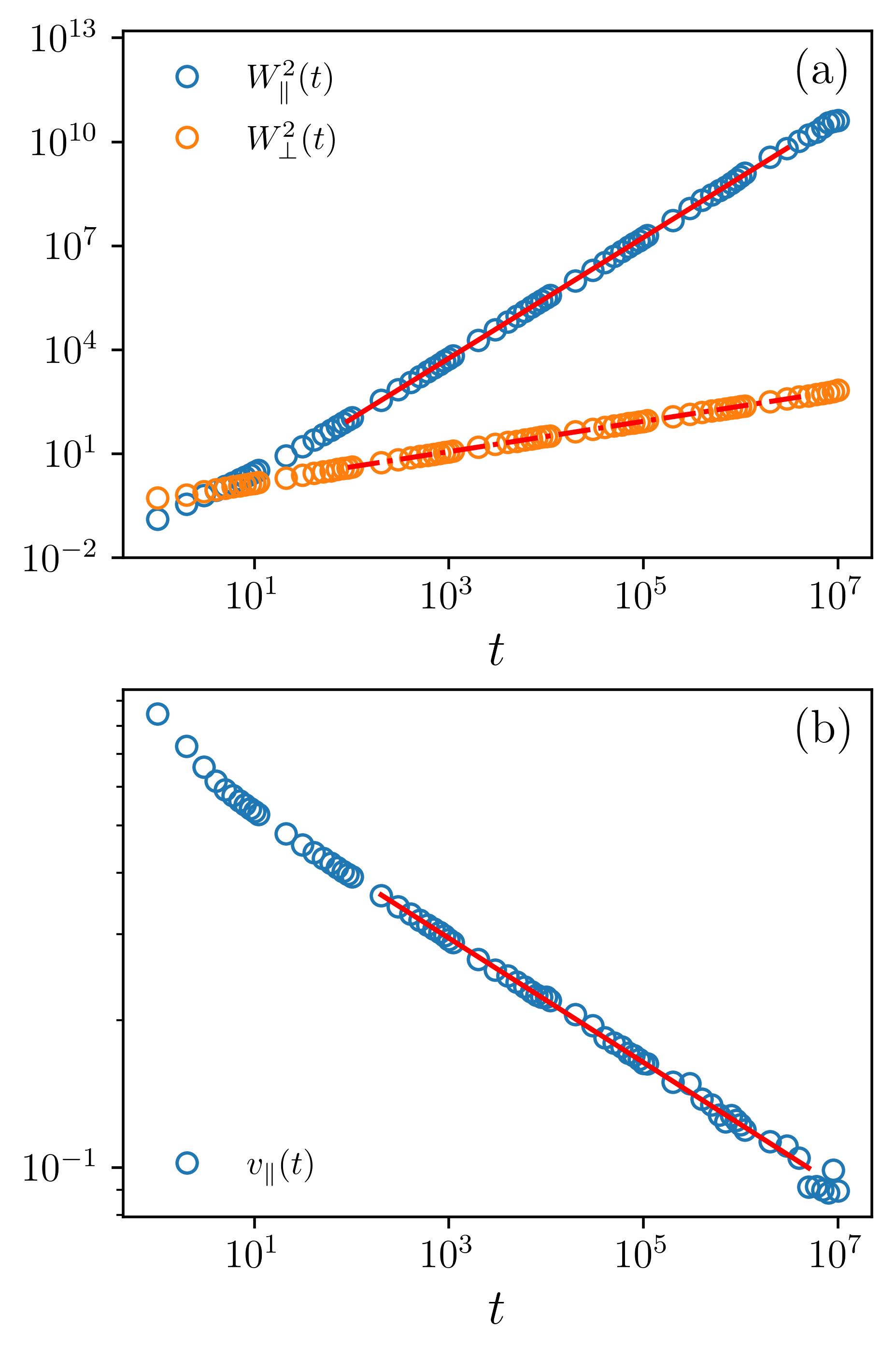}
    \caption{
    Non-steady relaxation of the roughness (a) and velocity (b) from a flat initial configuration at constant force $f = f_c \approx 0.765$, for a string with $L = 2^{18}$ monomers, averaged over 8 disorder realizations. Red lines show power-laws fits in the universal relaxation regime. In (a) we fit $W^2_\alpha(t)\sim t^{2\zeta_\alpha/z_\alpha}$ with $\alpha=\parallel,\perp$ to obtain $z_{\parallel} = 1.43 \pm 0.01$, $z_{\perp} = 2.27 \pm 0.05$. (b) From the relaxation of the mean velocity $v \sim t^{-\beta/\nu z_\parallel}=t^{-(2-\zeta_\parallel)\beta / z_\parallel}$ we obtain $\beta = 0.24 \pm 0.01$.
    }
    \label{fig:nonsteadyw2yv}
\end{figure}
\subsection{Relaxation from a flat initial condition in the quasistatic regime}

In the quasistatic protocol we start with a flat initial condition  ${\bf u}(z,t)=\{u_0,0\}$ such that $m^2(w-u_0)= f_c$. Since the flat string is uncorrelated from the disorder we also have $v(t=0)=f_c$.
As     observed for interfaces relaxing at depinning~\cite{FerreroBustingorryKolton2013},
before reaching the steady state  the string is in a  universal transient regime, which yields information about the   critical exponents of the steady-state depinning transition. In particular   $u_\perp$ and $u_\parallel$ each evolves with a  different dynamical length, $\ell_\parallel(t) \sim t^{1/z_\parallel}$, and $\ell_\perp(t) \sim t^{1/z_\perp}$, controlling the relaxational dynamics.

Since the interface is initially flat,   $W^2_\alpha(t=0)=0$. After a non-universal microscopic transient the global width reaches a universal transient regime described by $W^2_\alpha(t)\sim \ell_\alpha(t)^{2\zeta_\alpha}$, and hence
\begin{equation}
    W^2_\alpha(t)\sim t^{2\zeta_\alpha/z_\alpha}.
    \label{eq:w2nonsteady}
\end{equation}
This holds as long as  $\ell_\alpha(t) < \xi_m$.
On the other hand,  at long times it saturates 
as $W^2_\alpha(t)\sim \xi_m^{2\zeta_\alpha}\sim L^{2\zeta_\alpha}$ (the last relation holds provided $mL$ is kept fixed).
The power-law regime of Eq.~(\ref{eq:w2nonsteady}) is confirmed in Fig.~\ref{fig:nonsteadyw2yv}(a) were we show the evolution of $W^2_\parallel(t)$ and $W^2_\perp(t)$.
Using the known values of 
$\zeta_\parallel$ Eq.~(\ref{eq:roughnessexponentsparallel}) and $\zeta_\perp$ Eq.~(\ref{eq:roughnessexponentsperp}) and by fitting in the appropriate (intermediate)  range indicated in red, we get
\begin{eqnarray}
z_\parallel &=& 1.43  \pm 0.01,
\label{eq:dynamicalexponentspara}
\\
z_\perp &=& 2.27  \pm 0.05.
\label{eq:dynamicalexponentsperp}
\end{eqnarray}
This 
validates the relation  \eq{rel:zperpfromzparallel} 
\begin{equation}
z_\perp=z_\parallel+1/\nu,
\label{eq:dynexprelation}
\end{equation}
predicted in Ref.~\cite{ErtasKardar1996}, as 
$z_\parallel+1/\nu-z_\perp = -0.09 \pm 0.08$. 
For reference, the analytical values proposed in \cite{GrassbergerDharMohanty2016} combined with the scaling relation \eq{eq:dynexprelation} are
\bea
z_\parallel& =& \frac{10}7 = 1.42857,\\
z_\perp &=& \frac{61}{28} = 2.17857.
\eea
In the same universal regime  where 
Eq.~(\ref{eq:w2nonsteady}) holds,
the parallel center-of-mass velocity reaches a universal transient regime, where it vanishes as $v(t)\sim \ell_\parallel(t)^{-\beta/\nu}$, and hence
\begin{equation}
    v(t)\sim t^{-\beta/\nu z_\parallel}.
\end{equation}
In Fig.~\ref{fig:nonsteadyw2yv}(b) we show the fit to this regime and obtain, knowing $\nu$ from Eq.~(\ref{eq:nufromsts})
and $z_\parallel$ from Eq. (\ref{eq:dynamicalexponentspara}), 
\begin{equation}
    \beta= 0.24 \pm 0.01 .
    \label{eq:beta}
\end{equation}
This is indistinguishable from the result for the one-dimensional interface \cite{FerreroBustingorryKolton2013}. It  is compatible with the exact relation \cite{GrassbergerDharMohanty2016}
\begin{equation}
\beta=\nu(z_\parallel-\zeta_\parallel)   = \frac5{21} =  0.238095...
\end{equation}
To conclude, these results  validate the exponent relations due to the planar approximation~\cite{ErtasKardar1996} in the modified form of section \ref{Analytical results for the planar approximation: Improved EK model}.

\begin{figure}
    \centering
    \includegraphics[width=.95\columnwidth]{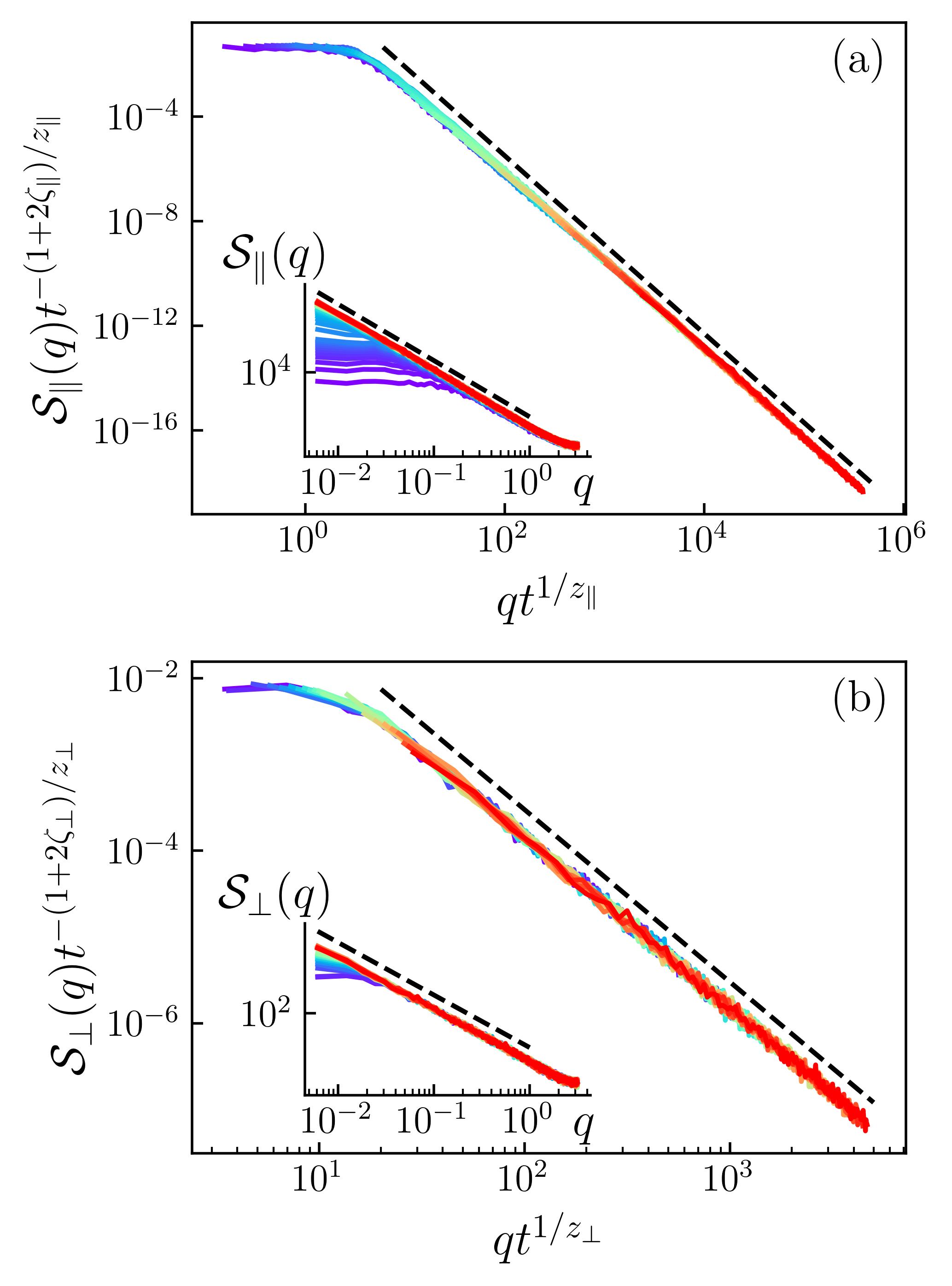}
    \caption{
    Relaxation of the structure factors
    for $L=1024$ monomers and $m^2 = 10^{-6}$, showing  ${\cal S}_\alpha(q)\sim q^{-(1+2\zeta_\alpha)}G_{\alpha}(q t^{1/z_\alpha})$ scaling, in both directions $\alpha=\parallel,\perp$. Dashed lines indicate power laws and insets show corresponding raw data.
    (a) Rescaled ${\cal S}_\parallel(q)$ from $t = 10^3$ (violet) to $t = 5 \times 10^7$  (red). The dashed line corresponds to $z_{\parallel} = 1.43$ and $\zeta_{\parallel} = 1.25$. 
    (b) Rescaled $S_\perp(q)$ from $t = 10^6$ (violet) to $t = 5\times 10^7$ (red). The dashed line corresponds to $z_{\perp} = 2.18$ and $\zeta_{\perp} = 0.5$.
}
    \label{fig:Sq_dinamical_para}
\end{figure}

Finally, a   detailed geometrical view of the  relaxation can be obtained from the structure factors. Using the exponents   obtained 
in Fig.~\ref{fig:Sq_dinamical_para} we show that when $\ell_\alpha(t)<\xi_m$ the evolution of the structure factors  accurately follows the scaling
\begin{eqnarray}
{\cal S}_\alpha(q,t)\sim q^{-(1+2\zeta_\alpha)} F_\alpha(q \ell_\alpha(t)), 
\end{eqnarray}
with $F(x)=\text{const}$ for $x \ll 1$, and $F(x)=x^{(1+2\zeta_\alpha)}$ for $x \gg 1$. Therefore, the string progressively becomes self affine with exponents $\zeta_\alpha$  up to the corresponding  scales $\ell_\alpha(t)$. For larger distances the  memory of the flat initial condition is preserved.

\subsection{Depinning avalanches}

\begin{figure}
    \centering
    \includegraphics[width=.99\columnwidth]{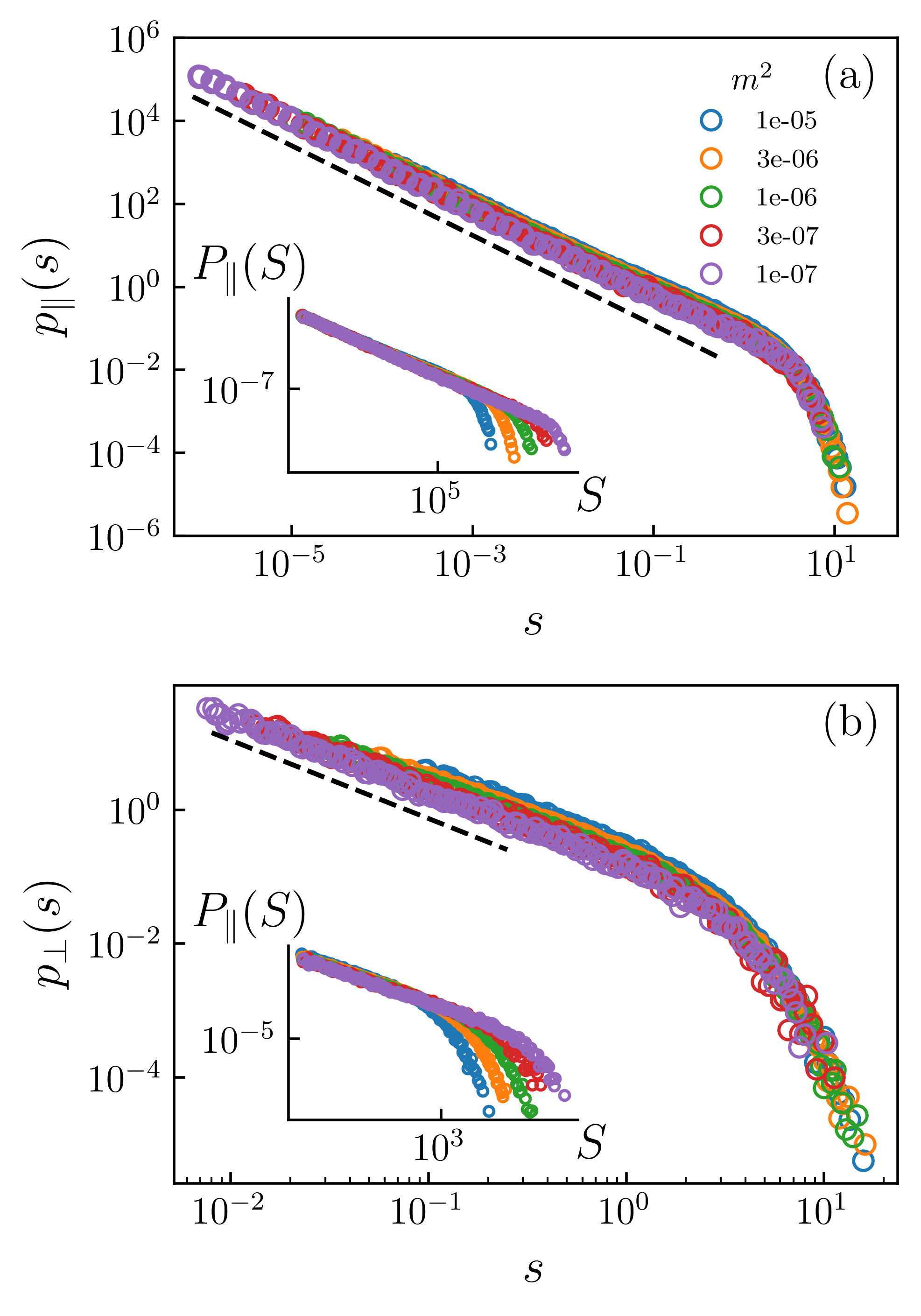}
    \caption{Rescaled distribution of center-of-mass jumps for different $m^2$ and $L = 4096$ monomers in the parallel (a) and perpendicular (b) directions, according to Eq. (\ref{eq:reducedavdist}). Dashed lines indicate power-law fits $p_{\alpha}(s) \propto s^{-\tau}$ well below the cut-offs. Insets show raw data.
    The dashed-line fits yield $\tau_\parallel = 1.09 \pm 0.03$ (a) and $\tau_\perp = 1.17 \pm 0.06$ (b).}
    \label{fig:ava_para_vs_m}
\end{figure}

We now describe the avalanche statistics in the quasistatic regime. We first compute the center-of-mass jumps, defined in Eq.~(\ref{eq:pdfavalanches}).
In the insets of Fig.~\ref{fig:ava_para_vs_m} we see that jumps in both directions fairly follow a power-law decay with a cut-off
\be
P_\alpha(S)\sim S^{-\tau_\alpha} G_{\alpha}(S/S_m^{\alpha}).
\ee
Here  
\be S_m^{\alpha} := \frac{ \langle S^2_\alpha \rangle}{2 \langle S_\alpha \rangle}
\ee grows with decreasing $m$, and $G_{\alpha}(x)$ are cut-off functions, such that $G_\alpha(x)\sim \text{const}$ for $x\ll 1$ and $G_\alpha(x)\to 0$ roughly exponentially for $x > 1$. For the quantitative numerical analysis it is convenient \cite{RossoLeDoussalWiese2009} to define 
\begin{eqnarray}
s&:=& \frac{S}{S_m},\\
p_\alpha(s)&:=& P_\alpha(S) \frac{S_m^2}{\langle S \rangle}.
\label{eq:reducedavdist}
\end{eqnarray}
\begin{figure}
    \centering
    \includegraphics[width=.99\columnwidth]{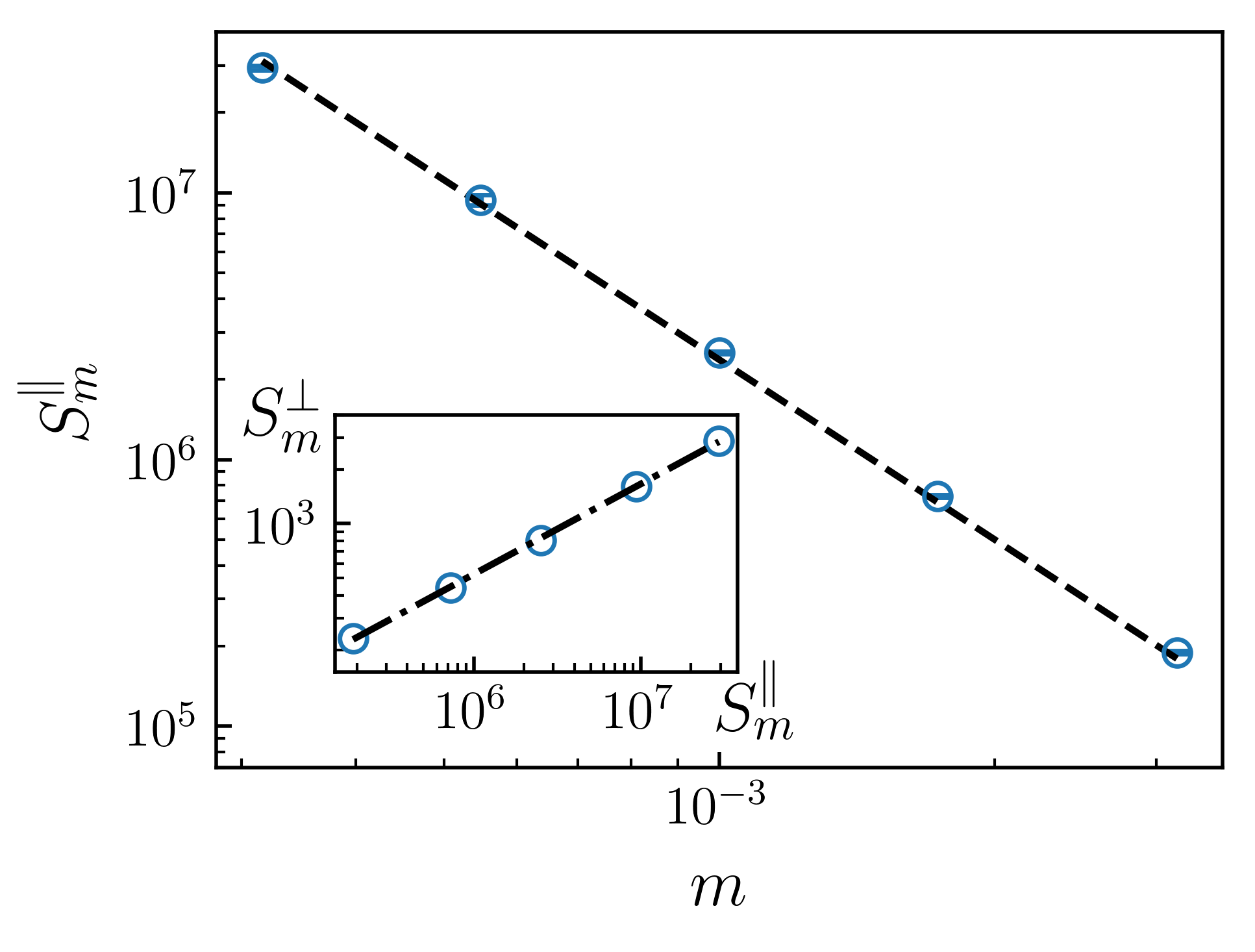}
    \caption{
    Dependence of the avalanche size cutoff $S_m^{\alpha}$ with the confinement factor 
    $m$. The dashed line corresponds to $S_m^{\parallel} \propto m^{-(D+\zeta_{\parallel})}$ with $D = 1$. Inset: the dot-dashed line shows that cut-offs in each direction are strongly correlated as $S_m^{\perp} \propto \sqrt{S_m^{\parallel}}$.
    }
    \label{fig:Sm_vs_m}
\end{figure}%
From the main panel and the inset of Fig.~\ref{fig:Sm_vs_m} we see that the cut-offs respectively scale as
\begin{eqnarray}
S_m^{\parallel} &\sim& m^{-(d+\zeta_\parallel)}, \qquad d=1,  
\label{eq:smparallel}
\\
S_m^{\perp} &\sim& \sqrt{S_m^\parallel}.
\label{eq:smperp}
\end{eqnarray}
In Fig.~\ref{fig:ava_para_vs_m} we show the master curves in the two directions, as obtained by rescaling those in the insets for different values of $m$.
The collapse for different $m$ is better in the parallel direction than in the perpendicular one,  probably due to the smaller range of sizes for the latter. Nevertheless, we can fit 
the avalanche exponents for $s<1$ in both cases, leading to
\begin{eqnarray}
\tau_\parallel&=&1.09\pm 0.03, 
\label{eq:tauparallel}
\\
\tau_\perp&=&1.17\pm 0.06.
\label{eq:tauperp}
\end{eqnarray}
The value of $\tau_\parallel$   is   consistent with the planar approximation, as numerical simulations of avalanches for 1-dimensional interfaces present an indistinguishable value for $\tau$~\cite{RossoLeDoussalWiese2009}. 
The scaling relation  $\tau = 2-2/(d+\zeta)$ of Narayan and Fisher \cite{NarayanFisher1993} (for $N=1$) with the exponents of \cite{GrassbergerDharMohanty2016} (see  Eq.~(\ref{eq:roughnessexponentsparallel})) yield 
\begin{equation}
    \tau_\parallel = 2 - \frac{2}{d+\zeta_{\parallel}}\stackrel{d\to 1}\longrightarrow \frac{10}9 = 1.11111...
    \label{eq:NF}
\end{equation}
For the scalar model ($N=1$) this scaling relation was conjectured by Narayan and Fisher~\cite{NarayanFisher1993} assuming a finite density of avalanches at the depinning threshold. It was rederived in Ref.~\cite{DobrinevskiLeDoussalWiese2014a} from FRG. This result significantly differs from the mean-field result $\tau_{\rm MF}=3/2$. Eq.~(\ref{eq:NF}) was tested numerically~\cite{RossoLeDoussalWiese2009} and analytically   via 1-loop FRG calculations~\cite{LeDoussal2009}.
\begin{figure}
\centering
\includegraphics[width=.99\columnwidth]{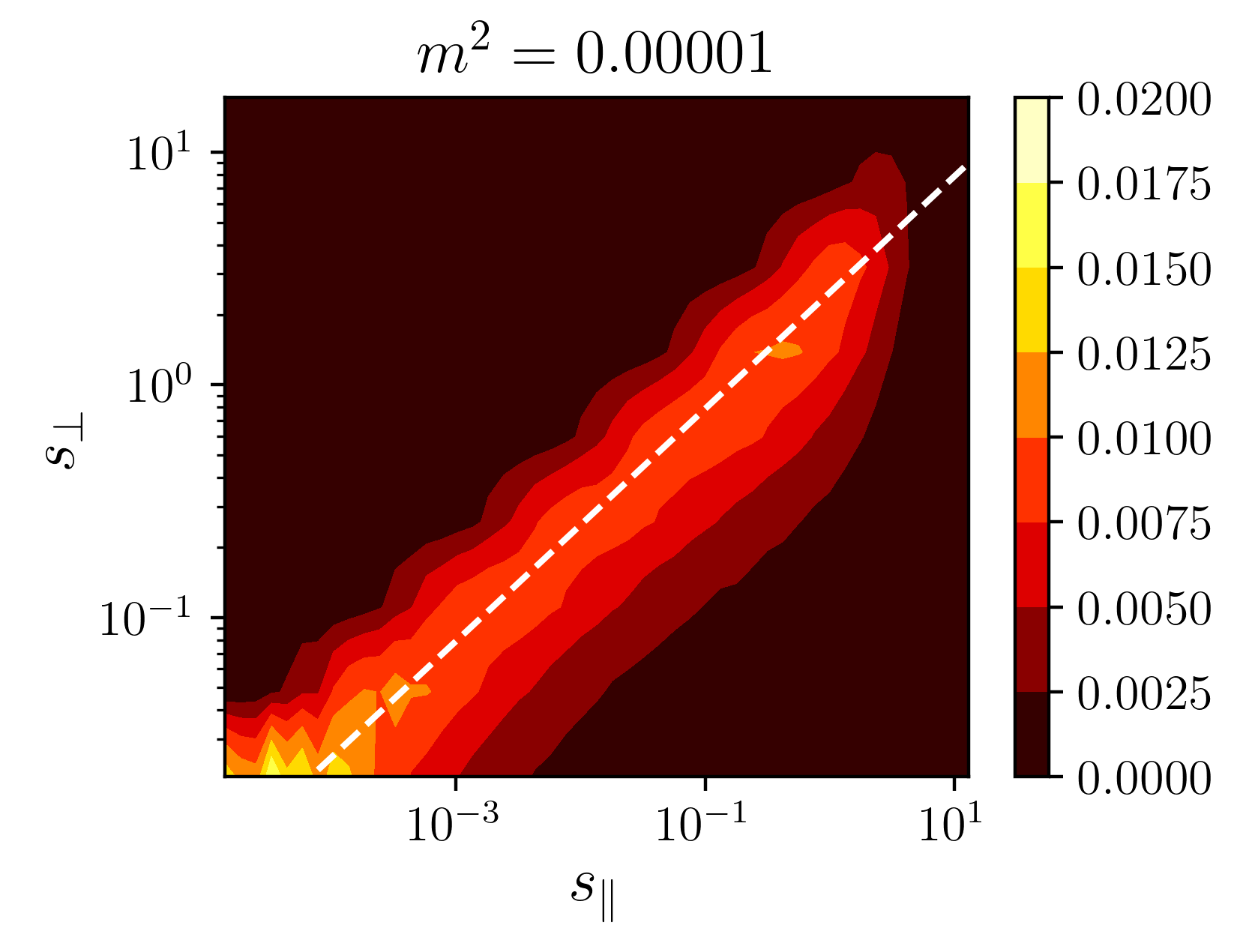}
\caption{Normalized count of events with center-of-mass jump sizes $(s_{\parallel}, s_{\perp})$, highlighting their strong correlation. The white dashed-line corresponds to $s_{\perp} \propto \sqrt{s_{\parallel}}$.}
\label{fig:jointavalanche}
\end{figure}

To understand the value of $\tau_{\perp}$, we remind \Eq{S-perp-from-S-parallel}, $\left < S_\perp^2 \right>|_{S_\parallel} = 2 \sigma  S_\parallel$.
From this we immediately obtain   the scaling relation (\ref{eq:smperp}) for $S_m^{\perp}$.
We have analyzed the joint pdf $p(s_{\parallel}, s_{\perp})$ for a long sequence of avalanches, see Fig.~\ref{fig:jointavalanche}. The strong correlation along the line $s_{\perp} \sim \sqrt{s_{\parallel}}$  confirms \Eq{S-perp-from-S-parallel}. 
This allows us to write  $P_\parallel(S_\parallel ) \rmd S_\parallel =  P_\perp({S}_\perp) \rmd S_\perp $ with $S_\perp \sim S_\parallel^{1/2}$.
Assuming that for small arguments $P_\parallel(S_\parallel)\sim S_\parallel^{-\tau_\parallel}$  and  $P_\perp(S_\perp)\sim S^{-\tau_\perp}$, we obtain that  

\begin{equation}
\label{taus-scaling}
\tau_\perp = 2\tau_\parallel-1 \stackrel{d\to 1}\longrightarrow \frac{11}9 = 1.22222...
\end{equation}
The numerically obtained value reported in Eq.~(\ref{eq:tauperp}) is 
$
\tau_\perp = 1.17 \pm 0.06
$, in fair agreement with the one predicted by the scaling relation \eq{taus-scaling}.

\subsection{Waiting-time distribution}
\begin{figure}
\centering
\includegraphics[width=.99\columnwidth]{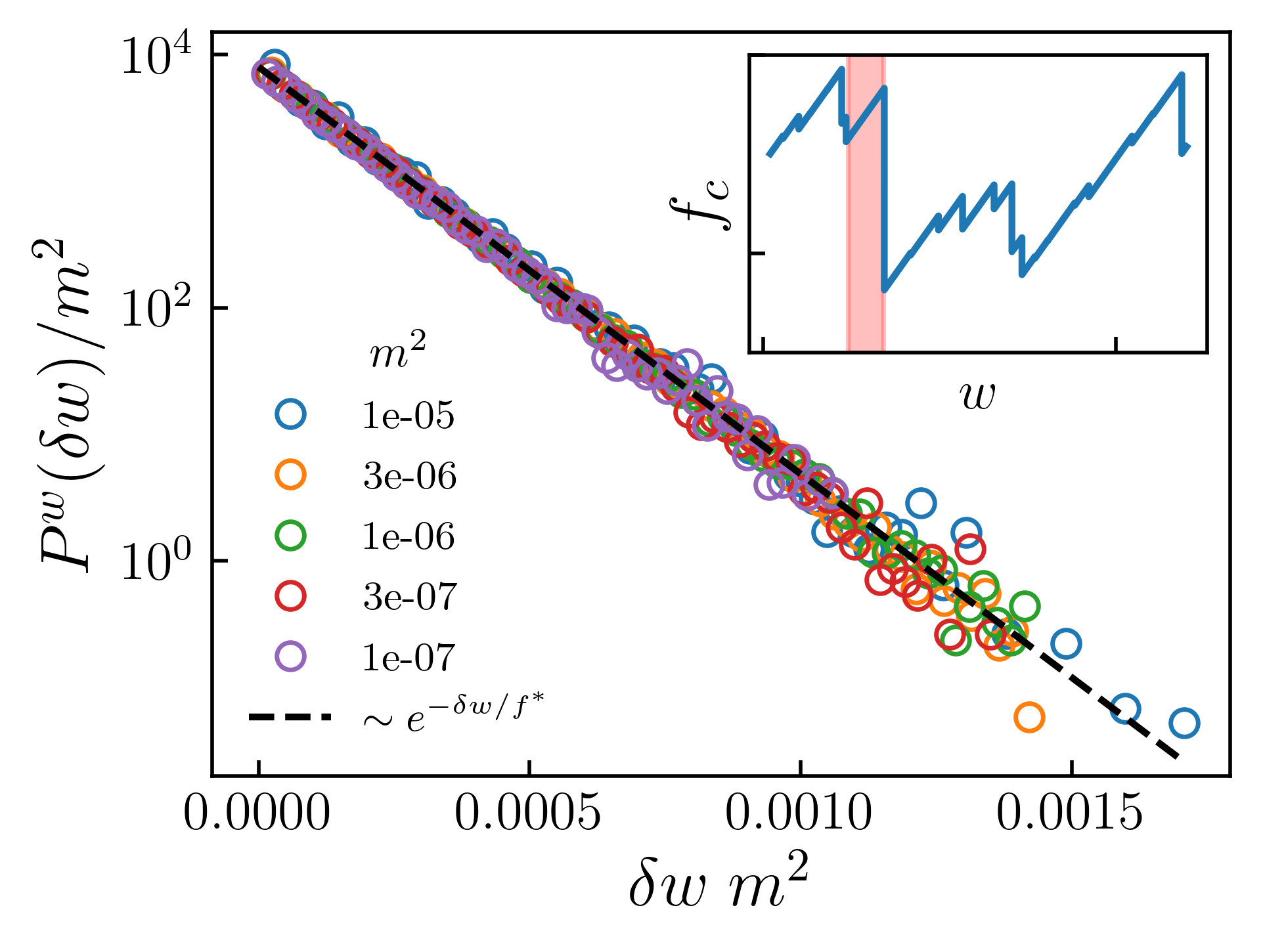}
\caption{Scaled ``waiting time'' distribution between consecutive avalanches in the parallel direction. 
The dashed-line indicates an exponential fit.}
\label{fig:parabolaposition}
\end{figure}

Consecutive avalanches are characterized by a ``waiting-time'' distribution $P^w$ defined in Eq.~(\ref{eq:waitingtimedist}).
In Fig.~\ref{fig:parabolaposition} 
we show that this distribution follows an exponential decay, already observed in Ref.~\cite{LeDoussalMiddletonWiese2008} ($N=1$, equilibrium), 
\bea
P^w(\delta w) 
&\approx& \frac{m^2}{\delta f^*} 
e^{-\delta w m^2/f^*}, 
\eea
with $f^* \approx 0.000 135$ a microscopic force.
Therefore our avalanches are characterized by a mean {\em waiting distance} $f^*/m^2$. If we choose $L= \xi_m\approx 1/m$,   the mean waiting distance diverges with system size as $\left< \delta w\right>\sim L^2$, implying a dominance of   large-stress accumulation periods needed to trigger  large avalanches  of size $S_\parallel \sim L^{1+\zeta_\parallel}$.   
Interestingly, a pure exponential distribution is also found for the ``Gumbel'' universality class of a driven particle in a short-range correlated random-force landscape \cite{LeDoussalWiese2009}. This   contrasts with the Weibull and Frechet universality classes.  
Nevertheless, the critical force distribution for 1-dimensional interfaces in a box of size $L \times L^\zeta$ is slightly different from Gumbel~\cite{BolechRosso2004, FedorenkoLeDoussalWiese2006}. Since in our case we   have  (in the parallel direction) an aspect ratio $\xi_m \times \xi_m^{\zeta}$ (with $\xi_m\approx 1/m$) it would be interesting to derive $P^w$ for such a distribution.

\subsection{The renormalized force-force correlator}
\begin{figure}
\centering
\includegraphics[width=.99\columnwidth]{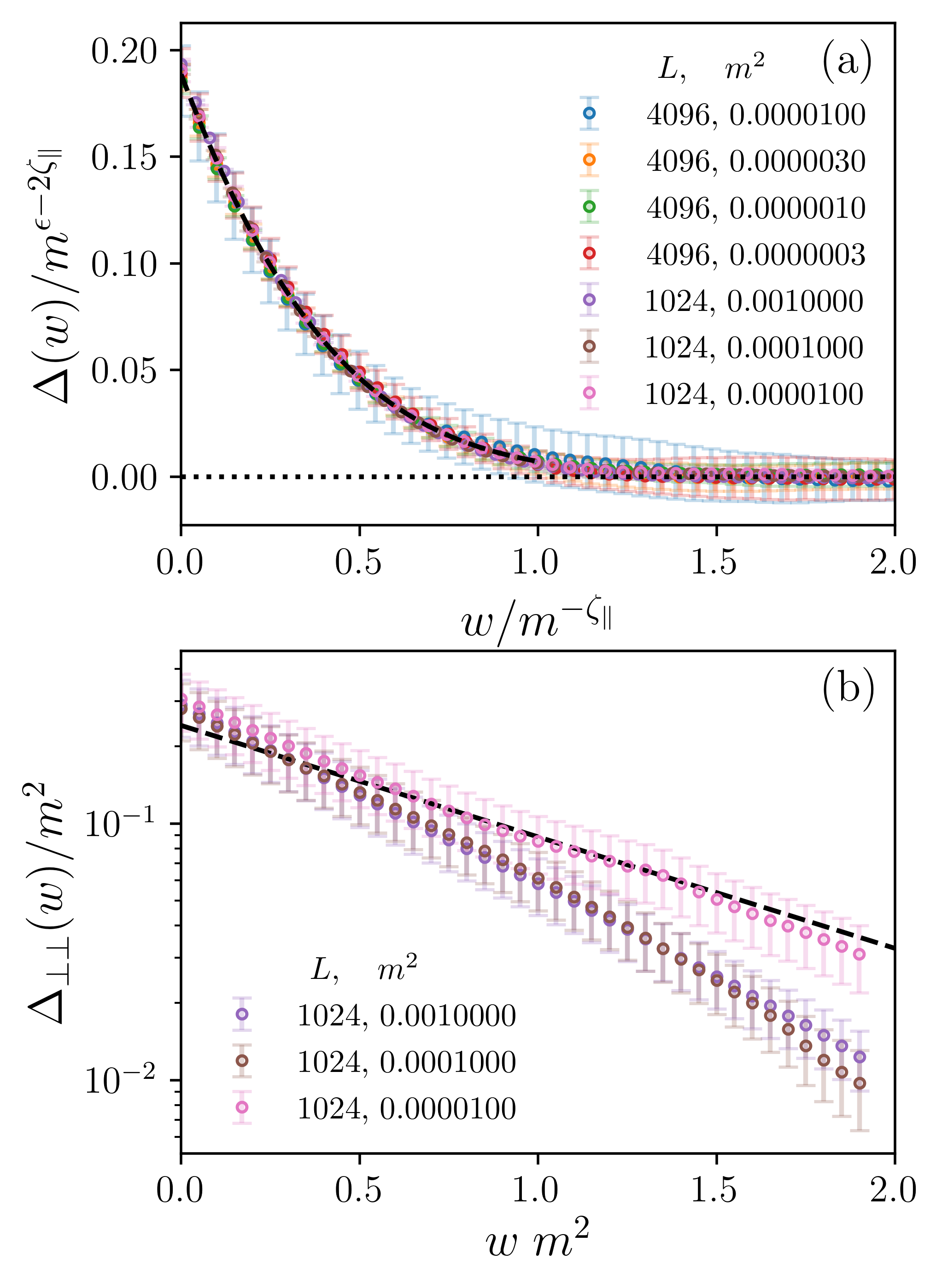}
\caption{(a) Scaled force-force correlator 
for components in the parallel direction (with $\epsilon=4-d=3$). 
The dashed line corresponds to a fit 
of $f(x) = c e^{-a x-b x^2-d x^3}$, yielding $a = 2.38\pm 0.07$, $b = 0.8\pm 0.2$, $c = 0.188\pm 0.001$ and $d = 0.09\pm 0.21$.
(b) Scaled force-force correlator 
for components in the perpendicular direction. The dashed-line corresponds to a fit to $g(x)=a'\exp(-x)$, 
$a'=0.27\pm0.02$.}
\label{fig:correlator}
\end{figure}

Finally, we discuss the force-force correlator of Eq.~(\ref{eq:correlator}), which is a central quantity in the renormalization group calculations. 
In Fig.~\ref{fig:correlator}(a) we show that the correlator $\Delta_\parallel \equiv  \Delta_{\parallel,\parallel}$ for different $m$ can be collapsed using $\Delta_\parallel(w) \approx m^{\epsilon-2\zeta_\parallel}\tilde{\Delta}_\parallel(w m^{\zeta_\parallel})$~\cite{LeDoussalWiese2007}, with $\epsilon = 4-d=3$ for our case and   $\zeta_\parallel=1.25$ from Eq.~(\ref{eq:roughnessexponentsparallel}). 

By fitting $\tilde\Delta_\parallel(x) = c e^{-a x-b x^2-d x^3}$ in the range $w \in[0,1]$ we obtain: $a = 2.38\pm 0.07$, $b = 0.8\pm 0.2$, $c = 0.188\pm 0.001$ and $d = 0.09\pm 0.21$. As a consequence $\frac{\tilde \Delta_\parallel(0) \tilde \Delta_\parallel''(0)}{\tilde \Delta_\parallel'(0^+)^2} = 0.71\pm 0.09$. Increasing the fit range to $w \in [0,1.6]$ this  value increases to $0.76\pm 0.06$. 
We thus estimate the scale-free universal ratio to be 
\be\label{84}
\frac{\Delta_\parallel(0) \Delta_\parallel''(0)}{\Delta_\parallel'(0^+)^2} \equiv \frac{\tilde \Delta_\parallel(0) \tilde\Delta_\parallel''(0)}{\tilde\Delta_\parallel'(0^+)^2}= 0.74\pm 0.07. 
\ee
This is   larger than the 1-loop value of $  2/3$ predicted in  Ref.~\cite{ErtasKardar1996}. It is   fairly close to  the value of $0.73(3)$ measured experimentally in 2-dimensional magnetic domain walls \cite{terBurgBohnDurinSommerWiese2021}. It can also be compared to the value predicted by FRG for short-range elasticity in the $N=1$ case \cite{terBurgBohnDurinSommerWiese2021,Wiese2021}, where one gets $2/3$ in $d=4$ (exact), $0.71$ in $d=3$ (2-loop) $0.75$ in $d=2$ (2-loop), $0.79$ in $d=1$ (2-loop) and $0.822$ (toy model in $d=0$).  Within error bars  the value of \Eq{84}  agrees with the one  predicted by the theory for $N=1$, and thus  confirms  the planar approximation.

In Fig.~\ref{fig:correlator}(b) we show that the correlator $\Delta_{\perp}\equiv\Delta_{\perp,\perp}$ can, for different $m$,    fairly well be collapsed, within the statistical error bars, 
using a master curve $\Delta_{\perp} (w) = m^2 \tilde{\Delta}_{\perp}(w m^2)$,
as anticipated in Eq.~(\ref{eq:Dperpperp}). 
By fitting the predicted exponential decay as  $\tilde{\Delta}_{\perp}(w)=\sigma \exp(-d w)$ for all curves combined, we obtain $\sigma=0.306\pm0.027$, $d=1.50\pm0.25$. 
The value of $\sigma$ is fairly close to the single-monomer standard deviation of perpendicular jumps, while the decay constant $d$ comes out   larger. However, we see that reducing $m$, the scaling function $\tilde \Delta(w)$ converges more and more to 
$\tilde \Delta_\perp(w)=\sigma \rme^{-w}$ predicted in \Eq{eq:Dperpperp}. The latter curve is shown in  black dashed  on Fig.~\ref{fig:correlator}(b).  It seems convergence is slow, and   the prediction \eq{eq:Dperpperp} is reached only asymptotically. We may therefore  suspect that the amplitude ratio \eq{84} has also  not yet converged. 
We  defer  an in-depth  analysis to  future work. 

\subsection{Crossover to the fast-flow regime}
\label{Crossover to the fast-flow regime}

\begin{figure}
    \centering
    \includegraphics[width=.95\columnwidth]{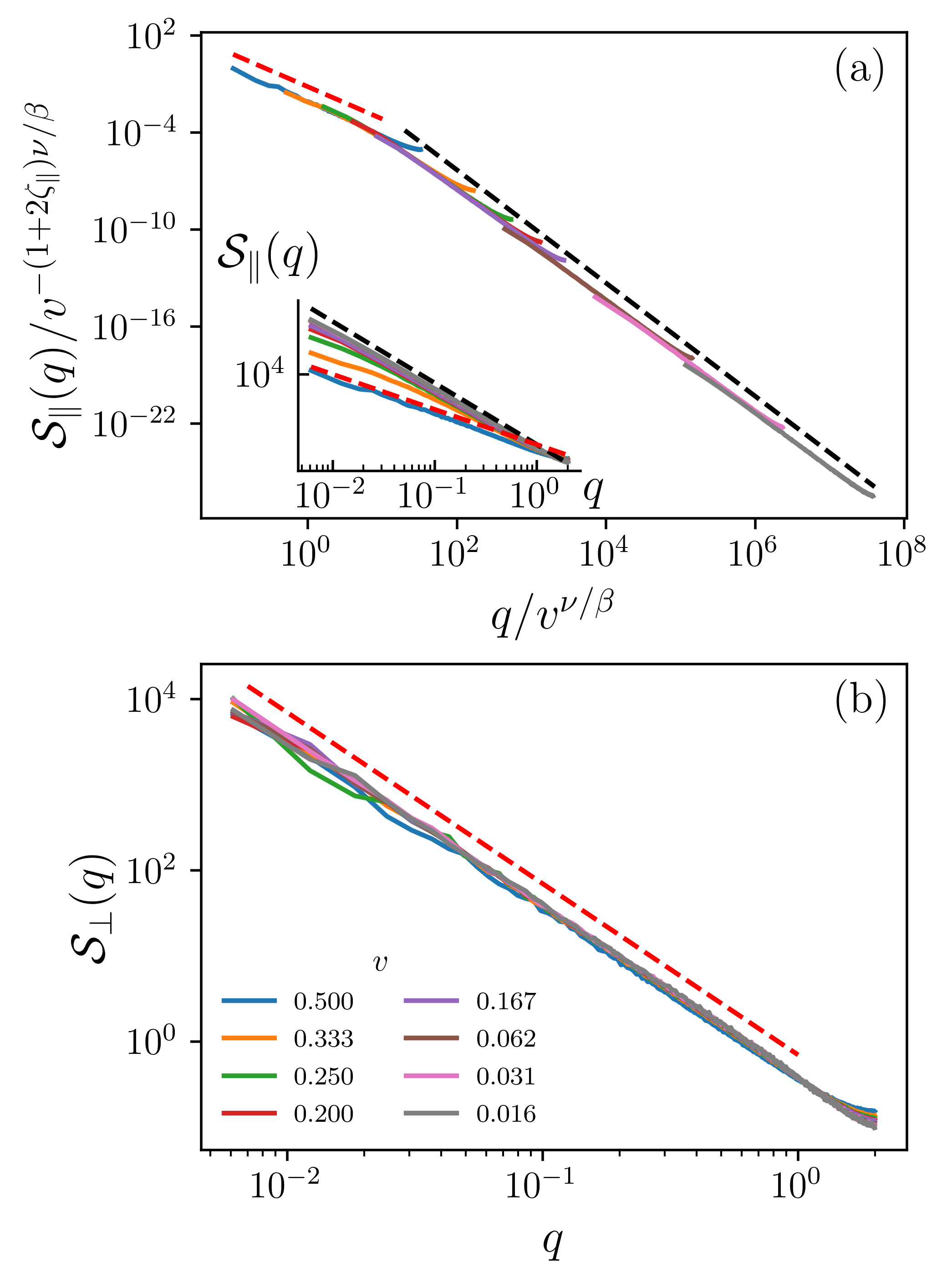}
    \caption{Steady-state structure factors at finite imposed velocities for $L = 1024$ and $m^2 = 0.00001$ for the longitudinal (a) and transversal directions (b).  
    (a) Rescaled structure in the parallel direction. 
    Red dashed lines indicate the fast-flow roughness exponent $\zeta_{\tt ff} = 1/2$, and the black dashed lines the depinning roughness exponent $\zeta_{\parallel} = 1.25$. 
    The crossover between the two regimes occurs  
    at the depinning correlation length  $\xi_f \sim v^{-\nu/\beta}$.     
    Inset shows raw non-scaled data.
    (b) The transverse structure factor appears to be independent of the imposed velocity and   described by a roughness exponent $\zeta_{\perp} = 0.5$ (red dashed line).}
    \label{fig:Sq_para_vs_v}
\end{figure}

At finite velocities just above the  depinning threshold, the steady-state correlation length is expected to diverges as $\xi_f \sim (f-f_c)^{-\nu} \sim v^{-\nu/\beta}$.
As for interfaces~\cite{KoltonRossoGiamarchiKrauth2009b}, $\xi_f$ is   a characteristic geometrical crossover length.
In Fig.~\ref{fig:Sq_para_vs_v}(a) we show that for intermediate and large length scales, and different small steady-state velocities 
\begin{eqnarray}
{\cal S}_\parallel(q) \sim q^{-(d+2\zeta_\parallel)}
{\tilde {\cal S}}_\parallel(q v^{-\nu/\beta}).
\end{eqnarray}
Here ${\tilde {\cal S}}_\parallel(x) = \text{const}$ for $x \gg 1$, while for $x\ll 1$  one has
 ${\tilde {\cal S}}_\parallel(x)\sim x^{2(\zeta_\parallel-\zeta_{\rm EW})}$, where $\zeta_{\rm EW}=(2-d)/2$ and $\zeta_\parallel=1.25$ as given by Eq.~(\ref{eq:roughnessexponentsparallel}).
From the renormalization group viewpoint 
this result is in agreement with the 
crossover of the depinning fixed point towards an Edwards-Wilkinson regime, 
as predicted for the  
FL~\cite{ErtasKardar1996} and for the interface~\cite{ChauveGiamarchiLeDoussal2000}. 
In other words, besides renormalizing the friction such that $v \sim (f-f_c)^{\beta}$,  pinning forces on the coarse-grained FL above $\xi_f$ are similar to thermal  noise. What was derived for $N=1$ remains valid in the planar approximation. 

In Fig.~\ref{fig:Sq_para_vs_v}(b) we show the structure factor in the perpendicular direction for different velocities near the depinning threshold. Remarkably, there are no   signatures of the correlation length $\xi_f\sim v^{-\nu/\beta}$. That is, for non-microscopic length scales we find that
\begin{eqnarray}
{\cal S}_\perp(q) \sim q^{-(1+2\zeta_\perp)} \sim q^{-2},
\end{eqnarray}
and we are not able to detect any geometrical crossover at $\xi_f(v)$, at variance with the clear crossover observed in $S_\parallel(q)$. The reason is  that the assumptions entering \Eq{EOM-trans} remain unchanged for large driving velocities  $v$. We thus only observe 
a crossover imposed by the confining potential at 
$q\xi_m\approx 1$. 

\begin{figure}[t]
\centering
\includegraphics[width=.99\columnwidth]{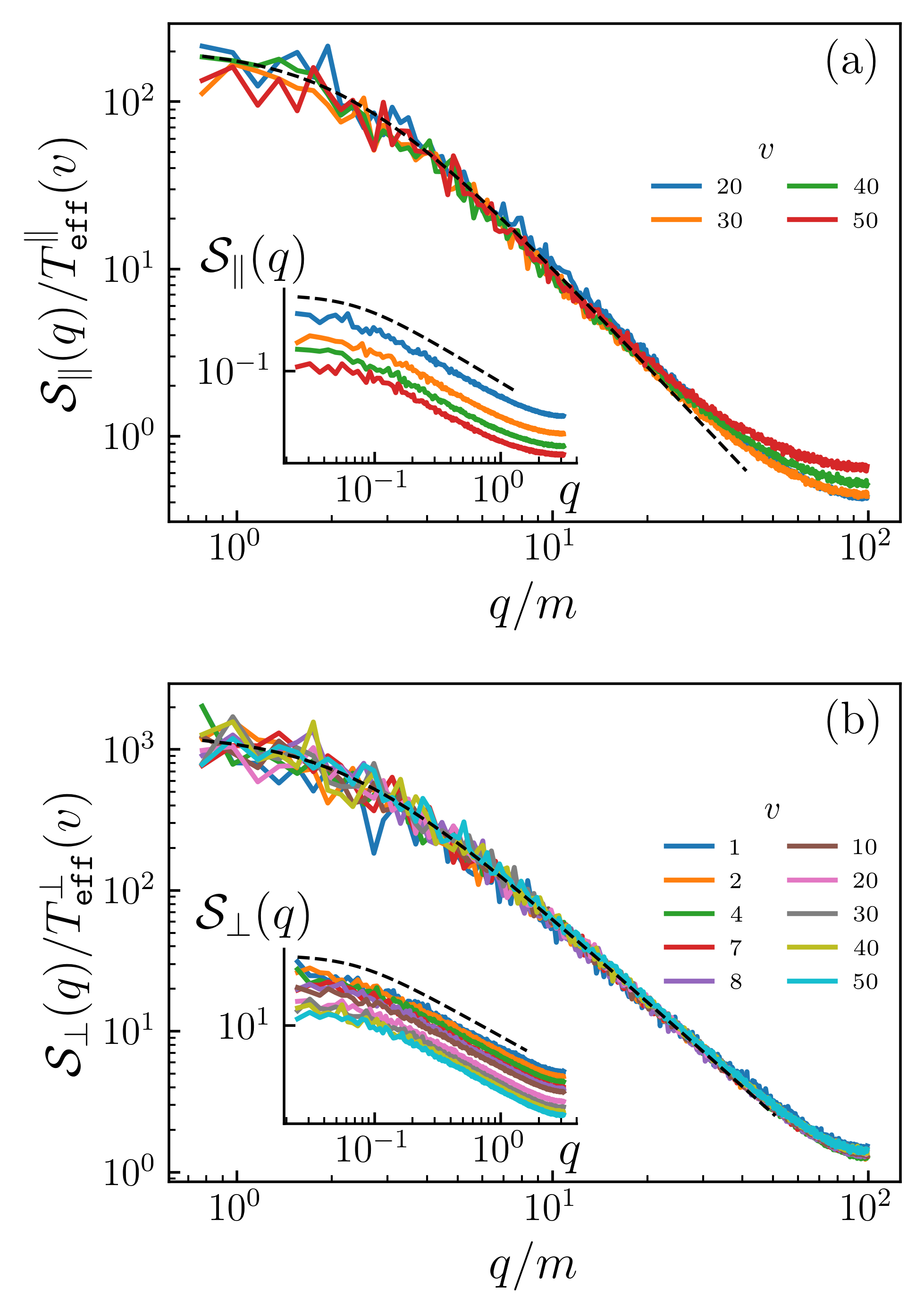}
\caption{Rescaled Structure factors at different large velocities for $m^2 = 0.001$ and $L = 1024$ in the longitudinal (a) and transverse (b) directions. In both cases raw-data (insets) can be collapsed by using effective temperatures $T^\parallel_{\tt eff}(v)$ (a) and $T^\perp_{\tt eff}(v)$ (b). Dashed-lines correspond to the purely ``thermal'' $\sim (c q^{2}+m^2)^{-1}$ dependence.
}
\label{fig:fastflowSofq}
\end{figure}%
At large velocities, in the {\em  fast-flow regime}, the depinning correlation length becomes small and  it is   expected that the pinning forces became a rapidly fluctuating uncorrelated noise acting on an otherwise flat moving elastic FL. We hence expect ${\cal S}_\alpha(q)\sim (c q^{2}+m^2)^{-1}$ at intermediate scales, $q \xi_m = q/m > 1$, and a crossover towards ${\cal S}_\alpha(q)\sim \text{const}$ for $q \xi_m = q/m < 1$. 
In the insets of Fig.~\ref{fig:fastflowSofq} we verify this by showing the structure factor as a function of $v$ (not necessarily small), for the longitudinal and perpendicular directions. 
At variance to   small $v$,  
${\cal S}_\perp(q)(cq^2+m^2)$ is $v$-dependent. 
Its 
behaviour at small $q$  motivates the study of effective temperatures. These are introduced from generalized fluctuation-dissipation theorems by using that the static linear response function in the $\alpha$ direction due to an external time-independent but $q$-dependent field in the $\alpha'$ direction is {\it exactly} given by 
\begin{eqnarray}
\chi_{\alpha,\alpha'}= \delta_{\alpha,\alpha'}\chi(q)\equiv
\frac{\delta_{\alpha,\alpha'}}{c q^2 + m^2}, 
\end{eqnarray}
due to the statistical tilt symmetry \cite{NarayanFisher1993}. 
In equilibrium ($v=0$) at a finite temperature $T$, the fluctuation-dissipation theorem implies that 
${\cal S}_\alpha(q)=T \chi(q)$.
For the driven, out-of-equilibrium FL at zero temperature we can  thus define   anisotropic effective temperatures $T_{\tt eff}^{\alpha}$ ($\alpha=\parallel,\perp$)
\begin{eqnarray}
T_{\tt eff}^{\alpha}(v) 
:= \frac{{\cal S}_\alpha(q)}{\chi(q)}=( c q^2 +m^2) {\cal S}_\alpha(q).
\end{eqnarray}%
At large   scales compared to the correlation length $\xi_f(v)$, and since ${\cal S}_\alpha(q)\sim (c q^{2}+m^2)^{-1}$, a single anisotropic effective temperature is sufficient for the whole regime.  This is expected to hold for 
$2\pi\xi_f^{-1} \gg q $. It remains valid for the  largest length scales,  and when  $q<m$, there is a crossover to a flat regime $S_\alpha(q)\simeq T_{\tt eff}(v) m^{-2}$, due to the confining potential. Therefore ${\cal S}_\alpha(q)/T^\alpha_{\tt eff}$ is independent of $v$ for large   length scales (small $q$) as shown in  Fig.~\ref{fig:fastflowSofq}, and $T^\alpha_{\tt eff}(v)$ describes large-scale fluctuations in general.

Since large length scales are   associated with a slow dynamics, these definitions may yield a {\it bona fide} temperature in a thermodynamic sense ~\cite{Cugliandolo2011}.
Using these definitions in the main panels of Fig.~\ref{fig:fastflowSofq} we show that ${\cal S}_\alpha(q)$ for different velocities $v$ can be distinguished  by $T_{\tt eff}^\alpha(v)$ for small enough $q$. For large $q$, only the parallel direction shows a deviation from the master curve, indicating that 
the parallel direction   retains genuine non-equilibrium features at short length scales.

\begin{figure}
\centering
\includegraphics[width=\columnwidth]{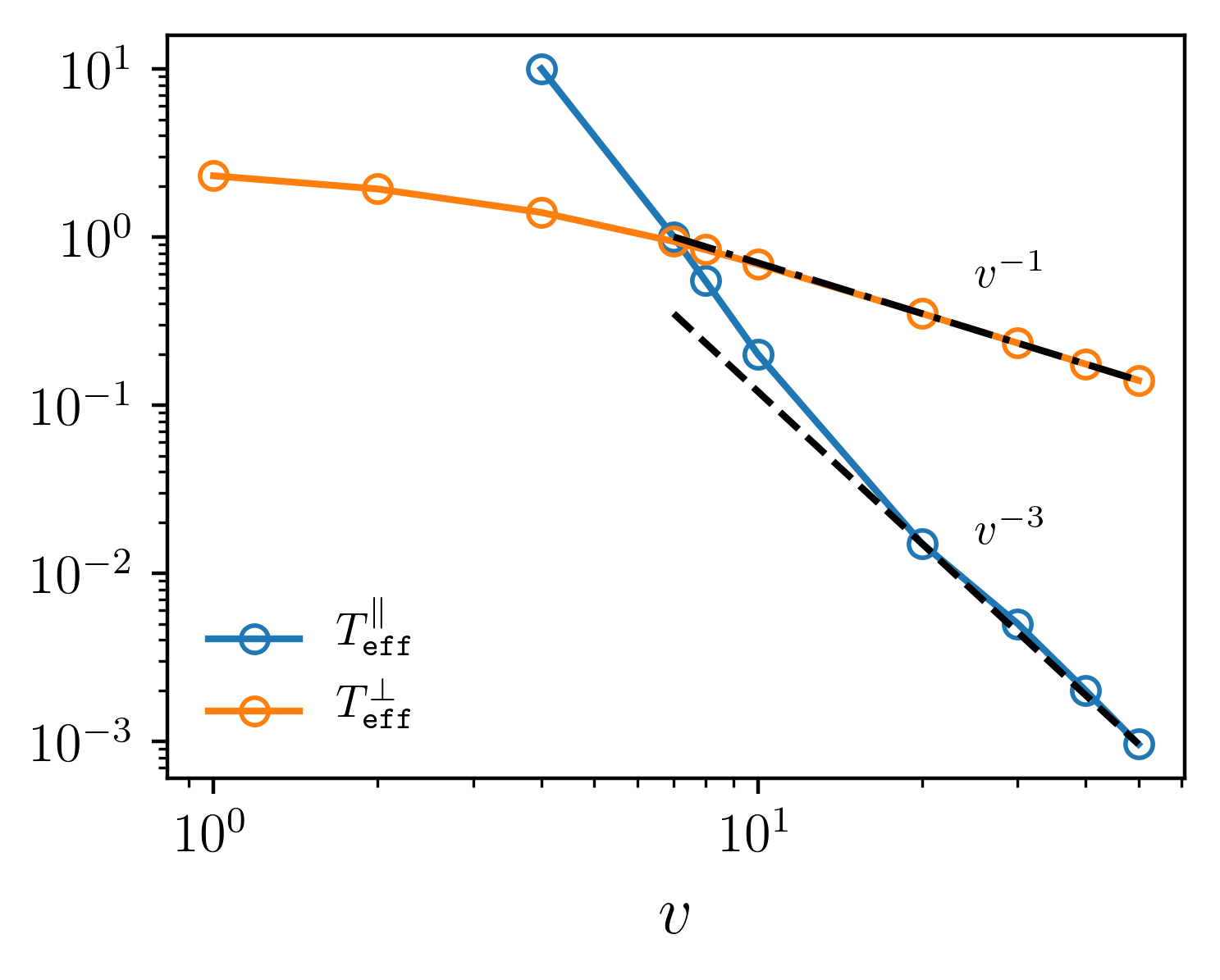}
\caption{Effective temperatures as a function of the velocity for $m^2 = 0.001$ and $L = 1024$. 
Dashed lines indicate asymptotic dependencies.}
\label{fig:teff}
\end{figure}

\begin{figure}
\centering
\includegraphics[width=\columnwidth]{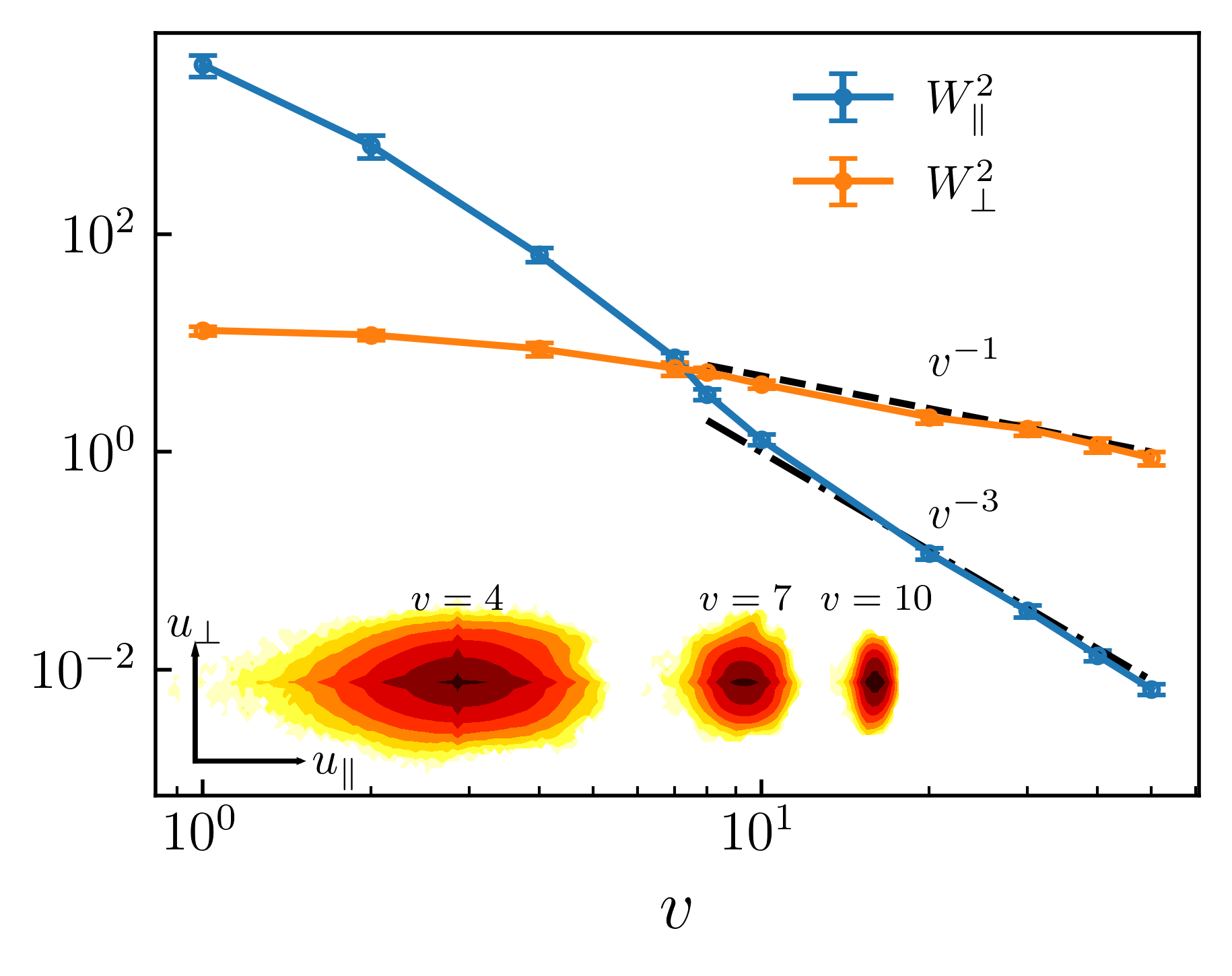}
\caption{Mean quadratic widths as a function of the velocity for $m^2 = 0.001$ and $L = 1024$.
Dashed-lines indicate asymptotic dependencies.}
\label{fig:widths}
\end{figure}

The velocity dependence of the two effective temperatures is shown in Fig.~\ref{fig:teff}.
They are direction dependent, and monotonously decrease with increasing $v$. They intersect at  a characteristic velocity of $v\approx 7$, above which $T^\perp_{\tt eff}>T^\parallel_{\tt eff}$.
At small $v$ the transversal temperature $T^\perp_{\tt eff}$  saturates at $T^\perp_{\tt eff}=\sigma \approx 1/3$, as discussed above, explaining why on  Fig.~\ref{fig:Sq_para_vs_v}(b)  no appreciable velocity dependence  is observed. The observed asymptotic forms are $T^\parallel_{\tt eff}\sim v^{-3}$ and $T^\perp_{\tt eff}\sim v^{-1}$.
The two-component Edwards-Wilkinson type of scaling with effective temperatures $T_{\tt eff}^\alpha$ of the structure factor at large velocities implies that the global width   scales as \be
\label{W2-fast-flow}
W_\alpha^2 \sim \frac{T_{\tt eff}^\alpha}{m^{2\zeta_{\rm EW}}} = \frac{T_{\tt eff}^\alpha}{m},
\ee
as verified in Fig.~\ref{fig:widths} for each direction,  as a function of   velocity.

\begin{figure}
\centering
\includegraphics[width=.99\columnwidth]{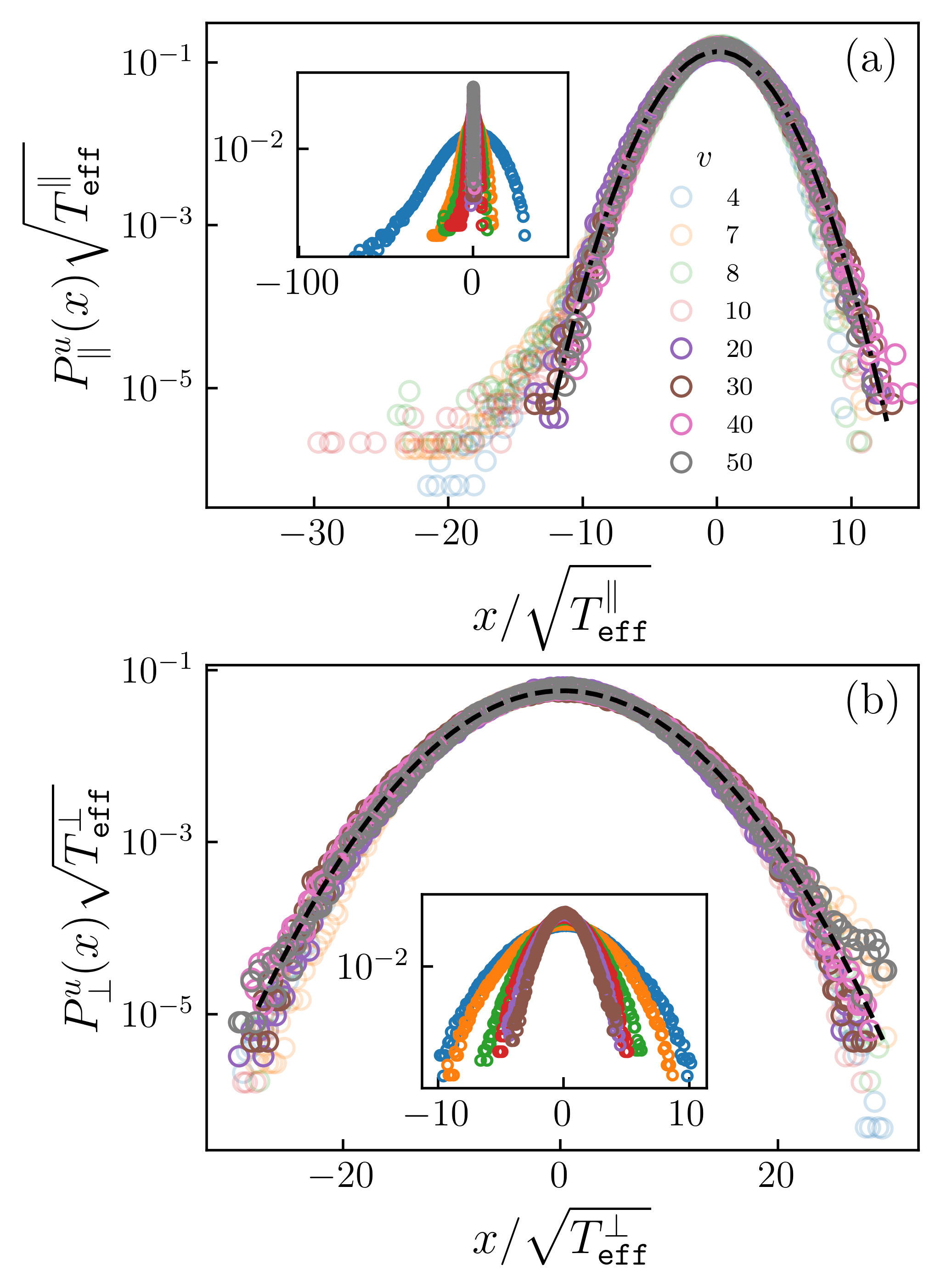}
\caption{Local displacement fluctuations around the center of mass in the parallel (a) and perpendicular directions (b). Lighter   symbols are used for   smaller velocities.
Parameters of the simulations are $m^2 = 0.001$ and $L = 1024$. Dashed-lines are Gaussian fits.}
\label{fig:pdfs}
\end{figure}

\begin{figure}
\centering
\includegraphics[width=.99\columnwidth]{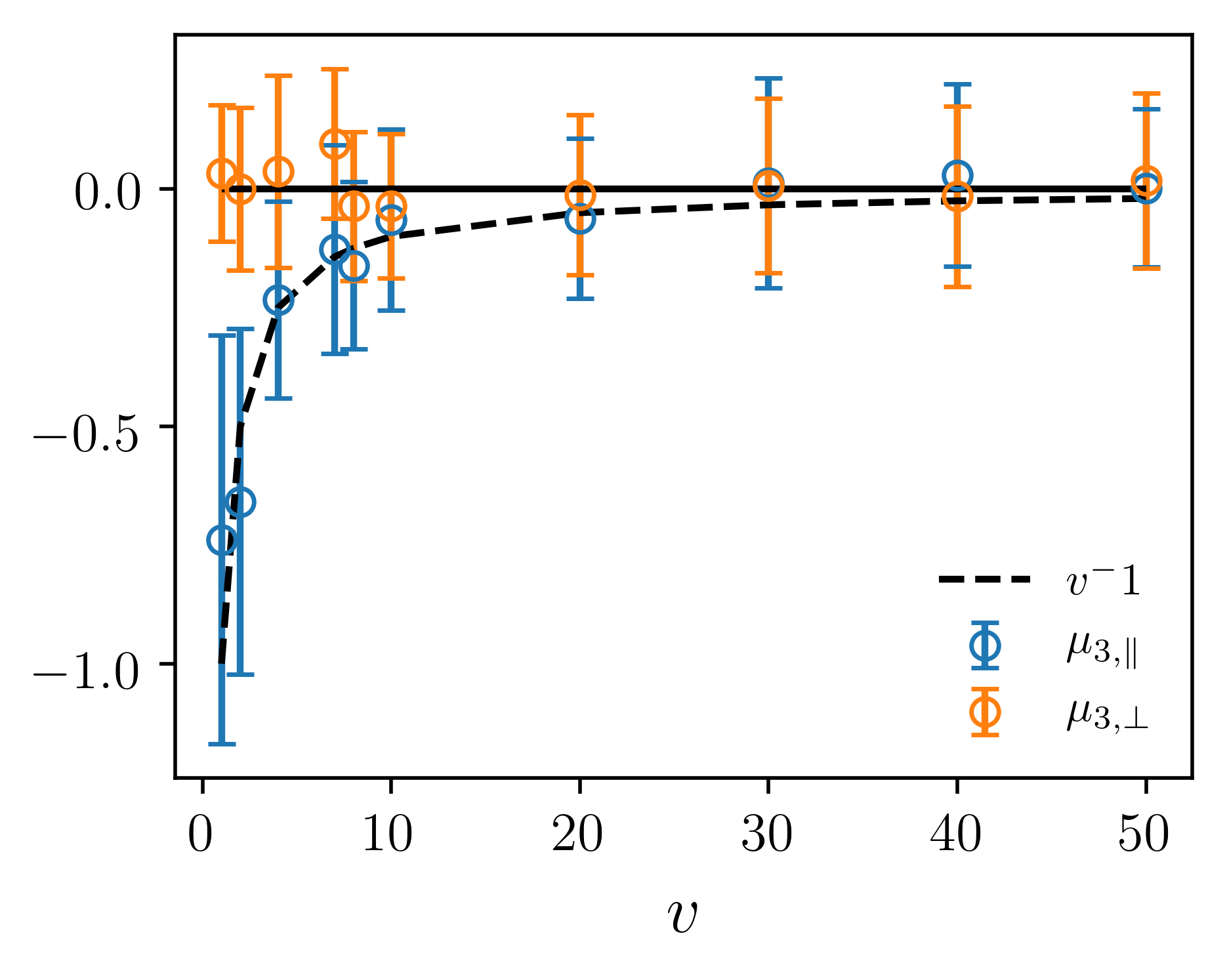}
\caption{
Skewness of the two components of the local displacement distribution as a function of the velocity for $m^2 = 0.001$ and $L = 1024$.}
\label{fig:skew}
\end{figure}

\begin{figure}
\centering
\includegraphics[width=.99\columnwidth]{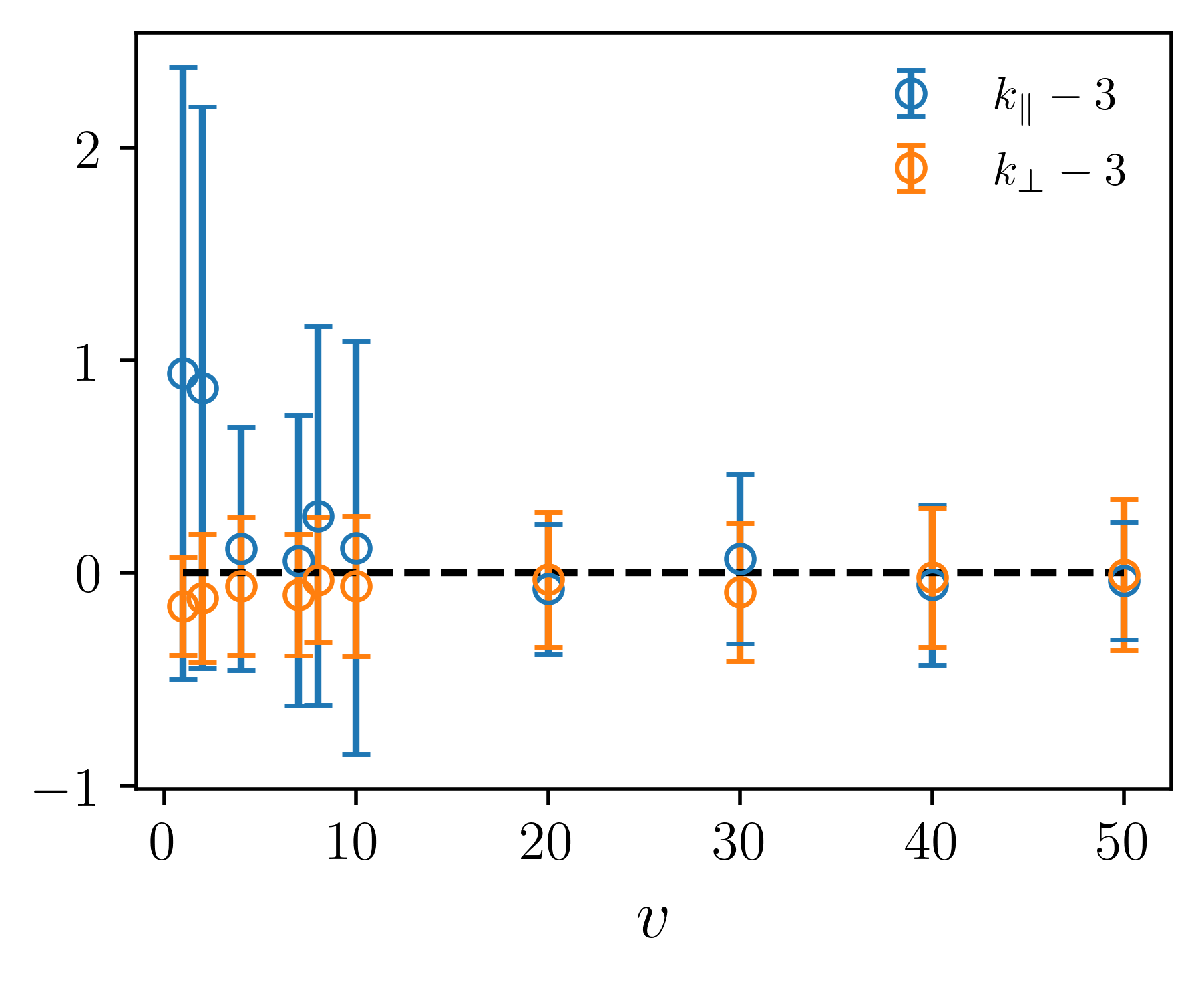}
\caption{Kurtosis of the local displacement fluctuations in both directions. The dashed-line correspond to a pure gaussian.}
\label{fig:kurt}
\end{figure}

The crossing of the effective temperatures is associated with the existence of an isotropic point for the global width, above which the FL tends to be   elongated in the transverse direction, in   contrast with the situation near depinning where they are elongated in the longitudinal direction. This was predicted in Ref.~\cite{NattermannScheidl2000} from general arguments.
To see this better, it is useful to compute the joint distribution function (\ref{eq:localdispjointpdf}) of local displacements, which gives us a top view of the FL fluctuations in the co-moving frame. 
In the inset of Fig.~\ref{fig:widths} we do not only see the change of aspect ratio and the reduction of the global width with increasing $v$, but we  also observe that the parallel-displacement distribution is asymmetric, with a more elongated   tail at smaller velocities, in contrast to the symmetric distribution in the perpendicular direction. To characterize it we show   
in Fig.~\ref{fig:pdfs}  the reduced distributions of  Eq.~(\ref{eq:localdisppdf}) for a large range of velocities.
Only for large $v$ do they converge towards a Gaussian,  
as can be seen in the inset of Fig.~\ref{fig:pdfs}(a): 
At low velocities (lighter colors) the distribution  has asymmetric tails. As shown in the main panel the variance is controlled by the velocity-dependent effective temperatures of Fig.~\ref{fig:teff}. 
These displacements translate into an appreciable skewness. In Fig.~\ref{fig:skew} we show the skewness, as a function of the velocity, defined as  
\be
\mu_{3,\alpha} = \frac{ \overline{\left<  \big[ u(x,t)-u_\alpha(t) \big]^3 \right>} } {W^2_\alpha(t)^{\frac 32}}.
\ee
As expected, the perpendicular direction has an undetectable skewness while the longitudinal one presents a negative skewness at small $v$. It roughly vanishes as $\sim 1/v$ for large velocities. In Fig.\ref{fig:kurt} we also show the kurtosis 
\be
k_{\alpha} = \frac{ \overline{\left<  \big[ u(x,t)-u_\alpha(t) \big]^4 \right>} } {W^4_\alpha(t)}.
\ee
Within error bars we find $k_\perp\approx 3$ and $k_\perp\approx 3$ at large velocities, consistent with an approximately gaussian shape. Only in the longitudinal direction at low velocities we observe a departure from gaussian, $k_\parallel > 3$, though with a large error bar.

The above observations are a strong indication for  the existence of a   {\em large-deviation} function, encountered for depinning already in  Ref.~\cite{LeDoussalPetkovicWiese2012}. Provided the limit exists, the 
 large-deviation function $F(x)$  is defined as 
\be
F(x) := - \lim _{v\to \infty}\frac{ \ln P(x v)}v .
\ee
Since our data do not allow to evaluate $F(x) $ precisely enough, we leave its determination for future work.

On the other hand, 
 the inset of Fig.~\ref{fig:pdfs}(b) shows that the perpendicular fluctuations are well approximated by a Gaussian with a velocity-controlled variance,
\begin{equation}
    P_\perp^{u}(x)\sim e^{-\frac{(x-u_\perp)^2}{2W^2_\perp(v)}}
\end{equation}
Since $  W^2_\perp(v) \approx T_{\tt eff}^\perp(v)/m$, see \Eq{W2-fast-flow}, 
the velocity dependence is exclusively controlled by the transverse effective temperature, as shown by the rescaled curves in the main panel of Fig.~\ref{fig:pdfs}(b).
These results show that   the perpendicular direction can     be described by an Edwards-Wilkinson equation with an effective temperature at all velocities. 
The longitudinal direction   shows genuine non-equilibrium effects well  beyond the depinning transition, which disappear roughly as $1/v$ for   large $v$. 

It is worth noting that these rare asymmetric parallel fluctuations may be   more pronounced   for   strong  pinning. For instance, when pinned by nanoparticles, the FL appears as a sequence of convex arcs in the direction of motion connecting localized pinned   pieces~\cite{koshelevkolton2011}, explicitly breaking the $\delta u_\parallel(z,t) \to -\delta u_\parallel(z,t)$ symmetry. This symmetry is however always broken at depinning \cite{SparfelWiese2021,Wiese2021}.
This kind of structure may explain both tails of the displacement distribution. 
Nevertheless, at very large velocities  both directions display anisotropic Gaussian fluctuations.
As discussed in the next section, the anisotropy of these fluctuations is rather sensible to whether the microscopic disorder is   RB or RF. 

\subsection{Random-field disorder}
So far we   discussed RB disorder acting on a vortex line, corresponding to short-range correlated pinning potentials. This type of disorder seems to be  the only one relevant in experiments, and in particular for  point disorder in bulk superconductors. While we do not know how to realize isotropic RF disorder  corresponding to  uncorrelated pinning forces, we nevertheless consider it here for comparison. We remind that the two types of disorder are differentiated by their correlators (see section \ref{sc:model}).
\begin{figure}
\centering
\includegraphics[width=.99\columnwidth]{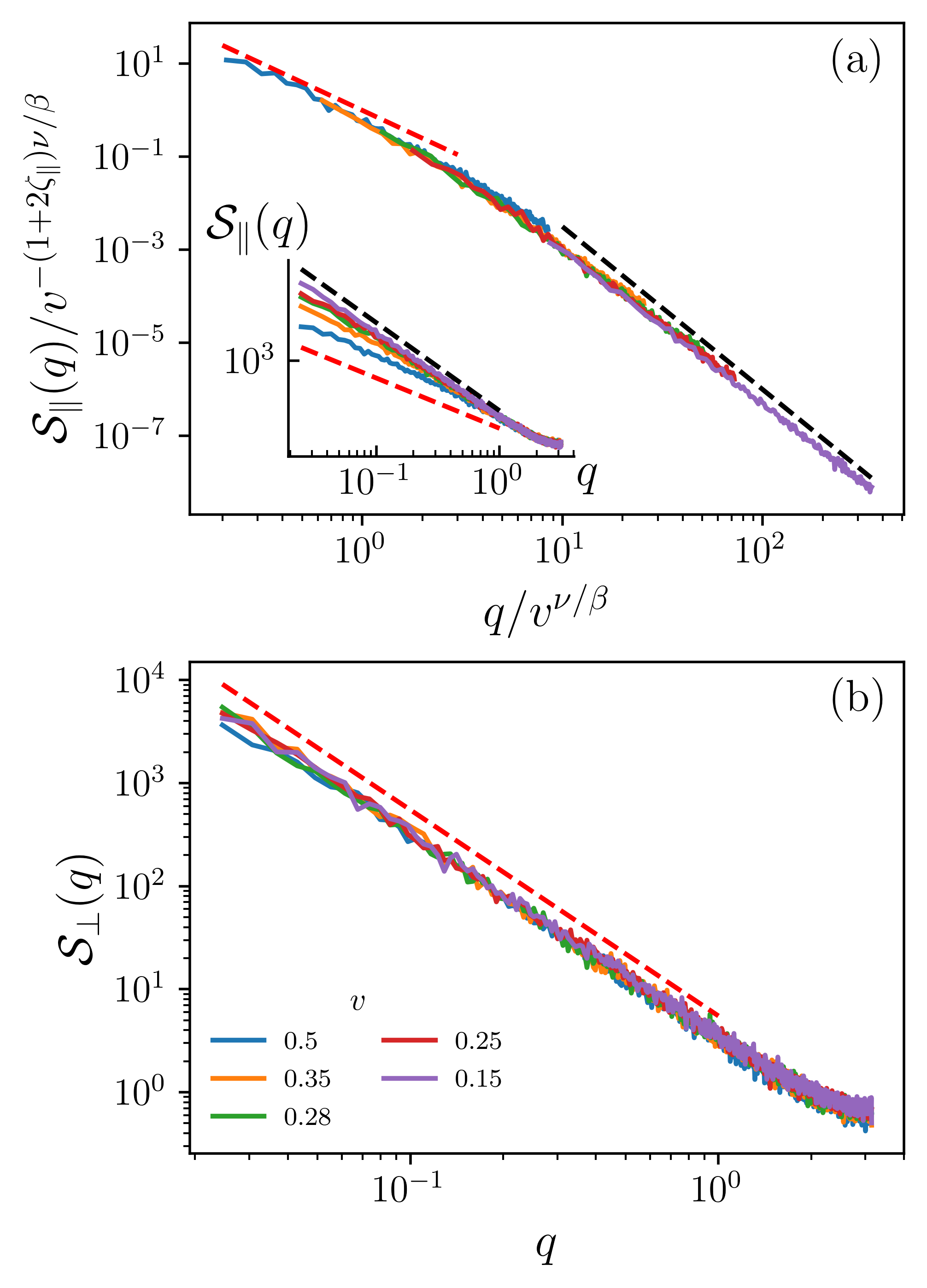}
\caption{
Steady-state structure factor at low velocities in the parallel (a) and perpendicular (b) directions for   RF     disorder. (a) The inset shows raw data and the main panel the scaled structure factor in the parallel direction using the exponents obtained in for   RB disorder. (b) Raw data in the perpendicular direction appears to be independent of $v$.
See Fig.~\ref{fig:Sq_para_vs_v} for a direct comparison with the RB case. 
}
\label{fig:div_RF}
\end{figure}

We first discuss the low-velocity regime near the depinning transition. In Fig.~\ref{fig:div_RF}(a) we show, using the same exponents of Table~\ref{table:criticalexps} obtained for the RB case,
that the steady-state structure factor scales as ${\cal S}_\parallel(q)\sim q^{-(1+2\zeta_\parallel)} G(q \xi)$,  $\xi \sim (f-f_c)^{-\nu}$, with $G(x)\sim x^{1+2\zeta_\parallel}$ for $x \gg 1$, and $G(x)\sim \text{const}$ for $x \ll 1$, 
provided $\xi < 1/m$ and $q > m$. 
Since the same scaling was shown in Fig.~\ref{fig:Sq_para_vs_v}(a) for the RB case, this result is again consistent  with the planar approximation, and with the finding  that RB and RF share the same depinning universality class \cite{LeDoussalWieseChauve2002,ChauveLeDoussalWiese2000a}. In Fig.~\ref{fig:div_RF}(b) we   show that ${\cal S}_\perp(q)\sim q^{-(1+2\zeta_\perp)}$, with no clear signature of $\xi$, as observed before in Fig.~\ref{fig:Sq_para_vs_v}(b) for the RB case. 
\begin{figure}[t]
\centering
\includegraphics[width=.99\columnwidth]{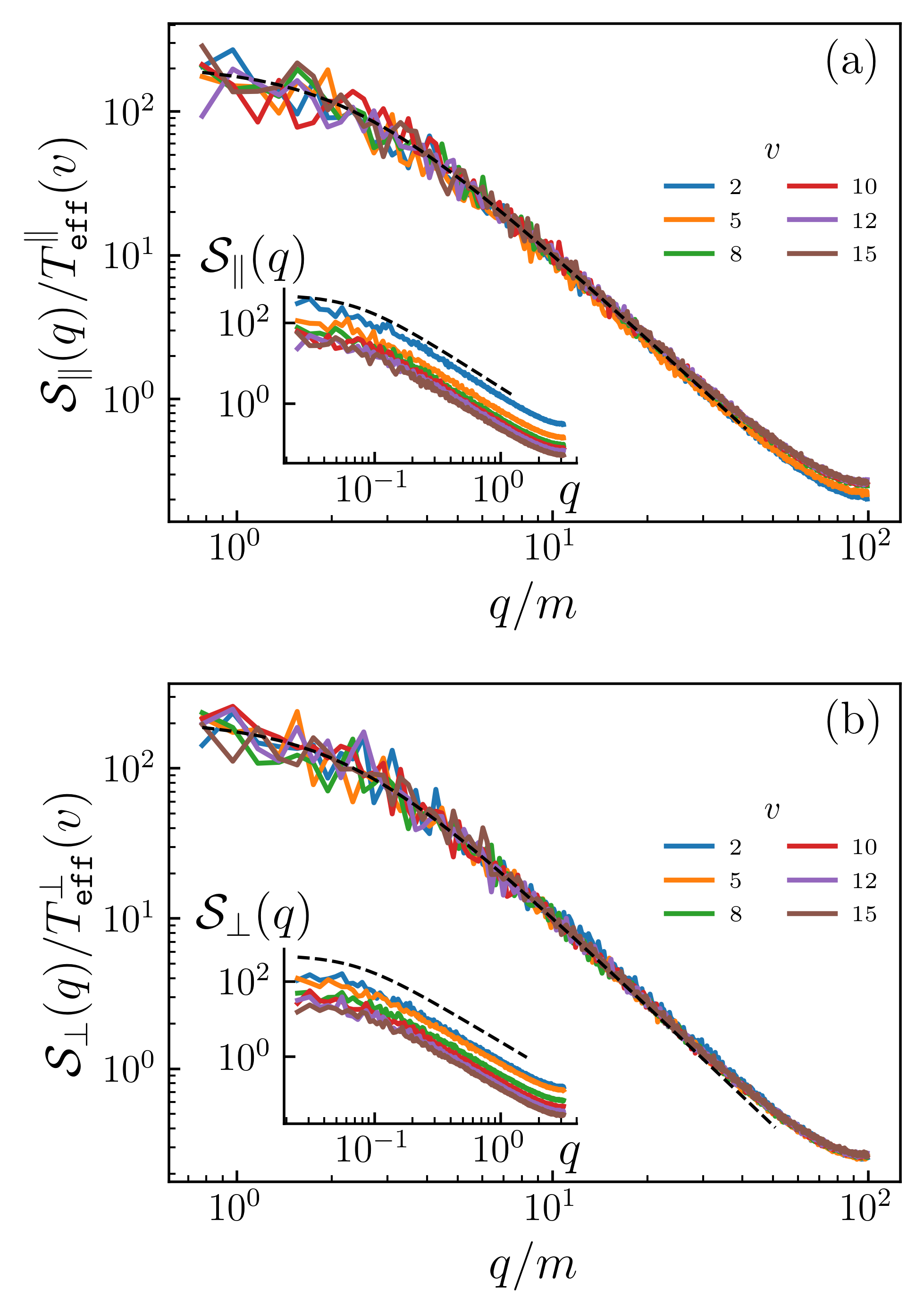}
\caption{
Rescaled structure factors at different (large) velocities for $m^2 = 0.001$ and $L = 1024$ in the longitudinal (a) and transverse (b) directions, for RF disorder. In both cases raw data (insets) can be collapsed by using the effective temperatures $T^\parallel_{\tt eff}(v)$ (a) and $T^\perp_{\tt eff}(v)$ (b). Dashed lines correspond to a purely ``thermal'' $\sim (c q^{2}+m^2)^{-1}$ dependence. 
See Fig.\ref{fig:fastflowSofq} for a   comparison to the RB case.
}
\label{fig:fastflowSofqRF}
\end{figure}%

The crossover to the fast-flow regime reveals some important differences between RB and RF.  
In Fig. \ref{fig:fastflowSofqRF} we show that 
effective temperatures in both directions are well-defined and  rescale the structure factor for an extended range of velocities.
This result can be compared directly to Fig. \ref{fig:fastflowSofq} for the RB case.
\begin{figure}
\centering
\includegraphics[width=.99\columnwidth]{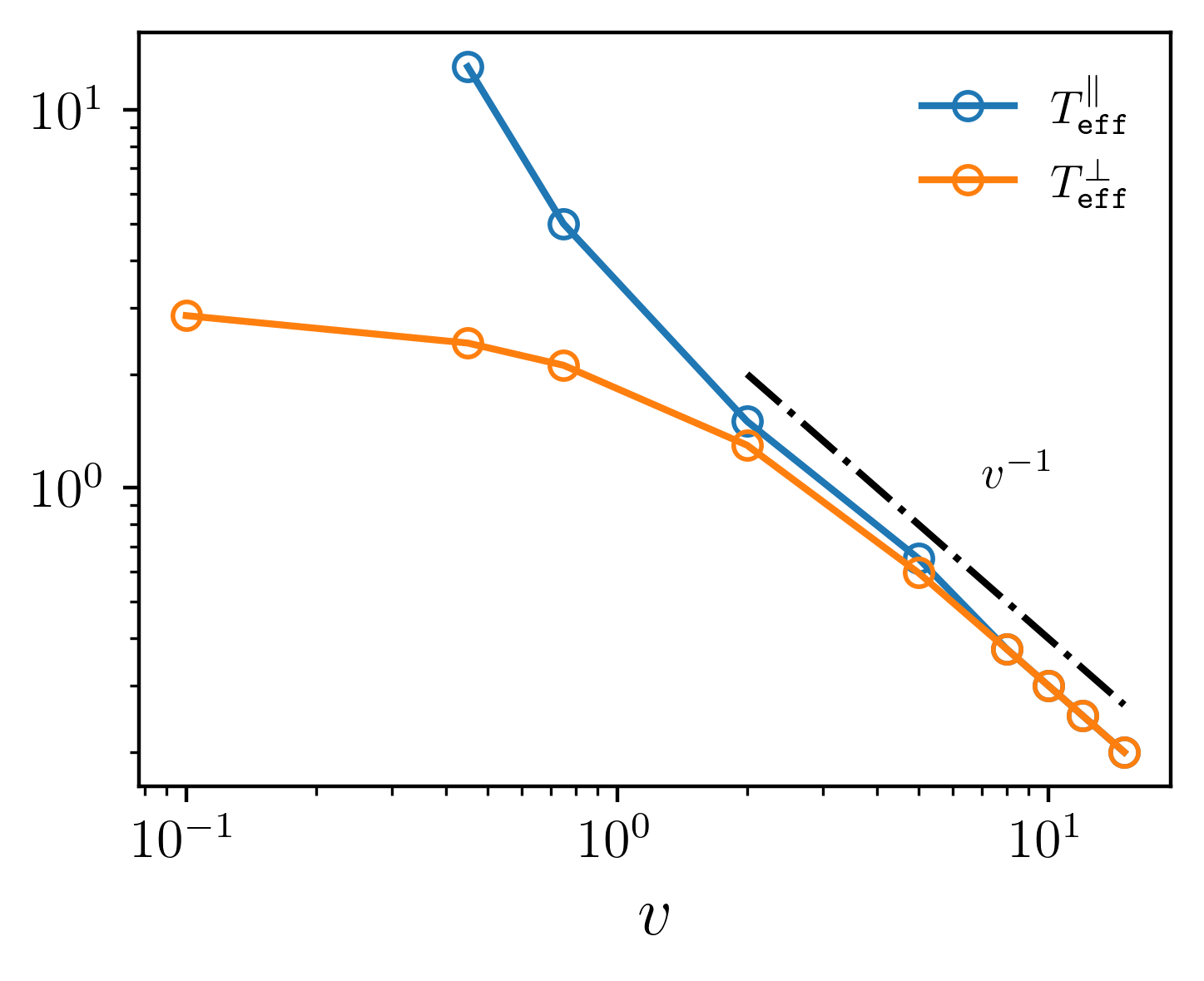}
\caption{Effective temperatures as a function of the mean velocity for RF disorder. Dashed-dotted lines indicate the asymptotic behavior.}
\label{fig:Teff_RF}
\end{figure}

\begin{figure}
\centering
\includegraphics[width=.99\columnwidth]{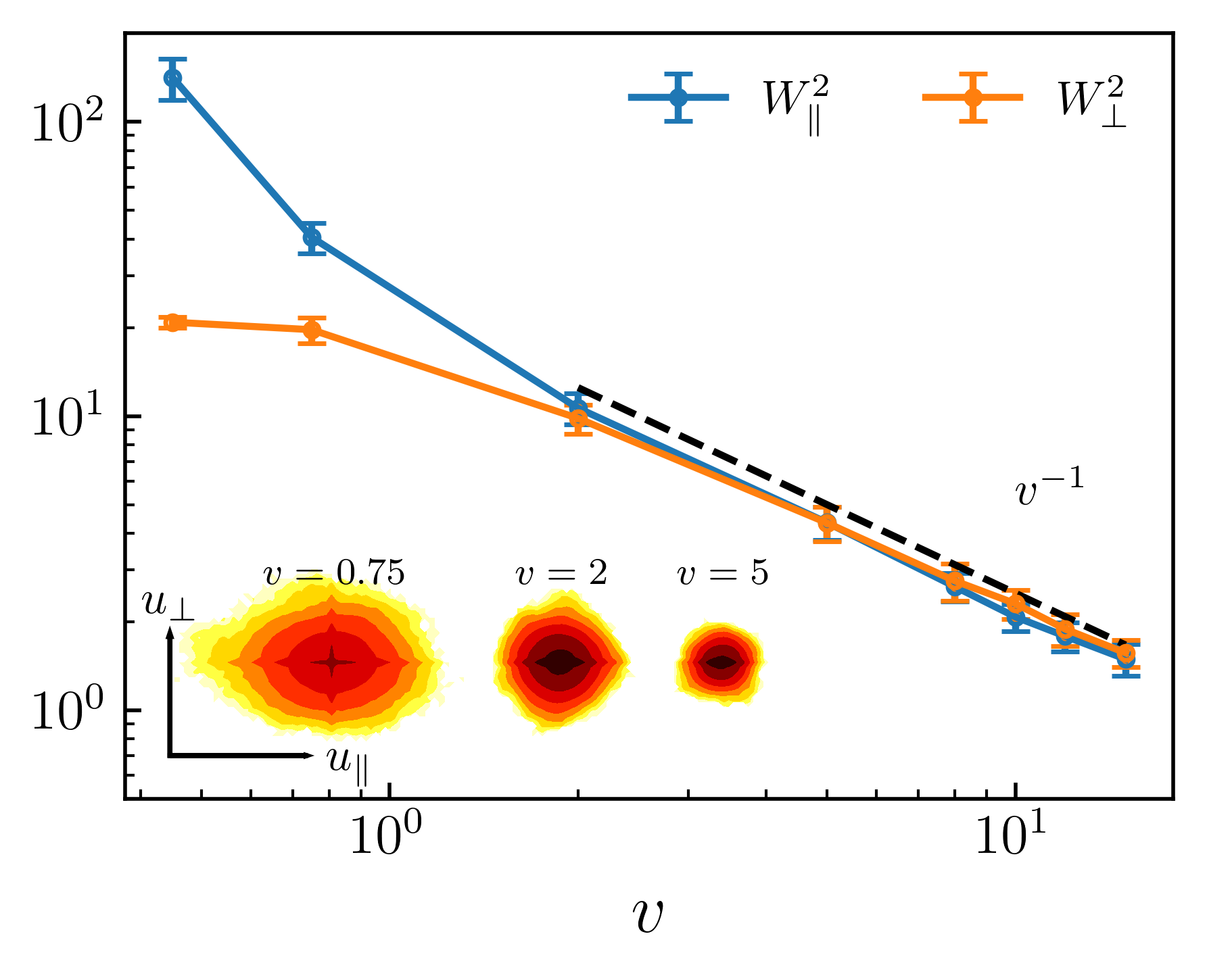}
\caption{Mean squared widths as a function of mean velocity for RF disorder. Dashed-dotted lines indicate the asymptotic behavior. The inset displays the joint local-displacement distribution for three velocities.}
\label{fig:wv_RF}
\end{figure}

In Figs.~\ref{fig:Teff_RF} and \ref{fig:wv_RF} for the anisotropic effective temperatures and global widths we  see that at intermediate velocities the RF case is qualitatively similar to the RB case, in the sense that $T_{\tt eff}^\parallel >  T_{\tt eff}^\perp$ and $W_{\parallel}^2 > W_{\perp}^2 \sim 1/v$. For small velocities, the local displacement distribution is asymmetric in the longitudinal direction, as was shown in Fig.~\ref{fig:pdfs} for the RB case. In contrast, for larger velocities the   curves for the different directions no longer cross at a characteristic velocity (see Figs.~\ref{fig:Teff_RF} and \ref{fig:wv_RF}), but directly merge into an isotropic decay at large velocities, with $T_{\tt eff}^\parallel \approx T_{\tt eff}^\perp \sim 1/v$ and $W_{\parallel}^2 \approx W_{\perp}^2 \sim 1/v$. Isotropic RF  disorder thus produces isotropic fluctuations at large velocities, in  contrast with the   anisotropic fluctuations in the RB case. 

\section{Discussion and conclusions}
\label{sc:conclusion}

We   studied depinning and flow of a flux lines with harmonic elasticity in an isotropic random medium with short-range correlated disorder. 
We   report novel phenomena, such as the asymmetry of local parallel displacements at low velocities, the inversion of the  aspect ratio of widths in the RB case, and important  differences between   RB and RF in the fast-flow regime. 

For   quasistatic driving   we   calculated several universal quantities. In Table \ref{table:criticalexps} we summarize the values of all  critical exponents that we measured, and the relations   between them according to our numerical tests. Some  critical exponents differ appreciably from previous reports. Our value $\beta= 0.24(1)$ differs   from $\beta \approx 0.31$ or $\beta=0$ given Ref.~\cite{ErtasKardar1996}, 
but   is indistinguishable from the one  for interfaces in two-dimensional random media~\cite{FerreMetaxasMouginJametGorchonJeudy2013}. 
The value $\zeta_{\parallel}\approx 1.25(1)$   contrasts with $\zeta_{\parallel}=1$ from Ref.~\cite{ErtasKardar1996}, agrees with the one reported in Ref.~\cite{koshelevkolton2011} for strong disorder, and     is indistinguishable from the one for one dimensional interfaces~\cite{Leschhorn1993,RossoHartmannKrauth2003,FerreMetaxasMouginJametGorchonJeudy2013}. This result is physically relevant as $\zeta_\parallel>1$ implies the breakdown of linear elasticity at large length scales. 
Some proposed scaling relations do not pass our numerical tests, particularly 
$\zeta_{\perp}=\zeta_{\parallel}-1/2$ \cite{ErtasKardar1996} and $\zeta_{\perp}=5\zeta_{\parallel}/2-2$~\cite{Kardar1998}.  Other relations predicted in Ref.~\cite{ErtasKardar1996} are     verified, as shown in Table \ref{table:criticalexps}. We     added the tested relation  for $\tau_\parallel$, which is identical to the one for interfaces in two dimensions~\cite{NarayanFisher1993,RossoLeDoussalWiese2009}. A new relation links $\tau_\perp$ to $\tau_\parallel$.
In spite of   differences in the above scaling relations, the main message is that  the Ertas-Kardar planar approximation is   working well, provided we use the   appropriate results     for the $(D,N)=(1,1)$ case~\cite{FerreroBustingorryKolton2013}, and correct the roughness exponent for the transversal direction to $\zeta_\perp = (2-d)/2$.
We explicitly verified that for $(D,N)=(1,2)$, microscopic  RB and RF disorder  lead to  a single RF universality class at depinning, a result we     expect from the planar approximation.

\begin{table}[b]
\vspace{0.2cm}
\centering
\begin{tabular}{||c | c | c |c ||} 
 \hline
 $\zeta_\parallel$ & $1.25 \pm 0.01$ & Fig.~\ref{fig:Sq_para_vs_m}(a) & \\
 $\zeta_\perp$ & $0.5 \pm 0.01$ & Fig.~\ref{fig:Sq_para_vs_m}(b) & \\ 
 $z_\parallel$ & $1.43\pm 0.01$  & Fig.~\ref{fig:nonsteadyw2yv}(a) & \\
 $z_\perp$ & $2.27\pm 0.05$ &  Fig.~\ref{fig:nonsteadyw2yv}(a) & $z_\perp=z_\parallel+1/\nu$\\
 $\nu$  & $1.33 \pm 0.02$ & Fig.~\ref{fig:Sq_para_vs_v}(a) & $\nu=1/(2-\zeta_\parallel)$\\
 $\beta$ & $0.24 \pm 0.01$ & Fig.~\ref{fig:nonsteadyw2yv}(b) & $\beta=\nu(z_\parallel-\zeta_\parallel)$\\
 $\tau_\parallel$ & $1.09\pm 0.03$ & Fig.~\ref{fig:ava_para_vs_m}(a) & $\tau_\parallel = 2-2/(1+\zeta_\parallel)$ \\
 $\tau_\perp$ & $1.17\pm 0.06$ & Fig.~\ref{fig:ava_para_vs_m}(b) & $\tau_\perp = 2\tau_\parallel-1$ \\
 \hline
\end{tabular}
\caption{Critical depinning exponents obtained in this work. We also indicate the figure where the exponent was fitted or tested and the scaling relations that hold within our numerical uncertainty. The same exponents and relations hold both for   RB and RF  disorder.}
\label{table:criticalexps}
\end{table}

For intermediate driving velocities we show that the transverse local fluctuations are   Gaussian, and that the structure of the elastic string is described by a single exponent $\zeta_\perp=1/2$, together with a well-defined effective temperature that tends to saturate at small velocities and   vanishes as $T_{\tt eff}^{\perp}\sim 1/v$ at large velocities. This supports the identification of a transverse ``shaking temperature'' in Ref.~\cite{NattermannScheidl2000} as a   limit of the transverse effective temperature we define, and which is valid for all finite velocities.
Local longitudinal  fluctuations are    skewed at low velocities and become Gaussian  at large ones, with an effective longitudinal temperature vanishing as $T_{\tt eff}^{\parallel}\sim 1/v^3$ for RB disorder. 
At large velocities the correlation length becomes   small and the interface is essentially flat in the driving direction (see Fig.~\ref{fig:widths}).
This   result is   inconsistent with the prediction of a zero ``longitudinal shaking temperature''   in Ref.~\cite{NattermannScheidl2000}. 
The difference may be attributed to the fact that the latter calculation neglects terms of order $\ca O(1/v^2)$ and for RB disorder   only contains the  leading term proportional to $\int_u \Delta(u)=0$. 
On the other hand, the $1/v^3$ behavior is  inconsistent with the prediction of an  ``Edwards-Wilkinson temperature'' proportional to $ 1/v$~\cite{ChauveGiamarchiLeDoussal2000}, if we assume that the planar approximation   holds in this regime. This discrepancy is due to the use of a $v$-independent RF disorder in the high-velocity regime, ignoring that the microscopic  disorder is RB, and that the driving velocity reduces the effects of the RG flow bringing it to RF.
In contrast, we
   confirmed numerically the  analytical expectation  that   microscopic RF disorder produces an isotropic effective temperature vanishing as $T_{\tt eff}^{\parallel}\approx T_{\tt eff}^{\perp} \sim 1/v$.

The same dependencies in the effective temperatures  are observed in the diffusion  of a single monomer driven in  2d RB disorder~\cite{kolton2006}.
This   suggests that the longitudinal effective temperature in the large-velocity regime is controlled by what happens for a single  monomer. 
We confirm that in the comoving frame the string can be  described as a two-component Edwards-Wilkinson line with uncorrelated noise controlled by $v$ as predicted in \cite{NattermannScheidl2000,ChauveGiamarchiLeDoussal2000},   with an   anisotropy that depends as discussed on the  microscopic disorder. These results   show that the  nature of the microscopic disorder can be detected by observing the anisotropic fluctuations in the fast-flow regime.

\begin{figure}[t]
    \centering
    \includegraphics[width=.95\columnwidth]{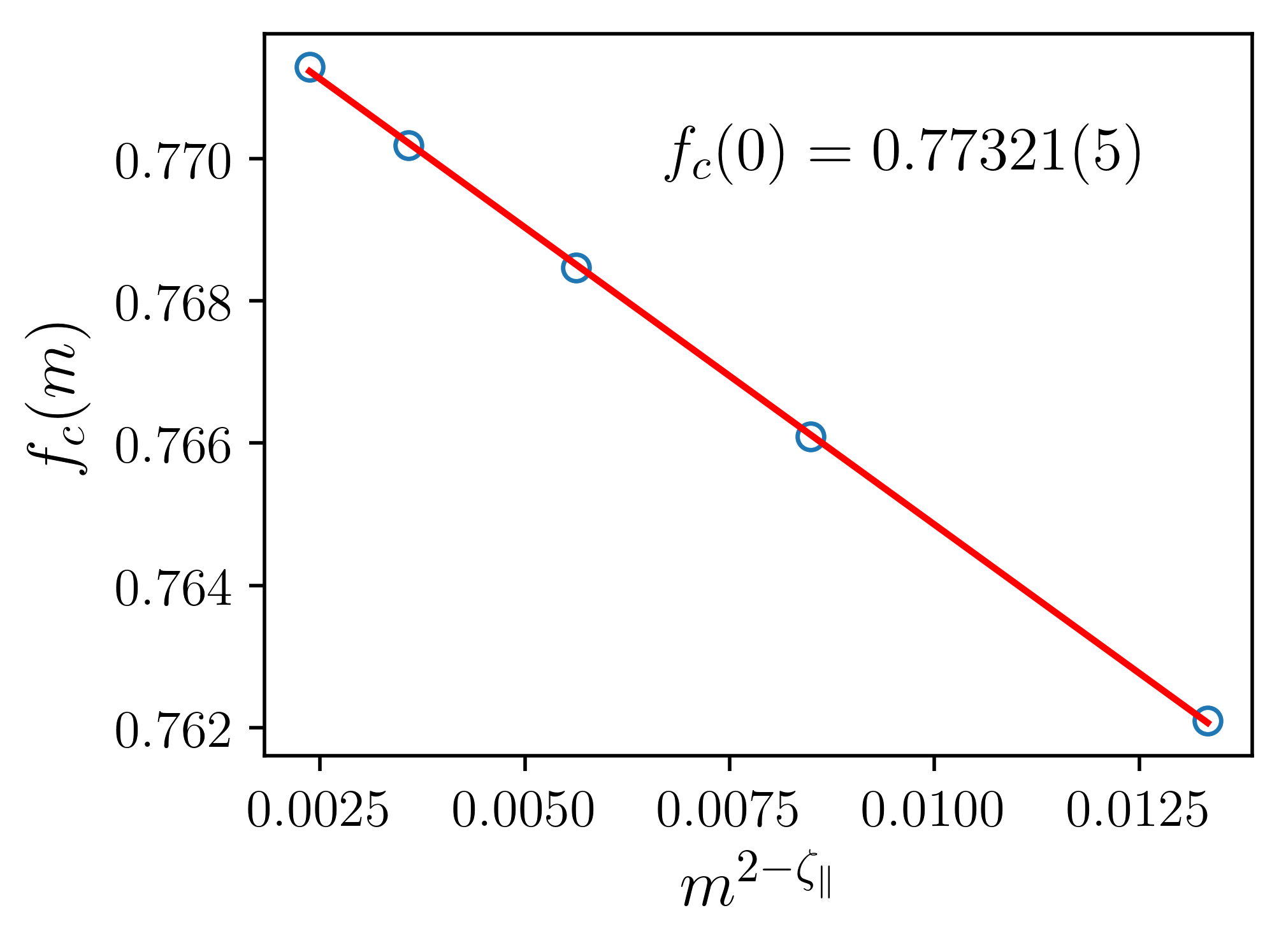}
    \caption{\label{fig:fc_vs_m} Critical force as a function of   mass  behaves as  $f_{c}(m) \sim m^2 u_m = m^{2-\zeta_\parallel}$. Extrapolated   to $m = 0$ it gives the zero-mass critical force $f_c(0) = 0.77321(5)$.}
    
\end{figure}

Our results should be relevant for flux lines or other elastic lines in random media, such as  polymers driven in random quenched media, or cracks. Since cracks have long-range elasticity, we expect the transversal roughness to be logarithmic. 
This agrees with \cite{RamanathanErtasFisher1997}, and was experimentally observed in \cite{DalmasLelargeVandembroucq2008}.

\begin{acknowledgements}
We thank L.~Ponson for discussions. We acknowledge financial support through grants 
PICT 2016-0069, PICT-2019-01991, and SIIP-Uncuyo 06/C578. This work used computational resources from CCAD - Universidad Nacional de Cordoba, and from the Physics Department-Centro Atomico Bariloche, both of which are part of SNCAD - MinCyT, República Argentina. 
\end{acknowledgements}

\appendix
\section{Force-controlled driving versus velocity-controlled driving}
\label{Force-controlled driving versus velocity-controlled driving}

    Using the velocity-controled driving we performed a set of simulations for different masses and   velocities. In Fig.~\ref{fig:Sq_para_vs_v} we show the depinning transition with exponents $\zeta_\parallel = 1.25$, $\nu = 1.33$ and $\beta = 0.33$. The critical force fluctuates around  $f_c(m) = m^2 \langle v t - u_\parallel \rangle$, and this scales with the mass as $f_c(m) \approx m^2 u_m = m^{2-\zeta_\parallel}$ as shown in Fig.~\ref{fig:fc_vs_m}. Fitting this relation,  we extract the zero-mass critical force  $f_c(0) = 0.77321(5)$.

\begin{figure}[t]
\centering
\includegraphics[width=\columnwidth]{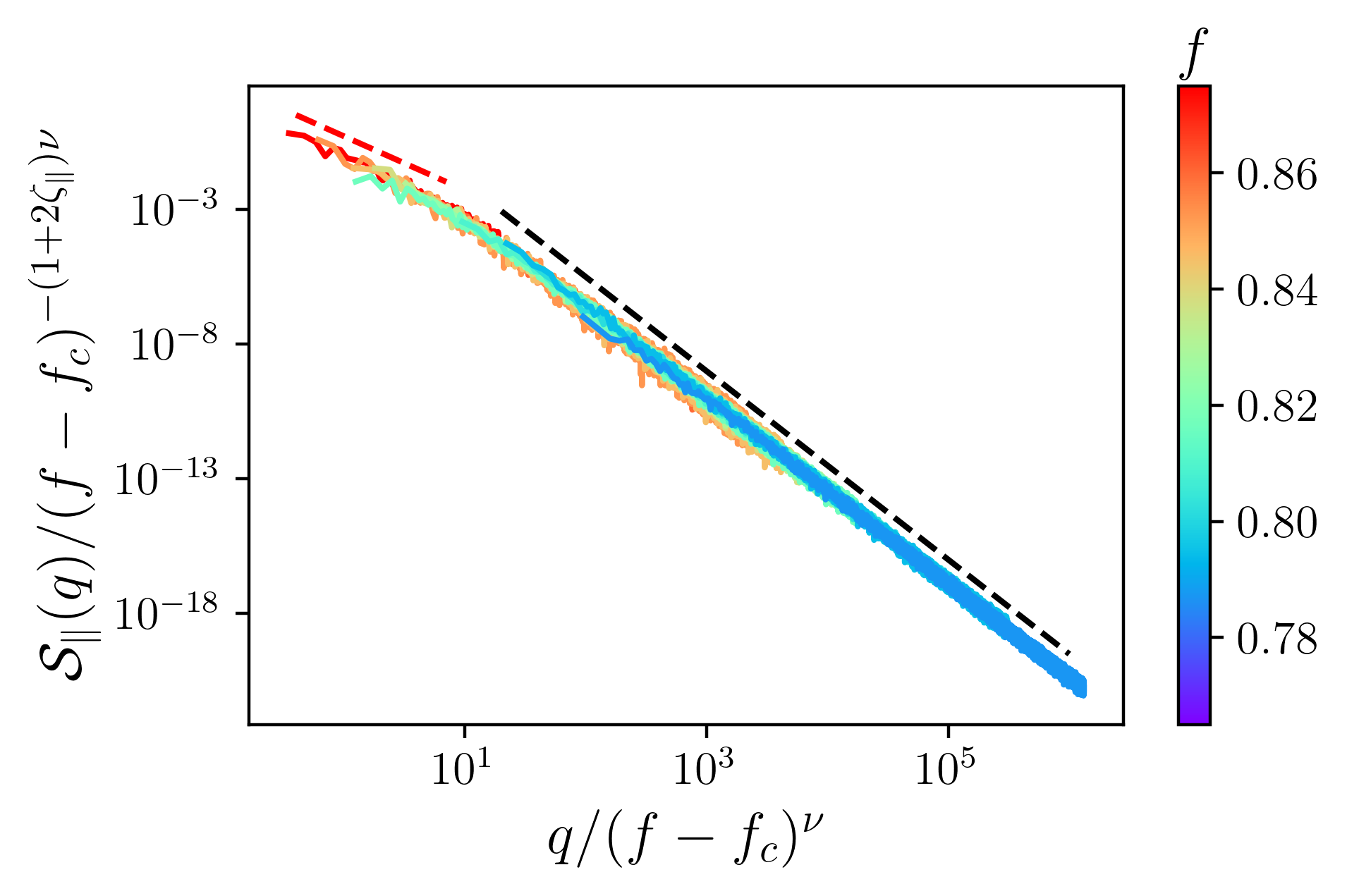}
\caption{Scaled steady-state structure factor for the constant-force ensemble according to the deppinning length $l_\parallel \sim v(f)^{-\frac{\nu}{\beta}} = [f-f_c(0)]^{-\nu}$ with $\zeta_\parallel = 1.25$ , $\nu = 1.33$ and zero-mass critical force $f_c(0) = 0.77321(5)$. Red dashed lines indicate the fast-flow
roughness exponent $\zeta_{\rm ff} = 0.5$, and the black dashed-lines the depinning roughness exponent $\zeta_\parallel = 1.25$.}
\label{fig:contant_force}
\end{figure}

With the same parameters we perform a set of simulations in the force-controled driving ensamble. 
In Fig. \ref{fig:contant_force} we show the structure factor for the parallel direction scaled acording to the depping length $l_\parallel \sim v(f)^{-\frac{\nu}{\beta}} = (f-f_c)^{-\nu}$ with the same scaling exponents $\zeta_\parallel$ , $\nu$ and critical force $f_c = 0.7656$. The distance to the zero-mass critical force is $|f_c(0)-f_c| \approx 0.0076$.

\section{Correlations of an Ornstein-Uhlenbeck process}
\label{Correlations of an Ornstein-Uhlenbeck process}
We wrote the equation of motion 
\bea
\partial _{u_\parallel} u_\perp(x, u_\parallel) &=& - m^2 u_\perp (x,u_\parallel) + \nabla^2 u_\perp(x,u_\parallel)  \nn\\
&& + \sqrt{2 \sigma  } \eta (x,t), \\
\left< \eta (x,u_\parallel) \eta(x',u'_\parallel)\right> &=& \delta^d(x-x') \delta(u_\parallel-u_\parallel').
\eea
Integrating over $x$, dividing by $L^d$ and replacing $u_\parallel \to w$ yields
\bea
\partial_{w} u_\perp(w) &=& - m^2 u_\perp(w) + \sqrt {\frac{  2\sigma}{L^{d}}} \eta(w),\\
\left< \eta(w) \eta(w') \right> &=& \delta(w-w').
\eea
This is an Ornstein-Uhlenbeck process,  solved by 
\be
u_\perp(w) = \int_{-\infty}^w \rmd w_1 \, \rme^{- m^2  (w-w_1)}\xi(w_1) .
\ee
It leads to force correlations 
\bea
&&\Delta_{\perp}(w-w') = m^4 L^d \left<  {u_\perp (w) u_\perp (w')} \right> \nn\\
&&= m^4 L^d \int \limits_{-\infty}^w \rmd w_1 \int\limits_{-\infty}^{w'} \rmd w_2\, \rme^{-m^2 (w+w' -w_1-w_2)} \left< \eta(w_1)\eta(w_2) \right>
\nn
\\
&&= 2\sigma m^4  \int_{-\infty}^{{\rm min}(w,w')} \rmd \tilde w \,\rme^{- m^2(w+w' - 2\tilde w)}  \nn
\\
&&=   \sigma m^2 \,\rme^{-m^2 |w-w'|}.
\eea

\section{Single-monomer diffusion in the commoving frame}
\label{Single monomer diffusion}

Let us consider an overdamped particle driven by a force $f$ in a one-dimensional space with quenched random forces:
\begin{equation}
    \eta \dot {x} = {F}({x})+ {f},
\end{equation}
where ${F}(x)$ is a short-range correlated quenched random force field such that
\begin{eqnarray}
\overline{{F}({x})}&=&0,\\
\overline{{F}({x}){F}({x'})}&=& f_0^2 g(|{x}-{x'}|/d_0).
\end{eqnarray}
Here $d_0$ is a characteristic length, 
$f_0$ a characteristic force amplitude, and $g(u)$ a rapidly decaying function of unit range and unit amplitude. 
Without loss of generality 
we can adimensionalize the equation of motion by measuring distances in units of $d_0$, forces in units of $f_0$ and time in units of $\tau_0=\eta d_0 /f_0$, such that 
\begin{equation}
    \dot x = F(x)+ f.
    \label{eq:monomereq}
\end{equation}
We   consider two toy models, one for  RF and  one for RB disorder, in which the force fields
 are piecewise constant.

\subsection{RF disorder}
To construct a RF disorder such that $\int_y g(y)>0$ we  take
\begin{eqnarray}
F(x)=R_{[x]}
\end{eqnarray}
where $[\dots]$ denotes the integer part. 
The $R_n$ are uniformely distributed 
random numbers in the interval $[-1,1]$
such that 
\begin{eqnarray}
\overline{R_n R_m} &=&\frac{\delta_{n,m}}{3} .
\end{eqnarray}
From Eq.~(\ref{eq:monomereq})
the time spent in the interval $n$
is 
\begin{eqnarray}
\Delta t_n = \frac{1}{R+f} , 
\end{eqnarray}
where $R \equiv R_n$, and hence
\begin{eqnarray}
\langle \Delta t\rangle &=&
\frac{1}{2} \int_{-1}^{1}
\frac{1}{R+f} \;\rmd R 
= \frac{1}{2}\ln \left(\frac{f+1}{f-1}\right), \\
\langle \Delta t^2 \rangle &=&
\frac{1}{2} \int_{-1}^{1}
\frac{1}{(R+f)^2} \; \rmd R 
= \frac{1}{f^2-1}.\qquad 
\end{eqnarray}
The mean velocity is  
\begin{eqnarray}
v = \frac{1}{\langle \Delta t \rangle}=
\frac{2}{\ln \left(\frac{f+1}{f-1}\right)}, 
\end{eqnarray}
displaying a   depinning transition at $f=1$, while for $f\gg 1$, $v\simeq f$ as expected. The (differential) mobility is  
\begin{eqnarray}
\mu:= \frac{\rmd v}{\rmd f} = 
\frac{4}{
(f^2-1) \ln \left(\frac{f+1}{f-1}\right)^2},
\end{eqnarray}
such that $\mu \to \infty$ when $f\to 1$ and 
$\mu\to 1$ when $f\to \infty$.
The diffusion constant in the commoving frame
is $D \equiv \langle [x-v t]^2 \rangle/t= \langle [1-v \Delta t]^2 \rangle$
and can     be expressed   in terms of $\langle\Delta t\rangle$ and $\langle\Delta t^2\rangle$ as 
\begin{eqnarray}
D = \frac{\langle\Delta t^2\rangle-\langle\Delta t\rangle^2}{\langle\Delta t\rangle^3}.
\label{eq:defDifu}
\end{eqnarray}
Using the generalized Einstein relation we   get the effective temperature as $T_{\tt eff}=D/\mu$.

We are interested in the $f\gg 1$ fast-flow behaviour of $D$ and $T_{\tt eff}$. Expanding in powers of $1/f$ we get
\begin{eqnarray}
D &\simeq& \frac{1}{3f} + \frac{7}{45 f^3 } + \mathcal{O}(f^{-5}),\\
T_{\tt eff}&\simeq& \frac{1}{3f}+\frac{2}{45f^3}
+\ca O(f^{-5}).
\end{eqnarray}
Recovering the physical dimensions, we   get at the lowest order, that $D \sim  (d_0^2/\tau_0) (f_0/f)=(f_0 d_0/\eta) (f_0/f)\sim d_0 v^2_0/v$, with $v_0=f_0/\eta$.

\subsection{RB disorder}
To model  RB disorder 
with $\int_y g(y)=0$, 
we define the random forces in Eq.~(\ref{eq:monomereq}) as
\begin{eqnarray}
    F(x) = R_{[x]} \sign(x-[x]-1/2)+f.
\end{eqnarray}
As above  $R_n$ are iid random variables, uniformly distributed in  $[-1,1]$.
By repeating the   procedure of the previous section we obtain   $\langle \Delta t \rangle$ as   for the RF case, 
and thus identical $v$ and $\mu$ as a function of $f$. However, $\langle \Delta t^2 \rangle$ is different,
\begin{eqnarray}
\langle \Delta t^2 \rangle &=&
\frac{1}{2} \int_{-1}^{1}
\left[ \frac1{ 2(f+R)}  + \frac1{2 (f-R)} \right] ^2\; \rmd R 
\nn\\
&=& 
\frac{1}{4 f}
 \left[\frac{2f}{f^2-1}+\ln \left(\frac{f+1}{f-1}\right)\right].
\end{eqnarray}
This leads to different asymptotic behaviors,
\begin{eqnarray}
D \simeq T_{\tt eff} \simeq \frac{4}{45 f^3}   + \mathcal{O}(f^{-5}). 
\end{eqnarray}
Recovering physical dimensions we get $D\sim d_0 v_0 (f_0/f)^3 \sim 
d_0 v_0^4/v^3$, with $v_0=f_0/\eta$.

\section{Notations}
\begin{table}
\caption{Glossary}
\begin{tabularx}{0.5\textwidth}{@{}XX@{}}
\toprule
  Symbols \\
  $d$ & Internal dimension\\
  $D$ & Space dimension\\
  $N$ & Number of displacement components\\
  $\alpha$ & Direction respect to the driving force $\alpha=\perp,\parallel$ \\
  $x$ & Internal coordinate \\
  $t$ & Time \\
  $v$ & Mean velocity in the direction of the drive\\
  $\zeta_\alpha$ & Depinning roughness exponents \\
  $\zeta_{\tt ff}$ & Fast flow roughness exponent \\
  $\beta$ & Velocity exponent\\
  $z_\alpha$ & Dynamical exponents \\
  $\nu$ & Depinning correlation length exponent\\
  $\tau_{\alpha}$ & Avalanche exponents\\
  ${\bf u}$ & Local displacement vector\\
  ${\bf F}_p$ & Local pinning force vector\\
  ${\bf f}$ & Driving force vector\\
  $T^{\alpha}_{\tt eff}$ & Effective temperatures \\
  $u_\alpha$ & Local displacement components \\
  $V$ & Random quenched potential \\
  $u$ & Center of mass displacement in the drive direction\\
  $f_c$ & Critical depinning force\\
  $m^2$ & Curvature of the driving parabolic potential\\
  $w$ & Center of the driving parabolic potential\\
  $\delta w$ & Discretization of $w$\\
  $\chi$ & Static linear response function associated with the structure factor \\
  $\mu_{3,\alpha}$ & Skewness parameter of local displacement fluctuations\\
  $k_{\alpha}$ & Kurtosis parameter of local displacement fluctuations\\
  $W_\alpha$ & Global widths \\
  $\Delta$ & Microscopic pinning force-force correlator\\
  $R$ & Microscopic pinning potential-potential correlator\\  
  $B_\alpha$ & Displacement correlation function\\
  $p_\alpha$ & Avalanche size reduced PDFs\\
  $S_\alpha$ & Structure factors\\
  ${\cal S}^\alpha_m$ & Avalanche size cutoffs\\
  $\xi_m$ & Confining potential characteristic length-scale\\  
  $P_\alpha$ & PDF of center of mass jumps in the $\alpha$ direction \\  
  $\delta u_\alpha$ & local displacement relative to center of mass\\  
  $\sigma$ & Effective noise intensity \\
  $f^*$ & Characteristic force in the waiting time distribution \\
  $P^w$ & Waiting time PDF \\
  $P^u$ & Local displacement PDF \\
  $\xi$ & Depinning correlation length \\
  RB & Random-Bond type of disorder \\
  RF & Random-Field type of disorder \\
  FRG & Functional Renormalization Group\\		
\bottomrule
\end{tabularx}
\end{table}

\typeout{}
\bibliography{citation,biblio}

\begin{thebibliography}{94}%
\makeatletter
\providecommand \@ifxundefined [1]{%
 \@ifx{#1\undefined}
}%
\providecommand \@ifnum [1]{%
 \ifnum #1\expandafter \@firstoftwo
 \else \expandafter \@secondoftwo
 \fi
}%
\providecommand \@ifx [1]{%
 \ifx #1\expandafter \@firstoftwo
 \else \expandafter \@secondoftwo
 \fi
}%
\providecommand \natexlab [1]{#1}%
\providecommand \enquote  [1]{``#1''}%
\providecommand \bibnamefont  [1]{#1}%
\providecommand \bibfnamefont [1]{#1}%
\providecommand \citenamefont [1]{#1}%
\providecommand \href@noop [0]{\@secondoftwo}%
\providecommand \href [0]{\begingroup \@sanitize@url \@href}%
\providecommand \@href[1]{\@@startlink{#1}\@@href}%
\providecommand \@@href[1]{\endgroup#1\@@endlink}%
\providecommand \@sanitize@url [0]{\catcode `\\12\catcode `\$12\catcode
  `\&12\catcode `\#12\catcode `\^12\catcode `\_12\catcode `\%12\relax}%
\providecommand \@@startlink[1]{}%
\providecommand \@@endlink[0]{}%
\providecommand \url  [0]{\begingroup\@sanitize@url \@url }%
\providecommand \@url [1]{\endgroup\@href {#1}{\urlprefix }}%
\providecommand \urlprefix  [0]{URL }%
\providecommand \Eprint [0]{\href }%
\providecommand \doibase [0]{https://doi.org/}%
\providecommand \selectlanguage [0]{\@gobble}%
\providecommand \bibinfo  [0]{\@secondoftwo}%
\providecommand \bibfield  [0]{\@secondoftwo}%
\providecommand \translation [1]{[#1]}%
\providecommand \BibitemOpen [0]{}%
\providecommand \bibitemStop [0]{}%
\providecommand \bibitemNoStop [0]{.\EOS\space}%
\providecommand \EOS [0]{\spacefactor3000\relax}%
\providecommand \BibitemShut  [1]{\csname bibitem#1\endcsname}%
\let\auto@bib@innerbib\@empty
\bibitem [{\citenamefont {Durin}\ and\ \citenamefont
  {Zapperi}(2006)}]{DurinZapperi2006b}%
  \BibitemOpen
  \bibfield  {author} {\bibinfo {author} {\bibfnamefont {G.}~\bibnamefont
  {Durin}}\ and\ \bibinfo {author} {\bibfnamefont {S.}~\bibnamefont
  {Zapperi}},\ }\bibfield  {title} {\bibinfo {title} {{The Barkhausen
  effect}},\ }in\ \href {http://arxiv.org/abs/cond-mat/0404512} {\emph
  {\bibinfo {booktitle} {The Science of Hysteresis}}},\ \bibinfo {editor}
  {edited by\ \bibinfo {editor} {\bibfnamefont {G.}~\bibnamefont {Bertotti}}\
  and\ \bibinfo {editor} {\bibfnamefont {I.}~\bibnamefont {Mayergoyz}}}\
  (\bibinfo {address} {Amsterdam},\ \bibinfo {year} {2006})\ p.~\bibinfo
  {pages} {51},\ \Eprint {https://arxiv.org/abs/cond-mat/0404512}
  {arXiv:cond-mat/0404512 [cond-mat]} \BibitemShut {NoStop}%
\bibitem [{\citenamefont {Ferr\'{e}}\ \emph {et~al.}(2013)\citenamefont
  {Ferr\'{e}}, \citenamefont {Metaxas}, \citenamefont {Mougin}, \citenamefont
  {Jamet}, \citenamefont {Gorchon},\ and\ \citenamefont
  {Jeudy}}]{FerreMetaxasMouginJametGorchonJeudy2013}%
  \BibitemOpen
  \bibfield  {author} {\bibinfo {author} {\bibfnamefont {J.}~\bibnamefont
  {Ferr\'{e}}}, \bibinfo {author} {\bibfnamefont {P.}~\bibnamefont {Metaxas}},
  \bibinfo {author} {\bibfnamefont {A.}~\bibnamefont {Mougin}}, \bibinfo
  {author} {\bibfnamefont {J.-P.}\ \bibnamefont {Jamet}}, \bibinfo {author}
  {\bibfnamefont {J.}~\bibnamefont {Gorchon}},\ and\ \bibinfo {author}
  {\bibfnamefont {V.}~\bibnamefont {Jeudy}},\ }\bibfield  {title} {\bibinfo
  {title} {Universal magnetic domain wall dynamics in the presence of weak
  disorder},\ }\href
  {https://doi.org/https://doi.org/10.1016/j.crhy.2013.08.001} {\bibfield
  {journal} {\bibinfo  {journal} {Comptes Rendus Physique}\ }\textbf {\bibinfo
  {volume} {14}},\ \bibinfo {pages} {651 } (\bibinfo {year} {2013})},\ \bibinfo
  {note} {disordered systems / Syst{\`e}mes d{\'e}sordonn{\'e}s}\BibitemShut
  {NoStop}%
\bibitem [{\citenamefont {Durin}\ \emph {et~al.}(2016)\citenamefont {Durin},
  \citenamefont {Bohn}, \citenamefont {Correa}, \citenamefont {Sommer},
  \citenamefont {Doussal},\ and\ \citenamefont
  {Wiese}}]{DurinBohnCorreaSommerLeDoussalWiese2016}%
  \BibitemOpen
  \bibfield  {author} {\bibinfo {author} {\bibfnamefont {G.}~\bibnamefont
  {Durin}}, \bibinfo {author} {\bibfnamefont {F.}~\bibnamefont {Bohn}},
  \bibinfo {author} {\bibfnamefont {M.}~\bibnamefont {Correa}}, \bibinfo
  {author} {\bibfnamefont {R.}~\bibnamefont {Sommer}}, \bibinfo {author}
  {\bibfnamefont {P.~L.}\ \bibnamefont {Doussal}},\ and\ \bibinfo {author}
  {\bibfnamefont {K.}~\bibnamefont {Wiese}},\ }\bibfield  {title} {\bibinfo
  {title} {Quantitative scaling of magnetic avalanches},\ }\href
  {https://doi.org/10.1103/PhysRevLett.117.087201} {\bibfield  {journal}
  {\bibinfo  {journal} {Phys. Rev. Lett.}\ }\textbf {\bibinfo {volume} {117}},\
  \bibinfo {pages} {087201} (\bibinfo {year} {2016})},\ \Eprint
  {https://arxiv.org/abs/arXiv:1601.01331} {arXiv:1601.01331} \BibitemShut
  {NoStop}%
\bibitem [{\citenamefont {Kleemann}(2007)}]{Kleemann2007}%
  \BibitemOpen
  \bibfield  {author} {\bibinfo {author} {\bibfnamefont {W.}~\bibnamefont
  {Kleemann}},\ }\bibfield  {title} {\bibinfo {title} {Universal domain wall
  dynamics in disordered ferroic materials},\ }\href
  {https://doi.org/10.1146/annurev.matsci.37.052506.084243} {\bibfield
  {journal} {\bibinfo  {journal} {Annu. Rev. Mater. Res.}\ }\textbf {\bibinfo
  {volume} {37}},\ \bibinfo {pages} {415} (\bibinfo {year} {2007})}\BibitemShut
  {NoStop}%
\bibitem [{\citenamefont {Paruch}\ and\ \citenamefont
  {Guyonnet}(2013)}]{ParuchGuyonnet2013}%
  \BibitemOpen
  \bibfield  {author} {\bibinfo {author} {\bibfnamefont {P.}~\bibnamefont
  {Paruch}}\ and\ \bibinfo {author} {\bibfnamefont {J.}~\bibnamefont
  {Guyonnet}},\ }\bibfield  {title} {\bibinfo {title} {Nanoscale studies of
  ferroelectric domain walls as pinned elastic interfaces},\ }\href
  {https://doi.org/10.1016/j.crhy.2013.08.004} {\bibfield  {journal} {\bibinfo
  {journal} {Comptes Rendus Physique}\ }\textbf {\bibinfo {volume} {14}},\
  \bibinfo {pages} {667} (\bibinfo {year} {2013})}\BibitemShut {NoStop}%
\bibitem [{\citenamefont {Bonamy}\ \emph {et~al.}(2008)\citenamefont {Bonamy},
  \citenamefont {Santucci},\ and\ \citenamefont
  {Ponson}}]{BonamySantucciPonson2008}%
  \BibitemOpen
  \bibfield  {author} {\bibinfo {author} {\bibfnamefont {D.}~\bibnamefont
  {Bonamy}}, \bibinfo {author} {\bibfnamefont {S.}~\bibnamefont {Santucci}},\
  and\ \bibinfo {author} {\bibfnamefont {L.}~\bibnamefont {Ponson}},\
  }\bibfield  {title} {\bibinfo {title} {Crackling dynamics in material failure
  as the signature of a self-organized dynamic phase transition},\ }\href
  {https://doi.org/10.1103/PhysRevLett.101.045501} {\bibfield  {journal}
  {\bibinfo  {journal} {Phys. Rev. Lett.}\ }\textbf {\bibinfo {volume} {101}},\
  \bibinfo {eid} {045501} (\bibinfo {year} {2008})}\BibitemShut {NoStop}%
\bibitem [{\citenamefont {Ponson}(2009)}]{Ponson2009}%
  \BibitemOpen
  \bibfield  {author} {\bibinfo {author} {\bibfnamefont {L.}~\bibnamefont
  {Ponson}},\ }\bibfield  {title} {\bibinfo {title} {Depinning transition in
  the failure of inhomogeneous brittle materials},\ }\href
  {https://doi.org/10.1103/PhysRevLett.103.055501} {\bibfield  {journal}
  {\bibinfo  {journal} {Phys. Rev. Lett.}\ }\textbf {\bibinfo {volume} {103}},\
  \bibinfo {pages} {055501} (\bibinfo {year} {2009})}\BibitemShut {NoStop}%
\bibitem [{\citenamefont {Le~Priol}\ \emph {et~al.}(2020)\citenamefont
  {Le~Priol}, \citenamefont {Chopin}, \citenamefont {Le~Doussal}, \citenamefont
  {Ponson},\ and\ \citenamefont
  {Rosso}}]{LePriolChopinLeDoussalPonsonRosso2020}%
  \BibitemOpen
  \bibfield  {author} {\bibinfo {author} {\bibfnamefont {C.}~\bibnamefont
  {Le~Priol}}, \bibinfo {author} {\bibfnamefont {J.}~\bibnamefont {Chopin}},
  \bibinfo {author} {\bibfnamefont {P.}~\bibnamefont {Le~Doussal}}, \bibinfo
  {author} {\bibfnamefont {L.}~\bibnamefont {Ponson}},\ and\ \bibinfo {author}
  {\bibfnamefont {A.}~\bibnamefont {Rosso}},\ }\bibfield  {title} {\bibinfo
  {title} {Universal scaling of the velocity field in crack front
  propagation},\ }\href {https://doi.org/10.1103/PhysRevLett.124.065501}
  {\bibfield  {journal} {\bibinfo  {journal} {Phys. Rev. Lett.}\ }\textbf
  {\bibinfo {volume} {124}},\ \bibinfo {pages} {065501} (\bibinfo {year}
  {2020})}\BibitemShut {NoStop}%
\bibitem [{\citenamefont {Moulinet}\ \emph {et~al.}(2004)\citenamefont
  {Moulinet}, \citenamefont {Rosso}, \citenamefont {Krauth},\ and\
  \citenamefont {Rolley}}]{MoulinetRossoKrauthRolley2004}%
  \BibitemOpen
  \bibfield  {author} {\bibinfo {author} {\bibfnamefont {S.}~\bibnamefont
  {Moulinet}}, \bibinfo {author} {\bibfnamefont {A.}~\bibnamefont {Rosso}},
  \bibinfo {author} {\bibfnamefont {W.}~\bibnamefont {Krauth}},\ and\ \bibinfo
  {author} {\bibfnamefont {E.}~\bibnamefont {Rolley}},\ }\bibfield  {title}
  {\bibinfo {title} {Width distribution of contact lines on a disordered
  substrate},\ }\href {https://doi.org/10.1103/PhysRevE.69.035103} {\bibfield
  {journal} {\bibinfo  {journal} {Phys. Rev. E}\ }\textbf {\bibinfo {volume}
  {69}},\ \bibinfo {pages} {035103} (\bibinfo {year} {2004})},\ \Eprint
  {https://arxiv.org/abs/cond-mat/0310173} {cond-mat/0310173} \BibitemShut
  {NoStop}%
\bibitem [{\citenamefont {Doussal}(2009)}]{LeDoussal2009}%
  \BibitemOpen
  \bibfield  {author} {\bibinfo {author} {\bibfnamefont {P.~L.}\ \bibnamefont
  {Doussal}},\ }\bibfield  {title} {\bibinfo {title} {Sinai model in presence
  of dilute absorbers},\ }\href@noop {} {\bibfield  {journal} {\bibinfo
  {journal} {J. Stat. Mech.}\ ,\ \bibinfo {pages} {P07032}} (\bibinfo {year}
  {2009})},\ \Eprint {https://arxiv.org/abs/arXiv:0906.0267} {arXiv:0906.0267}
  \BibitemShut {NoStop}%
\bibitem [{\citenamefont {Planet}\ \emph {et~al.}(2009)\citenamefont {Planet},
  \citenamefont {Santucci},\ and\ \citenamefont
  {Ort\'{\i}n}}]{PlanetSantucciOrtin2009}%
  \BibitemOpen
  \bibfield  {author} {\bibinfo {author} {\bibfnamefont {R.}~\bibnamefont
  {Planet}}, \bibinfo {author} {\bibfnamefont {S.}~\bibnamefont {Santucci}},\
  and\ \bibinfo {author} {\bibfnamefont {J.}~\bibnamefont {Ort\'{\i}n}},\
  }\bibfield  {title} {\bibinfo {title} {Avalanches and non-gaussian
  fluctuations of the global velocity of imbibition fronts},\ }\href
  {https://doi.org/10.1103/PhysRevLett.102.094502} {\bibfield  {journal}
  {\bibinfo  {journal} {Phys. Rev. Lett.}\ }\textbf {\bibinfo {volume} {102}},\
  \bibinfo {pages} {094502} (\bibinfo {year} {2009})}\BibitemShut {NoStop}%
\bibitem [{\citenamefont {Atis}\ \emph {et~al.}(2015)\citenamefont {Atis},
  \citenamefont {Dubey}, \citenamefont {Salin}, \citenamefont {Talon},
  \citenamefont {{Le Doussal}},\ and\ \citenamefont
  {Wiese}}]{AtisDubeySalinTalonLeDoussalWiese2015}%
  \BibitemOpen
  \bibfield  {author} {\bibinfo {author} {\bibfnamefont {S.}~\bibnamefont
  {Atis}}, \bibinfo {author} {\bibfnamefont {A.~K.}\ \bibnamefont {Dubey}},
  \bibinfo {author} {\bibfnamefont {D.}~\bibnamefont {Salin}}, \bibinfo
  {author} {\bibfnamefont {L.}~\bibnamefont {Talon}}, \bibinfo {author}
  {\bibfnamefont {P.}~\bibnamefont {{Le Doussal}}},\ and\ \bibinfo {author}
  {\bibfnamefont {K.~J.}\ \bibnamefont {Wiese}},\ }\bibfield  {title} {\bibinfo
  {title} {Experimental evidence for three universality classes for reaction
  fronts in disordered flows},\ }\href
  {https://doi.org/10.1103/PhysRevLett.114.234502} {\bibfield  {journal}
  {\bibinfo  {journal} {Phys. Rev. Lett.}\ }\textbf {\bibinfo {volume} {114}},\
  \bibinfo {pages} {234502} (\bibinfo {year} {2015})}\BibitemShut {NoStop}%
\bibitem [{\citenamefont {Bayart}\ \emph {et~al.}(2015)\citenamefont {Bayart},
  \citenamefont {Svetlizky},\ and\ \citenamefont
  {Fineberg}}]{BayartSvetlizkyFineberg2015}%
  \BibitemOpen
  \bibfield  {author} {\bibinfo {author} {\bibfnamefont {E.}~\bibnamefont
  {Bayart}}, \bibinfo {author} {\bibfnamefont {I.}~\bibnamefont {Svetlizky}},\
  and\ \bibinfo {author} {\bibfnamefont {J.}~\bibnamefont {Fineberg}},\
  }\bibfield  {title} {\bibinfo {title} {Fracture mechanics determine the
  lengths of interface ruptures that mediate frictional motion},\ }\href
  {http://dx.doi.org/10.1038/nphys3539} {\bibfield  {journal} {\bibinfo
  {journal} {Nature Physics}\ }\textbf {\bibinfo {volume} {12}},\ \bibinfo
  {pages} {166 EP} (\bibinfo {year} {2015})}\BibitemShut {NoStop}%
\bibitem [{\citenamefont {{Nicolas}}\ \emph {et~al.}(2017)\citenamefont
  {{Nicolas}}, \citenamefont {{Ferrero}}, \citenamefont {{Martens}},\ and\
  \citenamefont {{Barrat}}}]{NicolasFerreroMartensBarrat2017}%
  \BibitemOpen
  \bibfield  {author} {\bibinfo {author} {\bibfnamefont {A.}~\bibnamefont
  {{Nicolas}}}, \bibinfo {author} {\bibfnamefont {E.~E.}\ \bibnamefont
  {{Ferrero}}}, \bibinfo {author} {\bibfnamefont {K.}~\bibnamefont
  {{Martens}}},\ and\ \bibinfo {author} {\bibfnamefont {J.-L.}\ \bibnamefont
  {{Barrat}}},\ }\bibfield  {title} {\bibinfo {title} {Deformation and flow of
  amorphous solids: a review of mesoscale elastoplastic models},\ }\href@noop
  {} {\  (\bibinfo {year} {2017})},\ \Eprint
  {https://arxiv.org/abs/arXiv:1708.09194} {arXiv:1708.09194} \BibitemShut
  {NoStop}%
\bibitem [{\citenamefont {Sethna}\ \emph {et~al.}(2017)\citenamefont {Sethna},
  \citenamefont {Bierbaum}, \citenamefont {Dahmen}, \citenamefont {Goodrich},
  \citenamefont {Greer}, \citenamefont {Hayden}, \citenamefont {Kent-Dobias},
  \citenamefont {Lee}, \citenamefont {Liarte}, \citenamefont {Ni},
  \citenamefont {Quinn}, \citenamefont {Raju}, \citenamefont {Rocklin},
  \citenamefont {Shekhawat},\ and\ \citenamefont
  {Zapperi}}]{SethnaBierbaumDahmenGoodrichGreerHaydenKentDobiasLeeLiarteNiQuinnRajuRocklinShekhawatZapperi2017}%
  \BibitemOpen
  \bibfield  {author} {\bibinfo {author} {\bibfnamefont {J.~P.}\ \bibnamefont
  {Sethna}}, \bibinfo {author} {\bibfnamefont {M.~K.}\ \bibnamefont
  {Bierbaum}}, \bibinfo {author} {\bibfnamefont {K.~A.}\ \bibnamefont
  {Dahmen}}, \bibinfo {author} {\bibfnamefont {C.~P.}\ \bibnamefont
  {Goodrich}}, \bibinfo {author} {\bibfnamefont {J.~R.}\ \bibnamefont {Greer}},
  \bibinfo {author} {\bibfnamefont {L.~X.}\ \bibnamefont {Hayden}}, \bibinfo
  {author} {\bibfnamefont {J.~P.}\ \bibnamefont {Kent-Dobias}}, \bibinfo
  {author} {\bibfnamefont {E.~D.}\ \bibnamefont {Lee}}, \bibinfo {author}
  {\bibfnamefont {D.~B.}\ \bibnamefont {Liarte}}, \bibinfo {author}
  {\bibfnamefont {X.}~\bibnamefont {Ni}}, \bibinfo {author} {\bibfnamefont
  {K.~N.}\ \bibnamefont {Quinn}}, \bibinfo {author} {\bibfnamefont
  {A.}~\bibnamefont {Raju}}, \bibinfo {author} {\bibfnamefont {D.~Z.}\
  \bibnamefont {Rocklin}}, \bibinfo {author} {\bibfnamefont {A.}~\bibnamefont
  {Shekhawat}},\ and\ \bibinfo {author} {\bibfnamefont {S.}~\bibnamefont
  {Zapperi}},\ }\bibfield  {title} {\bibinfo {title} {Deformation of crystals:
  Connections with statistical physics},\ }\href
  {https://doi.org/10.1146/annurev-matsci-070115-032036} {\bibfield  {journal}
  {\bibinfo  {journal} {Annu. Rev. Mater. Res.}\ }\textbf {\bibinfo {volume}
  {47}},\ \bibinfo {pages} {217} (\bibinfo {year} {2017})}\BibitemShut
  {NoStop}%
\bibitem [{\citenamefont {Nattermann}\ and\ \citenamefont
  {Scheidl}(2000)}]{NattermannScheidl2000}%
  \BibitemOpen
  \bibfield  {author} {\bibinfo {author} {\bibfnamefont {T.}~\bibnamefont
  {Nattermann}}\ and\ \bibinfo {author} {\bibfnamefont {S.}~\bibnamefont
  {Scheidl}},\ }\bibfield  {title} {\bibinfo {title} {Vortex-glass phases in
  type-{II} superconductors},\ }\href {https://doi.org/10.1080/000187300412257}
  {\bibfield  {journal} {\bibinfo  {journal} {Adv. Phys.}\ }\textbf {\bibinfo
  {volume} {49}},\ \bibinfo {pages} {607} (\bibinfo {year} {2000})},\ \Eprint
  {https://arxiv.org/abs/cond-mat/0003052} {cond-mat/0003052} \BibitemShut
  {NoStop}%
\bibitem [{\citenamefont {Giamarchi}\ and\ \citenamefont
  {Bhattacharya}(2002)}]{GiamarchiBhattacharya2002}%
  \BibitemOpen
  \bibfield  {author} {\bibinfo {author} {\bibfnamefont {T.}~\bibnamefont
  {Giamarchi}}\ and\ \bibinfo {author} {\bibfnamefont {S.}~\bibnamefont
  {Bhattacharya}},\ }\bibfield  {title} {\bibinfo {title} {Vortex phases},\
  }in\ \href@noop {} {\emph {\bibinfo {booktitle} {2001 Cargese school on
  "Trends in high magnetic field science"}}}\ (\bibinfo  {publisher}
  {Springer-Verlag},\ \bibinfo {year} {2002})\BibitemShut {NoStop}%
\bibitem [{\citenamefont {{Le Doussal}}(2010)}]{ledoussal2010}%
  \BibitemOpen
  \bibfield  {author} {\bibinfo {author} {\bibfnamefont {P.}~\bibnamefont {{Le
  Doussal}}},\ }\bibfield  {title} {\bibinfo {title} {Novel phases of vortices
  in superconductors},\ }\href {https://doi.org/10.1142/S0217979210056384}
  {\bibfield  {journal} {\bibinfo  {journal} {Int. J. Mod. Phys. B}\ }\textbf
  {\bibinfo {volume} {24}},\ \bibinfo {pages} {3855} (\bibinfo {year}
  {2010})}\BibitemShut {NoStop}%
\bibitem [{\citenamefont {Kwok}\ \emph {et~al.}(2016)\citenamefont {Kwok},
  \citenamefont {Welp}, \citenamefont {Glatz}, \citenamefont {Koshelev},
  \citenamefont {Kihlstrom},\ and\ \citenamefont
  {Crabtree}}]{KwokWelpGlatzKoshelevKihlstromCrabtree2016}%
  \BibitemOpen
  \bibfield  {author} {\bibinfo {author} {\bibfnamefont {W.-K.}\ \bibnamefont
  {Kwok}}, \bibinfo {author} {\bibfnamefont {U.}~\bibnamefont {Welp}}, \bibinfo
  {author} {\bibfnamefont {A.}~\bibnamefont {Glatz}}, \bibinfo {author}
  {\bibfnamefont {A.~E.}\ \bibnamefont {Koshelev}}, \bibinfo {author}
  {\bibfnamefont {K.~J.}\ \bibnamefont {Kihlstrom}},\ and\ \bibinfo {author}
  {\bibfnamefont {G.~W.}\ \bibnamefont {Crabtree}},\ }\bibfield  {title}
  {\bibinfo {title} {Vortices in high-performance high-temperature
  superconductors},\ }\href {https://doi.org/10.1088/0034-4885/79/11/116501}
  {\bibfield  {journal} {\bibinfo  {journal} {Rep. Prog. Phys.}\ }\textbf
  {\bibinfo {volume} {79}},\ \bibinfo {pages} {116501} (\bibinfo {year}
  {2016})}\BibitemShut {NoStop}%
\bibitem [{\citenamefont {Thomann}\ \emph {et~al.}(2017)\citenamefont
  {Thomann}, \citenamefont {Geshkenbein},\ and\ \citenamefont
  {Blatter}}]{ThomannGeshkenbeinBlatter2017}%
  \BibitemOpen
  \bibfield  {author} {\bibinfo {author} {\bibfnamefont {A.~U.}\ \bibnamefont
  {Thomann}}, \bibinfo {author} {\bibfnamefont {V.~B.}\ \bibnamefont
  {Geshkenbein}},\ and\ \bibinfo {author} {\bibfnamefont {G.}~\bibnamefont
  {Blatter}},\ }\bibfield  {title} {\bibinfo {title} {Vortex dynamics in
  type-ii superconductors under strong pinning conditions},\ }\href
  {https://doi.org/10.1103/PhysRevB.96.144516} {\bibfield  {journal} {\bibinfo
  {journal} {Phys. Rev. B}\ }\textbf {\bibinfo {volume} {96}},\ \bibinfo
  {pages} {144516} (\bibinfo {year} {2017})}\BibitemShut {NoStop}%
\bibitem [{\citenamefont {Sadovskyy}\ \emph {et~al.}(2019)\citenamefont
  {Sadovskyy}, \citenamefont {Koshelev}, \citenamefont {Kwok}, \citenamefont
  {Welp},\ and\ \citenamefont {Glatz}}]{SadovskyyKoshelevKwokWelpGlatz2019}%
  \BibitemOpen
  \bibfield  {author} {\bibinfo {author} {\bibfnamefont {I.~A.}\ \bibnamefont
  {Sadovskyy}}, \bibinfo {author} {\bibfnamefont {A.~E.}\ \bibnamefont
  {Koshelev}}, \bibinfo {author} {\bibfnamefont {W.-K.}\ \bibnamefont {Kwok}},
  \bibinfo {author} {\bibfnamefont {U.}~\bibnamefont {Welp}},\ and\ \bibinfo
  {author} {\bibfnamefont {A.}~\bibnamefont {Glatz}},\ }\bibfield  {title}
  {\bibinfo {title} {Targeted evolution of pinning landscapes for large
  superconducting critical currents},\ }\href
  {https://doi.org/10.1073/pnas.1817417116} {\bibfield  {journal} {\bibinfo
  {journal} {PNAS}\ }\textbf {\bibinfo {volume} {116}},\ \bibinfo {pages}
  {10291} (\bibinfo {year} {2019})}\BibitemShut {NoStop}%
\bibitem [{\citenamefont {Eley}\ \emph {et~al.}(2021)\citenamefont {Eley},
  \citenamefont {Glatz},\ and\ \citenamefont {Willa}}]{EleyGlatzWilla2021}%
  \BibitemOpen
  \bibfield  {author} {\bibinfo {author} {\bibfnamefont {S.}~\bibnamefont
  {Eley}}, \bibinfo {author} {\bibfnamefont {A.}~\bibnamefont {Glatz}},\ and\
  \bibinfo {author} {\bibfnamefont {R.}~\bibnamefont {Willa}},\ }\bibfield
  {title} {\bibinfo {title} {Challenges and transformative opportunities in
  superconductor vortex physics},\ }\href {https://doi.org/10.1063/5.0055611}
  {\bibfield  {journal} {\bibinfo  {journal} {J. Appl. Phys.}\ }\textbf
  {\bibinfo {volume} {130}},\ \bibinfo {pages} {050901} (\bibinfo {year}
  {2021})}\BibitemShut {NoStop}%
\bibitem [{\citenamefont {Schulz}\ \emph {et~al.}(2012)\citenamefont {Schulz},
  \citenamefont {Ritz}, \citenamefont {Bauer}, \citenamefont {Halder},
  \citenamefont {Wagner}, \citenamefont {Franz}, \citenamefont {Pfleiderer},
  \citenamefont {Everschor}, \citenamefont {Garst},\ and\ \citenamefont
  {Rosch}}]{SchulzRitzBauerHalderWagnerFranzPfleidererEverschorGarst2012}%
  \BibitemOpen
  \bibfield  {author} {\bibinfo {author} {\bibfnamefont {T.}~\bibnamefont
  {Schulz}}, \bibinfo {author} {\bibfnamefont {R.}~\bibnamefont {Ritz}},
  \bibinfo {author} {\bibfnamefont {A.}~\bibnamefont {Bauer}}, \bibinfo
  {author} {\bibfnamefont {M.}~\bibnamefont {Halder}}, \bibinfo {author}
  {\bibfnamefont {M.}~\bibnamefont {Wagner}}, \bibinfo {author} {\bibfnamefont
  {C.}~\bibnamefont {Franz}}, \bibinfo {author} {\bibfnamefont
  {C.}~\bibnamefont {Pfleiderer}}, \bibinfo {author} {\bibfnamefont
  {K.}~\bibnamefont {Everschor}}, \bibinfo {author} {\bibfnamefont
  {M.}~\bibnamefont {Garst}},\ and\ \bibinfo {author} {\bibfnamefont
  {A.}~\bibnamefont {Rosch}},\ }\bibfield  {title} {\bibinfo {title} {Emergent
  electrodynamics of skyrmions in a chiral magnet},\ }\href
  {http://dx.doi.org/10.1038/nphys2231} {\bibfield  {journal} {\bibinfo
  {journal} {Nature Physics}\ }\textbf {\bibinfo {volume} {8}},\ \bibinfo
  {pages} {301 EP} (\bibinfo {year} {2012})}\BibitemShut {NoStop}%
\bibitem [{\citenamefont {Jagla}\ and\ \citenamefont
  {Kolton}(2010)}]{JaglaKolton2010}%
  \BibitemOpen
  \bibfield  {author} {\bibinfo {author} {\bibfnamefont {E.~A.}\ \bibnamefont
  {Jagla}}\ and\ \bibinfo {author} {\bibfnamefont {A.~B.}\ \bibnamefont
  {Kolton}},\ }\bibfield  {title} {\bibinfo {title} {A mechanism for spatial
  and temporal earthquake clustering},\ }\bibfield  {journal} {\bibinfo
  {journal} {J. Geophys. Res. Solid Earth}\ }\textbf {\bibinfo {volume}
  {115}},\ \href {https://doi.org/10.1029/2009JB006974} {10.1029/2009JB006974}
  (\bibinfo {year} {2010})\BibitemShut {NoStop}%
\bibitem [{\citenamefont {Jagla}\ \emph {et~al.}(2014)\citenamefont {Jagla},
  \citenamefont {Landes},\ and\ \citenamefont {Rosso}}]{JaglaLandesRosso2014}%
  \BibitemOpen
  \bibfield  {author} {\bibinfo {author} {\bibfnamefont {E.~A.}\ \bibnamefont
  {Jagla}}, \bibinfo {author} {\bibfnamefont {F.~P.}\ \bibnamefont {Landes}},\
  and\ \bibinfo {author} {\bibfnamefont {A.}~\bibnamefont {Rosso}},\ }\bibfield
   {title} {\bibinfo {title} {Viscoelastic effects in avalanche dynamics: A key
  to earthquake statistics},\ }\href
  {https://doi.org/10.1103/PhysRevLett.112.174301} {\bibfield  {journal}
  {\bibinfo  {journal} {Phys. Rev. Lett.}\ }\textbf {\bibinfo {volume} {112}},\
  \bibinfo {pages} {174301} (\bibinfo {year} {2014})}\BibitemShut {NoStop}%
\bibitem [{\citenamefont {Kardar}(1998)}]{Kardar1998}%
  \BibitemOpen
  \bibfield  {author} {\bibinfo {author} {\bibfnamefont {M.}~\bibnamefont
  {Kardar}},\ }\bibfield  {title} {\bibinfo {title} {Nonequilibrium dynamics of
  interfaces and lines},\ }\href
  {https://doi.org/10.1016/S0370-1573(98)00007-6} {\bibfield  {journal}
  {\bibinfo  {journal} {Phys. Rep.}\ }\textbf {\bibinfo {volume} {301}},\
  \bibinfo {pages} {85} (\bibinfo {year} {1998})},\ \Eprint
  {https://arxiv.org/abs/cond-mat/9704172} {cond-mat/9704172} \BibitemShut
  {NoStop}%
\bibitem [{\citenamefont {Fisher}(1998)}]{Fisher1998}%
  \BibitemOpen
  \bibfield  {author} {\bibinfo {author} {\bibfnamefont {D.~S.}\ \bibnamefont
  {Fisher}},\ }\bibfield  {title} {\bibinfo {title} {Collective transport in
  random media: from superconductors to earthquakes},\ }\href
  {https://doi.org/10.1016/S0370-1573(98)00008-8} {\bibfield  {journal}
  {\bibinfo  {journal} {Phys. Rep.}\ }\textbf {\bibinfo {volume} {301}},\
  \bibinfo {pages} {113} (\bibinfo {year} {1998})},\ \Eprint
  {https://arxiv.org/abs/cond-mat/9711179} {cond-mat/9711179} \BibitemShut
  {NoStop}%
\bibitem [{\citenamefont {Wiese}(2021)}]{Wiese2021}%
  \BibitemOpen
  \bibfield  {author} {\bibinfo {author} {\bibfnamefont {K.}~\bibnamefont
  {Wiese}},\ }\bibfield  {title} {\bibinfo {title} {Theory and experiments for
  disordered elastic manifolds, depinning, avalanches, and sandpiles},\
  }\href@noop {} {\bibfield  {journal} {\bibinfo  {journal} {ROPP}\ }\textbf
  {\bibinfo {volume} {accepted}} (\bibinfo {year} {2021})},\ \Eprint
  {https://arxiv.org/abs/arXiv:2102.01215} {arXiv:2102.01215} \BibitemShut
  {NoStop}%
\bibitem [{Note1()}]{Note1}%
  \BibitemOpen
  \bibinfo {note} {See ~\cite {NattermannScheidl2000} for a general
  description.}\BibitemShut {Stop}%
\bibitem [{\citenamefont {Nattermann}\ \emph {et~al.}(1992)\citenamefont
  {Nattermann}, \citenamefont {Stepanow}, \citenamefont {Tang},\ and\
  \citenamefont {Leschhorn}}]{NattermannStepanowTangLeschhorn1992}%
  \BibitemOpen
  \bibfield  {author} {\bibinfo {author} {\bibfnamefont {T.}~\bibnamefont
  {Nattermann}}, \bibinfo {author} {\bibfnamefont {S.}~\bibnamefont
  {Stepanow}}, \bibinfo {author} {\bibfnamefont {L.-H.}\ \bibnamefont {Tang}},\
  and\ \bibinfo {author} {\bibfnamefont {H.}~\bibnamefont {Leschhorn}},\
  }\bibfield  {title} {\bibinfo {title} {Dynamics of interface depinning in a
  disordered medium},\ }\href {https://doi.org/10.1051/jp2:1992214} {\bibfield
  {journal} {\bibinfo  {journal} {J. Phys. II (France)}\ }\textbf {\bibinfo
  {volume} {2}},\ \bibinfo {pages} {1483} (\bibinfo {year} {1992})}\BibitemShut
  {NoStop}%
\bibitem [{\citenamefont {Narayan}\ and\ \citenamefont
  {Fisher}(1993)}]{NarayanFisher1993}%
  \BibitemOpen
  \bibfield  {author} {\bibinfo {author} {\bibfnamefont {O.}~\bibnamefont
  {Narayan}}\ and\ \bibinfo {author} {\bibfnamefont {D.}~\bibnamefont
  {Fisher}},\ }\bibfield  {title} {\bibinfo {title} {Threshold critical
  dynamics of driven interfaces in random media},\ }\href
  {https://doi.org/10.1103/PhysRevB.48.7030} {\bibfield  {journal} {\bibinfo
  {journal} {Phys. Rev. B}\ }\textbf {\bibinfo {volume} {48}},\ \bibinfo
  {pages} {7030} (\bibinfo {year} {1993})}\BibitemShut {NoStop}%
\bibitem [{\citenamefont {Leschhorn}\ \emph {et~al.}(1997)\citenamefont
  {Leschhorn}, \citenamefont {Nattermann}, \citenamefont {Stepanow},\ and\
  \citenamefont {Tang}}]{LeschhornNattermannStepanowTang1997}%
  \BibitemOpen
  \bibfield  {author} {\bibinfo {author} {\bibfnamefont {H.}~\bibnamefont
  {Leschhorn}}, \bibinfo {author} {\bibfnamefont {T.}~\bibnamefont
  {Nattermann}}, \bibinfo {author} {\bibfnamefont {S.}~\bibnamefont
  {Stepanow}},\ and\ \bibinfo {author} {\bibfnamefont {L.-H.}\ \bibnamefont
  {Tang}},\ }\bibfield  {title} {\bibinfo {title} {Driven interface depinning
  in a disordered medium},\ }\href {https://doi.org/10.1002/andp.19975090102}
  {\bibfield  {journal} {\bibinfo  {journal} {Annalen der Physik}\ }\textbf
  {\bibinfo {volume} {509}},\ \bibinfo {pages} {1} (\bibinfo {year} {1997})},\
  \Eprint {https://arxiv.org/abs/arXiv:cond-mat/9603114}
  {arXiv:cond-mat/9603114} \BibitemShut {NoStop}%
\bibitem [{\citenamefont {Chauve}\ \emph {et~al.}(2000)\citenamefont {Chauve},
  \citenamefont {Giamarchi},\ and\ \citenamefont
  {Doussal}}]{ChauveGiamarchiLeDoussal2000}%
  \BibitemOpen
  \bibfield  {author} {\bibinfo {author} {\bibfnamefont {P.}~\bibnamefont
  {Chauve}}, \bibinfo {author} {\bibfnamefont {T.}~\bibnamefont {Giamarchi}},\
  and\ \bibinfo {author} {\bibfnamefont {P.~L.}\ \bibnamefont {Doussal}},\
  }\bibfield  {title} {\bibinfo {title} {Creep and depinning in disordered
  media},\ }\href {https://doi.org/10.1103/PhysRevB.62.6241} {\bibfield
  {journal} {\bibinfo  {journal} {Phys. Rev. B}\ }\textbf {\bibinfo {volume}
  {62}},\ \bibinfo {pages} {6241} (\bibinfo {year} {2000})},\ \Eprint
  {https://arxiv.org/abs/cond-mat/0002299} {cond-mat/0002299} \BibitemShut
  {NoStop}%
\bibitem [{\citenamefont {Kolton}\ \emph {et~al.}(2013)\citenamefont {Kolton},
  \citenamefont {Bustingorry}, \citenamefont {Ferrero},\ and\ \citenamefont
  {Rosso}}]{KoltonBustingorryFerreroRosso2013}%
  \BibitemOpen
  \bibfield  {author} {\bibinfo {author} {\bibfnamefont {A.}~\bibnamefont
  {Kolton}}, \bibinfo {author} {\bibfnamefont {S.}~\bibnamefont {Bustingorry}},
  \bibinfo {author} {\bibfnamefont {E.}~\bibnamefont {Ferrero}},\ and\ \bibinfo
  {author} {\bibfnamefont {A.}~\bibnamefont {Rosso}},\ }\bibfield  {title}
  {\bibinfo {title} {Uniqueness of the thermodynamic limit for driven
  disordered elastic interfaces},\ }\href
  {https://doi.org/10.1088/1742-5468/2013/12/p12004} {\bibfield  {journal}
  {\bibinfo  {journal} {J. Stat. Mech.}\ }\textbf {\bibinfo {volume} {2013}},\
  \bibinfo {pages} {P12004} (\bibinfo {year} {2013})},\ \Eprint
  {https://arxiv.org/abs/arXiv:1308.4329} {arXiv:1308.4329} \BibitemShut
  {NoStop}%
\bibitem [{\citenamefont {Cao}\ \emph {et~al.}(2018)\citenamefont {Cao},
  \citenamefont {Bouzat}, \citenamefont {Kolton},\ and\ \citenamefont
  {Rosso}}]{CaoBouzatKoltonRosso2018}%
  \BibitemOpen
  \bibfield  {author} {\bibinfo {author} {\bibfnamefont {X.}~\bibnamefont
  {Cao}}, \bibinfo {author} {\bibfnamefont {S.}~\bibnamefont {Bouzat}},
  \bibinfo {author} {\bibfnamefont {A.~B.}\ \bibnamefont {Kolton}},\ and\
  \bibinfo {author} {\bibfnamefont {A.}~\bibnamefont {Rosso}},\ }\bibfield
  {title} {\bibinfo {title} {Localization of soft modes at the depinning
  transition},\ }\href {https://doi.org/10.1103/PhysRevE.97.022118} {\bibfield
  {journal} {\bibinfo  {journal} {Phys. Rev. E}\ }\textbf {\bibinfo {volume}
  {97}},\ \bibinfo {pages} {022118} (\bibinfo {year} {2018})}\BibitemShut
  {NoStop}%
\bibitem [{\citenamefont {Chauve}\ \emph {et~al.}(2001)\citenamefont {Chauve},
  \citenamefont {Doussal},\ and\ \citenamefont
  {Wiese}}]{ChauveLeDoussalWiese2000a}%
  \BibitemOpen
  \bibfield  {author} {\bibinfo {author} {\bibfnamefont {P.}~\bibnamefont
  {Chauve}}, \bibinfo {author} {\bibfnamefont {P.~L.}\ \bibnamefont
  {Doussal}},\ and\ \bibinfo {author} {\bibfnamefont {K.}~\bibnamefont
  {Wiese}},\ }\bibfield  {title} {\bibinfo {title} {Renormalization of pinned
  elastic systems: How does it work beyond one loop?},\ }\href
  {https://doi.org/10.1103/PhysRevLett.86.1785} {\bibfield  {journal} {\bibinfo
   {journal} {Phys. Rev. Lett.}\ }\textbf {\bibinfo {volume} {86}},\ \bibinfo
  {pages} {1785} (\bibinfo {year} {2001})},\ \Eprint
  {https://arxiv.org/abs/cond-mat/0006056} {cond-mat/0006056} \BibitemShut
  {NoStop}%
\bibitem [{\citenamefont {Doussal}\ \emph {et~al.}(2002)\citenamefont
  {Doussal}, \citenamefont {Wiese},\ and\ \citenamefont
  {Chauve}}]{LeDoussalWieseChauve2002}%
  \BibitemOpen
  \bibfield  {author} {\bibinfo {author} {\bibfnamefont {P.~L.}\ \bibnamefont
  {Doussal}}, \bibinfo {author} {\bibfnamefont {K.}~\bibnamefont {Wiese}},\
  and\ \bibinfo {author} {\bibfnamefont {P.}~\bibnamefont {Chauve}},\
  }\bibfield  {title} {\bibinfo {title} {2-loop functional renormalization
  group analysis of the depinning transition},\ }\href
  {https://doi.org/10.1103/PhysRevB.66.174201} {\bibfield  {journal} {\bibinfo
  {journal} {Phys. Rev. B}\ }\textbf {\bibinfo {volume} {66}},\ \bibinfo
  {pages} {174201} (\bibinfo {year} {2002})},\ \Eprint
  {https://arxiv.org/abs/cond-mat/0205108} {cond-mat/0205108} \BibitemShut
  {NoStop}%
\bibitem [{\citenamefont {Fedorenko}\ and\ \citenamefont
  {Stepanow}(2003)}]{FedorenkoStepanow2003}%
  \BibitemOpen
  \bibfield  {author} {\bibinfo {author} {\bibfnamefont {A.}~\bibnamefont
  {Fedorenko}}\ and\ \bibinfo {author} {\bibfnamefont {S.}~\bibnamefont
  {Stepanow}},\ }\bibfield  {title} {\bibinfo {title} {Universal energy
  distribution for interfaces in a random-field environment},\ }\href
  {https://doi.org/10.1103/PhysRevE.68.056115} {\bibfield  {journal} {\bibinfo
  {journal} {Phys. Rev. E}\ }\textbf {\bibinfo {volume} {68}},\ \bibinfo
  {pages} {056115} (\bibinfo {year} {2003})}\BibitemShut {NoStop}%
\bibitem [{\citenamefont {Leschhorn}(1993)}]{Leschhorn1993}%
  \BibitemOpen
  \bibfield  {author} {\bibinfo {author} {\bibfnamefont {H.}~\bibnamefont
  {Leschhorn}},\ }\bibfield  {title} {\bibinfo {title} {Interface depinning in
  a disordered medium — numerical results},\ }\href
  {https://doi.org/https://doi.org/10.1016/0378-4371(93)90161-V} {\bibfield
  {journal} {\bibinfo  {journal} {Physica A: Statistical Mechanics and its
  Applications}\ }\textbf {\bibinfo {volume} {195}},\ \bibinfo {pages} {324}
  (\bibinfo {year} {1993})}\BibitemShut {NoStop}%
\bibitem [{\citenamefont {Roters}\ \emph {et~al.}(1999)\citenamefont {Roters},
  \citenamefont {Hucht}, \citenamefont {L\"ubeck}, \citenamefont {Nowak},\ and\
  \citenamefont {Usadel}}]{RotersHuchtLubeckNowakUsadel1999}%
  \BibitemOpen
  \bibfield  {author} {\bibinfo {author} {\bibfnamefont {L.}~\bibnamefont
  {Roters}}, \bibinfo {author} {\bibfnamefont {A.}~\bibnamefont {Hucht}},
  \bibinfo {author} {\bibfnamefont {S.}~\bibnamefont {L\"ubeck}}, \bibinfo
  {author} {\bibfnamefont {U.}~\bibnamefont {Nowak}},\ and\ \bibinfo {author}
  {\bibfnamefont {K.}~\bibnamefont {Usadel}},\ }\bibfield  {title} {\bibinfo
  {title} {Depinning transition and thermal fluctuations in the random-field
  {Ising} model},\ }\href {https://doi.org/10.1103/PhysRevE.60.5202} {\bibfield
   {journal} {\bibinfo  {journal} {Phys. Rev. E}\ }\textbf {\bibinfo {volume}
  {60}},\ \bibinfo {pages} {5202} (\bibinfo {year} {1999})}\BibitemShut
  {NoStop}%
\bibitem [{\citenamefont {Rosso}\ \emph {et~al.}(2003)\citenamefont {Rosso},
  \citenamefont {Hartmann},\ and\ \citenamefont
  {Krauth}}]{RossoHartmannKrauth2003}%
  \BibitemOpen
  \bibfield  {author} {\bibinfo {author} {\bibfnamefont {A.}~\bibnamefont
  {Rosso}}, \bibinfo {author} {\bibfnamefont {A.~K.}\ \bibnamefont
  {Hartmann}},\ and\ \bibinfo {author} {\bibfnamefont {W.}~\bibnamefont
  {Krauth}},\ }\bibfield  {title} {\bibinfo {title} {Depinning of elastic
  manifolds},\ }\href {https://doi.org/10.1103/PhysRevE.67.021602} {\bibfield
  {journal} {\bibinfo  {journal} {Phys. Rev. E}\ }\textbf {\bibinfo {volume}
  {67}},\ \bibinfo {pages} {021602} (\bibinfo {year} {2003})}\BibitemShut
  {NoStop}%
\bibitem [{\citenamefont {Rosso}\ \emph {et~al.}(2007)\citenamefont {Rosso},
  \citenamefont {{Le~Doussal}},\ and\ \citenamefont
  {Wiese}}]{RossoLeDoussalWiese2007}%
  \BibitemOpen
  \bibfield  {author} {\bibinfo {author} {\bibfnamefont {A.}~\bibnamefont
  {Rosso}}, \bibinfo {author} {\bibfnamefont {P.}~\bibnamefont
  {{Le~Doussal}}},\ and\ \bibinfo {author} {\bibfnamefont {K.}~\bibnamefont
  {Wiese}},\ }\bibfield  {title} {\bibinfo {title} {Numerical calculation of
  the functional renormalization group fixed-point functions at the depinning
  transition},\ }\href {https://doi.org/10.1103/PhysRevB.75.220201} {\bibfield
  {journal} {\bibinfo  {journal} {Phys. Rev. B}\ }\textbf {\bibinfo {volume}
  {75}},\ \bibinfo {pages} {220201} (\bibinfo {year} {2007})},\ \Eprint
  {https://arxiv.org/abs/cond-mat/0610821} {cond-mat/0610821} \BibitemShut
  {NoStop}%
\bibitem [{\citenamefont {Ferrero}\ \emph {et~al.}(2013)\citenamefont
  {Ferrero}, \citenamefont {Bustingorry},\ and\ \citenamefont
  {Kolton}}]{FerreroBustingorryKolton2013}%
  \BibitemOpen
  \bibfield  {author} {\bibinfo {author} {\bibfnamefont {E.}~\bibnamefont
  {Ferrero}}, \bibinfo {author} {\bibfnamefont {S.}~\bibnamefont
  {Bustingorry}},\ and\ \bibinfo {author} {\bibfnamefont {A.}~\bibnamefont
  {Kolton}},\ }\bibfield  {title} {\bibinfo {title} {Non-steady relaxation and
  critical exponents at the depinning transition},\ }\href
  {https://doi.org/10.1103/PhysRevE.87.032122} {\bibfield  {journal} {\bibinfo
  {journal} {Phys. Rev. E}\ }\textbf {\bibinfo {volume} {87}},\ \bibinfo
  {pages} {032122} (\bibinfo {year} {2013})},\ \Eprint
  {https://arxiv.org/abs/arXiv:1211.7275} {arXiv:1211.7275} \BibitemShut
  {NoStop}%
\bibitem [{\citenamefont {Ramanathan}\ and\ \citenamefont
  {Fisher}(1998)}]{RamanathanFisher1998}%
  \BibitemOpen
  \bibfield  {author} {\bibinfo {author} {\bibfnamefont {S.}~\bibnamefont
  {Ramanathan}}\ and\ \bibinfo {author} {\bibfnamefont {D.~S.}\ \bibnamefont
  {Fisher}},\ }\bibfield  {title} {\bibinfo {title} {Onset of propagation of
  planar cracks in heterogeneous media},\ }\href
  {https://doi.org/10.1103/PhysRevB.58.6026} {\bibfield  {journal} {\bibinfo
  {journal} {Phys. Rev. B}\ }\textbf {\bibinfo {volume} {58}},\ \bibinfo
  {pages} {6026} (\bibinfo {year} {1998})}\BibitemShut {NoStop}%
\bibitem [{\citenamefont {Zapperi}\ \emph {et~al.}(1998)\citenamefont
  {Zapperi}, \citenamefont {Cizeau}, \citenamefont {Durin},\ and\ \citenamefont
  {Stanley}}]{ZapperiCizeauDurinStanley1998}%
  \BibitemOpen
  \bibfield  {author} {\bibinfo {author} {\bibfnamefont {S.}~\bibnamefont
  {Zapperi}}, \bibinfo {author} {\bibfnamefont {P.}~\bibnamefont {Cizeau}},
  \bibinfo {author} {\bibfnamefont {G.}~\bibnamefont {Durin}},\ and\ \bibinfo
  {author} {\bibfnamefont {H.}~\bibnamefont {Stanley}},\ }\bibfield  {title}
  {\bibinfo {title} {Dynamics of a ferromagnetic domain wall: Avalanches,
  depinning transition, and the {Barkhausen} effect},\ }\href
  {https://doi.org/10.1103/PhysRevB.58.6353} {\bibfield  {journal} {\bibinfo
  {journal} {Phys. Rev. B}\ }\textbf {\bibinfo {volume} {58}},\ \bibinfo
  {pages} {6353} (\bibinfo {year} {1998})}\BibitemShut {NoStop}%
\bibitem [{\citenamefont {Rosso}\ and\ \citenamefont
  {Krauth}(2002)}]{RossoKrauth2002}%
  \BibitemOpen
  \bibfield  {author} {\bibinfo {author} {\bibfnamefont {A.}~\bibnamefont
  {Rosso}}\ and\ \bibinfo {author} {\bibfnamefont {W.}~\bibnamefont {Krauth}},\
  }\bibfield  {title} {\bibinfo {title} {Roughness at the depinning threshold
  for a long-range elastic string},\ }\href
  {https://doi.org/10.1103/PhysRevE.65.025101} {\bibfield  {journal} {\bibinfo
  {journal} {Phys. Rev. E}\ }\textbf {\bibinfo {volume} {65}},\ \bibinfo
  {pages} {025101} (\bibinfo {year} {2002})}\BibitemShut {NoStop}%
\bibitem [{\citenamefont {Duemmer}\ and\ \citenamefont
  {Krauth}(2007)}]{DuemmerKrauth2007}%
  \BibitemOpen
  \bibfield  {author} {\bibinfo {author} {\bibfnamefont {O.}~\bibnamefont
  {Duemmer}}\ and\ \bibinfo {author} {\bibfnamefont {W.}~\bibnamefont
  {Krauth}},\ }\bibfield  {title} {\bibinfo {title} {Depinning exponents of the
  driven long-range elastic string},\ }\href
  {https://doi.org/10.1088/1742-5468/2007/01/p01019} {\bibfield  {journal}
  {\bibinfo  {journal} {Journal of Statistical Mechanics: Theory and
  Experiment}\ }\textbf {\bibinfo {volume} {2007}},\ \bibinfo {pages} {P01019}
  (\bibinfo {year} {2007})}\BibitemShut {NoStop}%
\bibitem [{\citenamefont {Laurson}\ \emph {et~al.}(2013)\citenamefont
  {Laurson}, \citenamefont {Illa}, \citenamefont {Santucci}, \citenamefont
  {{Tore Tallakstad}}, \citenamefont {M{\aa}l{\o}y},\ and\ \citenamefont
  {Alava}}]{LaursonIllaSantucciToreTallakstadMaloyAlava2013}%
  \BibitemOpen
  \bibfield  {author} {\bibinfo {author} {\bibfnamefont {L.}~\bibnamefont
  {Laurson}}, \bibinfo {author} {\bibfnamefont {X.}~\bibnamefont {Illa}},
  \bibinfo {author} {\bibfnamefont {S.}~\bibnamefont {Santucci}}, \bibinfo
  {author} {\bibfnamefont {K.}~\bibnamefont {{Tore Tallakstad}}}, \bibinfo
  {author} {\bibfnamefont {K.~J.}\ \bibnamefont {M{\aa}l{\o}y}},\ and\ \bibinfo
  {author} {\bibfnamefont {M.~J.}\ \bibnamefont {Alava}},\ }\bibfield  {title}
  {\bibinfo {title} {Evolution of the average avalanche shape with the
  universality class},\ }\href {http://dx.doi.org/10.1038/ncomms3927}
  {\bibfield  {journal} {\bibinfo  {journal} {Nature Communications}\ }\textbf
  {\bibinfo {volume} {4}},\ \bibinfo {pages} {3927} (\bibinfo {year}
  {2013})}\BibitemShut {NoStop}%
\bibitem [{\citenamefont {Boltz}\ and\ \citenamefont
  {Kierfeld}(2014)}]{BoltzKierfeld2014}%
  \BibitemOpen
  \bibfield  {author} {\bibinfo {author} {\bibfnamefont {H.-H.}\ \bibnamefont
  {Boltz}}\ and\ \bibinfo {author} {\bibfnamefont {J.}~\bibnamefont
  {Kierfeld}},\ }\bibfield  {title} {\bibinfo {title} {Depinning of stiff
  directed lines in random media},\ }\href
  {https://doi.org/10.1103/PhysRevE.90.012101} {\bibfield  {journal} {\bibinfo
  {journal} {Phys. Rev. E}\ }\textbf {\bibinfo {volume} {90}},\ \bibinfo
  {pages} {012101} (\bibinfo {year} {2014})}\BibitemShut {NoStop}%
\bibitem [{\citenamefont {Tang}\ \emph {et~al.}(1995)\citenamefont {Tang},
  \citenamefont {Kardar},\ and\ \citenamefont {Dhar}}]{TangKardarDhar1995}%
  \BibitemOpen
  \bibfield  {author} {\bibinfo {author} {\bibfnamefont {L.-H.}\ \bibnamefont
  {Tang}}, \bibinfo {author} {\bibfnamefont {M.}~\bibnamefont {Kardar}},\ and\
  \bibinfo {author} {\bibfnamefont {D.}~\bibnamefont {Dhar}},\ }\bibfield
  {title} {\bibinfo {title} {Driven depinning in anisotropic media},\ }\href
  {https://doi.org/10.1103/PhysRevLett.74.920} {\bibfield  {journal} {\bibinfo
  {journal} {Phys. Rev. Lett.}\ }\textbf {\bibinfo {volume} {74}},\ \bibinfo
  {pages} {920} (\bibinfo {year} {1995})}\BibitemShut {NoStop}%
\bibitem [{\citenamefont {Fedorenko}\ \emph
  {et~al.}(2006{\natexlab{a}})\citenamefont {Fedorenko}, \citenamefont
  {{Le~Doussal}},\ and\ \citenamefont {Wiese}}]{FedorenkoLeDoussalWiese2006b}%
  \BibitemOpen
  \bibfield  {author} {\bibinfo {author} {\bibfnamefont {A.}~\bibnamefont
  {Fedorenko}}, \bibinfo {author} {\bibfnamefont {P.}~\bibnamefont
  {{Le~Doussal}}},\ and\ \bibinfo {author} {\bibfnamefont {K.}~\bibnamefont
  {Wiese}},\ }\bibfield  {title} {\bibinfo {title} {Statics and dynamics of
  elastic manifolds in media with long-range correlated disorder},\ }\href
  {https://doi.org/10.1103/PhysRevE.74.061109} {\bibfield  {journal} {\bibinfo
  {journal} {Phys. Rev. E}\ }\textbf {\bibinfo {volume} {74}},\ \bibinfo
  {pages} {061109} (\bibinfo {year} {2006}{\natexlab{a}})},\ \Eprint
  {https://arxiv.org/abs/cond-mat/0609234} {cond-mat/0609234} \BibitemShut
  {NoStop}%
\bibitem [{\citenamefont {Bustingorry}\ \emph {et~al.}(2010)\citenamefont
  {Bustingorry}, \citenamefont {Kolton},\ and\ \citenamefont
  {Giamarchi}}]{BustingorryKoltonGiamarchi2010}%
  \BibitemOpen
  \bibfield  {author} {\bibinfo {author} {\bibfnamefont {S.}~\bibnamefont
  {Bustingorry}}, \bibinfo {author} {\bibfnamefont {A.~B.}\ \bibnamefont
  {Kolton}},\ and\ \bibinfo {author} {\bibfnamefont {T.}~\bibnamefont
  {Giamarchi}},\ }\bibfield  {title} {\bibinfo {title} {Random-manifold to
  random-periodic depinning of an elastic interface},\ }\href
  {https://doi.org/10.1103/PhysRevB.82.094202} {\bibfield  {journal} {\bibinfo
  {journal} {Phys. Rev. B}\ }\textbf {\bibinfo {volume} {82}},\ \bibinfo
  {pages} {094202} (\bibinfo {year} {2010})}\BibitemShut {NoStop}%
\bibitem [{\citenamefont {Amaral}\ \emph {et~al.}(1994)\citenamefont {Amaral},
  \citenamefont {Barabsi},\ and\ \citenamefont
  {Stanley}}]{AmaralBarabsiStanley1994}%
  \BibitemOpen
  \bibfield  {author} {\bibinfo {author} {\bibfnamefont {L.~A.~N.}\
  \bibnamefont {Amaral}}, \bibinfo {author} {\bibfnamefont {A.~L.}\
  \bibnamefont {Barabsi}},\ and\ \bibinfo {author} {\bibfnamefont {H.~E.}\
  \bibnamefont {Stanley}},\ }\bibfield  {title} {\bibinfo {title} {Universality
  classes for interface growth with quenched disorder},\ }\href
  {https://doi.org/10.1103/PhysRevLett.73.62} {\bibfield  {journal} {\bibinfo
  {journal} {Phys. Rev. Lett.}\ }\textbf {\bibinfo {volume} {73}},\ \bibinfo
  {pages} {62} (\bibinfo {year} {1994})}\BibitemShut {NoStop}%
\bibitem [{\citenamefont {Rosso}\ and\ \citenamefont
  {Krauth}(2001{\natexlab{a}})}]{RossoKrauth2001b}%
  \BibitemOpen
  \bibfield  {author} {\bibinfo {author} {\bibfnamefont {A.}~\bibnamefont
  {Rosso}}\ and\ \bibinfo {author} {\bibfnamefont {W.}~\bibnamefont {Krauth}},\
  }\bibfield  {title} {\bibinfo {title} {Origin of the roughness exponent in
  elastic strings at the depinning threshold},\ }\href
  {https://doi.org/10.1103/PhysRevLett.87.187002} {\bibfield  {journal}
  {\bibinfo  {journal} {Phys. Rev. Lett.}\ }\textbf {\bibinfo {volume} {87}},\
  \bibinfo {pages} {187002} (\bibinfo {year} {2001}{\natexlab{a}})},\ \Eprint
  {https://arxiv.org/abs/cond-mat/0104198} {cond-mat/0104198} \BibitemShut
  {NoStop}%
\bibitem [{\citenamefont {Goodman}\ and\ \citenamefont
  {Teitel}(2004)}]{GoodmanTeitel2004}%
  \BibitemOpen
  \bibfield  {author} {\bibinfo {author} {\bibfnamefont {T.}~\bibnamefont
  {Goodman}}\ and\ \bibinfo {author} {\bibfnamefont {S.}~\bibnamefont
  {Teitel}},\ }\bibfield  {title} {\bibinfo {title} {Roughness of a tilted
  anharmonic string at depinning},\ }\href
  {https://doi.org/10.1103/PhysRevE.69.062105} {\bibfield  {journal} {\bibinfo
  {journal} {Phys. Rev. E}\ }\textbf {\bibinfo {volume} {69}},\ \bibinfo
  {pages} {062105} (\bibinfo {year} {2004})}\BibitemShut {NoStop}%
\bibitem [{\citenamefont {{Le Doussal}}\ and\ \citenamefont
  {Wiese}(2003)}]{LedoussalWiese2003}%
  \BibitemOpen
  \bibfield  {author} {\bibinfo {author} {\bibfnamefont {P.}~\bibnamefont {{Le
  Doussal}}}\ and\ \bibinfo {author} {\bibfnamefont {K.~J.}\ \bibnamefont
  {Wiese}},\ }\bibfield  {title} {\bibinfo {title} {Functional renormalization
  group for anisotropic depinning and relation to branching processes},\ }\href
  {https://doi.org/10.1103/PhysRevE.67.016121} {\bibfield  {journal} {\bibinfo
  {journal} {Phys. Rev. E}\ }\textbf {\bibinfo {volume} {67}},\ \bibinfo
  {pages} {016121} (\bibinfo {year} {2003})}\BibitemShut {NoStop}%
\bibitem [{\citenamefont {Chen}\ \emph {et~al.}(2015)\citenamefont {Chen},
  \citenamefont {Zapperi},\ and\ \citenamefont
  {Sethna}}]{ChenZapperiSethna2015}%
  \BibitemOpen
  \bibfield  {author} {\bibinfo {author} {\bibfnamefont {Y.~J.}\ \bibnamefont
  {Chen}}, \bibinfo {author} {\bibfnamefont {S.}~\bibnamefont {Zapperi}},\ and\
  \bibinfo {author} {\bibfnamefont {J.~P.}\ \bibnamefont {Sethna}},\ }\bibfield
   {title} {\bibinfo {title} {Crossover behavior in interface depinning},\
  }\href {https://doi.org/10.1103/PhysRevE.92.022146} {\bibfield  {journal}
  {\bibinfo  {journal} {Phys. Rev. E}\ }\textbf {\bibinfo {volume} {92}},\
  \bibinfo {pages} {022146} (\bibinfo {year} {2015})}\BibitemShut {NoStop}%
\bibitem [{\citenamefont {Arag{\'o}n}\ \emph {et~al.}(2016)\citenamefont
  {Arag{\'o}n}, \citenamefont {Kolton}, \citenamefont {Doussal}, \citenamefont
  {Wiese},\ and\ \citenamefont {Jagla}}]{AragonKoltonLeDoussalWieseJagla2016}%
  \BibitemOpen
  \bibfield  {author} {\bibinfo {author} {\bibfnamefont {L.~E.}\ \bibnamefont
  {Arag{\'o}n}}, \bibinfo {author} {\bibfnamefont {A.~B.}\ \bibnamefont
  {Kolton}}, \bibinfo {author} {\bibfnamefont {P.~L.}\ \bibnamefont {Doussal}},
  \bibinfo {author} {\bibfnamefont {K.~J.}\ \bibnamefont {Wiese}},\ and\
  \bibinfo {author} {\bibfnamefont {E.~A.}\ \bibnamefont {Jagla}},\ }\bibfield
  {title} {\bibinfo {title} {Avalanches in tip-driven interfaces in random
  media},\ }\href {http://stacks.iop.org/0295-5075/113/i=1/a=10002} {\bibfield
  {journal} {\bibinfo  {journal} {EPL (Europhysics Letters)}\ }\textbf
  {\bibinfo {volume} {113}},\ \bibinfo {pages} {10002} (\bibinfo {year}
  {2016})}\BibitemShut {NoStop}%
\bibitem [{\citenamefont {Glatz}\ \emph {et~al.}(2003)\citenamefont {Glatz},
  \citenamefont {Nattermann},\ and\ \citenamefont
  {Pokrovsky}}]{GlatzNattermannPokrovsky2003}%
  \BibitemOpen
  \bibfield  {author} {\bibinfo {author} {\bibfnamefont {A.}~\bibnamefont
  {Glatz}}, \bibinfo {author} {\bibfnamefont {T.}~\bibnamefont {Nattermann}},\
  and\ \bibinfo {author} {\bibfnamefont {V.}~\bibnamefont {Pokrovsky}},\
  }\bibfield  {title} {\bibinfo {title} {Domain wall depinning in random media
  by ac fields},\ }\href {https://doi.org/10.1103/PhysRevLett.90.047201}
  {\bibfield  {journal} {\bibinfo  {journal} {Phys. Rev. Lett.}\ }\textbf
  {\bibinfo {volume} {90}},\ \bibinfo {pages} {047201} (\bibinfo {year}
  {2003})}\BibitemShut {NoStop}%
\bibitem [{\citenamefont {ter Burg}\ \emph {et~al.}(2021)\citenamefont {ter
  Burg}, \citenamefont {Bohn}, \citenamefont {Durin}, \citenamefont {Sommer},\
  and\ \citenamefont {Wiese}}]{terBurgBohnDurinSommerWiese2021}%
  \BibitemOpen
  \bibfield  {author} {\bibinfo {author} {\bibfnamefont {C.}~\bibnamefont {ter
  Burg}}, \bibinfo {author} {\bibfnamefont {F.}~\bibnamefont {Bohn}}, \bibinfo
  {author} {\bibfnamefont {F.}~\bibnamefont {Durin}}, \bibinfo {author}
  {\bibfnamefont {R.}~\bibnamefont {Sommer}},\ and\ \bibinfo {author}
  {\bibfnamefont {K.}~\bibnamefont {Wiese}},\ }\bibfield  {title} {\bibinfo
  {title} {Force correlations in disordered magnets},\ }\href@noop {} {\
  (\bibinfo {year} {2021})},\ \Eprint {https://arxiv.org/abs/arXiv:2109.01197}
  {arXiv:2109.01197 [cond-mat.dis-nn]} \BibitemShut {NoStop}%
\bibitem [{\citenamefont {Albornoz}\ \emph {et~al.}(2021)\citenamefont
  {Albornoz}, \citenamefont {Ferrero}, \citenamefont {Kolton}, \citenamefont
  {Jeudy}, \citenamefont {Bustingorry},\ and\ \citenamefont
  {Curiale}}]{AlbornozFerreroKoltonJeudyBustingorryCuriale2021}%
  \BibitemOpen
  \bibfield  {author} {\bibinfo {author} {\bibfnamefont {L.}~\bibnamefont
  {Albornoz}}, \bibinfo {author} {\bibfnamefont {E.}~\bibnamefont {Ferrero}},
  \bibinfo {author} {\bibfnamefont {A.}~\bibnamefont {Kolton}}, \bibinfo
  {author} {\bibfnamefont {V.}~\bibnamefont {Jeudy}}, \bibinfo {author}
  {\bibfnamefont {S.}~\bibnamefont {Bustingorry}},\ and\ \bibinfo {author}
  {\bibfnamefont {J.}~\bibnamefont {Curiale}},\ }\bibfield  {title} {\bibinfo
  {title} {Universal critical exponents of the magnetic domain wall depinning
  transition},\ }\href@noop {} {\  (\bibinfo {year} {2021})},\ \Eprint
  {https://arxiv.org/abs/arXiv:2101.06555} {arXiv:2101.06555} \BibitemShut
  {NoStop}%
\bibitem [{\citenamefont {Ertas}\ and\ \citenamefont
  {Kardar}(1996)}]{ErtasKardar1996}%
  \BibitemOpen
  \bibfield  {author} {\bibinfo {author} {\bibfnamefont {D.}~\bibnamefont
  {Ertas}}\ and\ \bibinfo {author} {\bibfnamefont {M.}~\bibnamefont {Kardar}},\
  }\bibfield  {title} {\bibinfo {title} {Anisotropic scaling in threshold
  critical dynamics of driven directed lines},\ }\href
  {https://doi.org/10.1103/PhysRevB.53.3520} {\bibfield  {journal} {\bibinfo
  {journal} {Phys. Rev.}\ }\textbf {\bibinfo {volume} {B 53}},\ \bibinfo
  {pages} {3520} (\bibinfo {year} {1996})}\BibitemShut {NoStop}%
\bibitem [{\citenamefont {Koshelev}\ and\ \citenamefont
  {Kolton}(2011)}]{koshelevkolton2011}%
  \BibitemOpen
  \bibfield  {author} {\bibinfo {author} {\bibfnamefont {A.~E.}\ \bibnamefont
  {Koshelev}}\ and\ \bibinfo {author} {\bibfnamefont {A.~B.}\ \bibnamefont
  {Kolton}},\ }\bibfield  {title} {\bibinfo {title} {Theory and simulations on
  strong pinning of vortex lines by nanoparticles},\ }\href
  {https://doi.org/10.1103/PhysRevB.84.104528} {\bibfield  {journal} {\bibinfo
  {journal} {Phys. Rev. B}\ }\textbf {\bibinfo {volume} {84}},\ \bibinfo
  {pages} {104528} (\bibinfo {year} {2011})}\BibitemShut {NoStop}%
\bibitem [{\citenamefont {Civale}(2019)}]{Civale2019}%
  \BibitemOpen
  \bibfield  {author} {\bibinfo {author} {\bibfnamefont {L.}~\bibnamefont
  {Civale}},\ }\bibfield  {title} {\bibinfo {title} {Pushing the limits for the
  highest critical currents in superconductors},\ }\href
  {https://doi.org/10.1073/pnas.1905568116} {\bibfield  {journal} {\bibinfo
  {journal} {PNAS}\ }\textbf {\bibinfo {volume} {116}},\ \bibinfo {pages}
  {10201} (\bibinfo {year} {2019})}\BibitemShut {NoStop}%
\bibitem [{\citenamefont {Middleton}(1992)}]{Middleton1992}%
  \BibitemOpen
  \bibfield  {author} {\bibinfo {author} {\bibfnamefont {A.}~\bibnamefont
  {Middleton}},\ }\bibfield  {title} {\bibinfo {title} {Asymptotic uniqueness
  of the sliding state for charge-density waves},\ }\href
  {https://doi.org/10.1103/PhysRevLett.68.670} {\bibfield  {journal} {\bibinfo
  {journal} {Phys. Rev. Lett.}\ }\textbf {\bibinfo {volume} {68}},\ \bibinfo
  {pages} {670} (\bibinfo {year} {1992})}\BibitemShut {NoStop}%
\bibitem [{\citenamefont {Dobrinevski}\ \emph {et~al.}(2012)\citenamefont
  {Dobrinevski}, \citenamefont {{Le Doussal}},\ and\ \citenamefont
  {Wiese}}]{DobrinevskiLeDoussalWiese2011b}%
  \BibitemOpen
  \bibfield  {author} {\bibinfo {author} {\bibfnamefont {A.}~\bibnamefont
  {Dobrinevski}}, \bibinfo {author} {\bibfnamefont {P.}~\bibnamefont {{Le
  Doussal}}},\ and\ \bibinfo {author} {\bibfnamefont {K.}~\bibnamefont
  {Wiese}},\ }\bibfield  {title} {\bibinfo {title} {Non-stationary dynamics of
  the {Alessandro-Beatrice-Bertotti-Montorsi} model},\ }\href
  {https://doi.org/10.1103/PhysRevE.85.031105} {\bibfield  {journal} {\bibinfo
  {journal} {Phys. Rev. E}\ }\textbf {\bibinfo {volume} {85}},\ \bibinfo
  {pages} {031105} (\bibinfo {year} {2012})},\ \Eprint
  {https://arxiv.org/abs/arXiv:1112.6307} {arXiv:1112.6307} \BibitemShut
  {NoStop}%
\bibitem [{\citenamefont {Dobrinevski}\ \emph {et~al.}(2014)\citenamefont
  {Dobrinevski}, \citenamefont {{Le Doussal}},\ and\ \citenamefont
  {Wiese}}]{DobrinevskiLeDoussalWiese2014a}%
  \BibitemOpen
  \bibfield  {author} {\bibinfo {author} {\bibfnamefont {A.}~\bibnamefont
  {Dobrinevski}}, \bibinfo {author} {\bibfnamefont {P.}~\bibnamefont {{Le
  Doussal}}},\ and\ \bibinfo {author} {\bibfnamefont {K.}~\bibnamefont
  {Wiese}},\ }\bibfield  {title} {\bibinfo {title} {Avalanche shape and
  exponents beyond mean-field theory},\ }\href
  {https://doi.org/10.1209/0295-5075/108/66002} {\bibfield  {journal} {\bibinfo
   {journal} {EPL}\ }\textbf {\bibinfo {volume} {108}},\ \bibinfo {pages}
  {66002} (\bibinfo {year} {2014})},\ \Eprint
  {https://arxiv.org/abs/arXiv:1407.7353} {arXiv:1407.7353} \BibitemShut
  {NoStop}%
\bibitem [{\citenamefont {Rosso}\ and\ \citenamefont
  {Krauth}(2001{\natexlab{b}})}]{RossoKrauth2001a}%
  \BibitemOpen
  \bibfield  {author} {\bibinfo {author} {\bibfnamefont {A.}~\bibnamefont
  {Rosso}}\ and\ \bibinfo {author} {\bibfnamefont {W.}~\bibnamefont {Krauth}},\
  }\bibfield  {title} {\bibinfo {title} {{Monte Carlo} dynamics of driven
  strings in disordered media},\ }\href
  {https://doi.org/10.1103/PhysRevB.65.012202} {\bibfield  {journal} {\bibinfo
  {journal} {Phys. Rev. B}\ }\textbf {\bibinfo {volume} {65}},\ \bibinfo
  {pages} {012202} (\bibinfo {year} {2001}{\natexlab{b}})},\ \Eprint
  {https://arxiv.org/abs/cond-mat/0102017} {cond-mat/0102017} \BibitemShut
  {NoStop}%
\bibitem [{\citenamefont {Leschhorn}\ and\ \citenamefont
  {Tang}(1993)}]{LeschhornTang1993}%
  \BibitemOpen
  \bibfield  {author} {\bibinfo {author} {\bibfnamefont {H.}~\bibnamefont
  {Leschhorn}}\ and\ \bibinfo {author} {\bibfnamefont {L.-H.}\ \bibnamefont
  {Tang}},\ }\bibfield  {title} {\bibinfo {title} {Comment on ``{Elastic}
  string in a random potential''},\ }\href
  {https://doi.org/10.1103/PhysRevLett.70.2973} {\bibfield  {journal} {\bibinfo
   {journal} {Phys. Rev. Lett.}\ }\textbf {\bibinfo {volume} {70}},\ \bibinfo
  {pages} {2973} (\bibinfo {year} {1993})}\BibitemShut {NoStop}%
\bibitem [{\citenamefont {Grassberger}\ \emph {et~al.}(2016)\citenamefont
  {Grassberger}, \citenamefont {Dhar},\ and\ \citenamefont
  {Mohanty}}]{GrassbergerDharMohanty2016}%
  \BibitemOpen
  \bibfield  {author} {\bibinfo {author} {\bibfnamefont {P.}~\bibnamefont
  {Grassberger}}, \bibinfo {author} {\bibfnamefont {D.}~\bibnamefont {Dhar}},\
  and\ \bibinfo {author} {\bibfnamefont {P.~K.}\ \bibnamefont {Mohanty}},\
  }\bibfield  {title} {\bibinfo {title} {Oslo model, hyperuniformity, and the
  quenched {Edwards-Wilkinson} model},\ }\href
  {https://doi.org/10.1103/PhysRevE.94.042314} {\bibfield  {journal} {\bibinfo
  {journal} {Phys. Rev. E}\ }\textbf {\bibinfo {volume} {94}},\ \bibinfo
  {pages} {042314} (\bibinfo {year} {2016})}\BibitemShut {NoStop}%
\bibitem [{\citenamefont {Shapira}\ and\ \citenamefont
  {Wiese}()}]{ShapiraWieseUnpublished}%
  \BibitemOpen
  \bibfield  {author} {\bibinfo {author} {\bibfnamefont {A.}~\bibnamefont
  {Shapira}}\ and\ \bibinfo {author} {\bibfnamefont {K.}~\bibnamefont
  {Wiese}},\ }\href@noop {} {}\bibinfo {howpublished} {unpublished}\BibitemShut
  {NoStop}%
\bibitem [{\citenamefont {Koshelev}\ and\ \citenamefont
  {Vinokur}(1994)}]{KoshelevVinokur1994}%
  \BibitemOpen
  \bibfield  {author} {\bibinfo {author} {\bibfnamefont {A.~E.}\ \bibnamefont
  {Koshelev}}\ and\ \bibinfo {author} {\bibfnamefont {V.~M.}\ \bibnamefont
  {Vinokur}},\ }\bibfield  {title} {\bibinfo {title} {Dynamic melting of the
  vortex lattice},\ }\href {https://doi.org/10.1103/PhysRevLett.73.3580}
  {\bibfield  {journal} {\bibinfo  {journal} {Phys. Rev. Lett.}\ }\textbf
  {\bibinfo {volume} {73}},\ \bibinfo {pages} {3580} (\bibinfo {year}
  {1994})}\BibitemShut {NoStop}%
\bibitem [{\citenamefont {Kolton}\ \emph {et~al.}(2002)\citenamefont {Kolton},
  \citenamefont {Exartier}, \citenamefont {Cugliandolo}, \citenamefont
  {Dom\'{\i}nguez},\ and\ \citenamefont
  {Gr\o{}nbech-Jensen}}]{KoltonExartierCugliandoloDominguezGronbechJensen2002}%
  \BibitemOpen
  \bibfield  {author} {\bibinfo {author} {\bibfnamefont {A.~B.}\ \bibnamefont
  {Kolton}}, \bibinfo {author} {\bibfnamefont {R.}~\bibnamefont {Exartier}},
  \bibinfo {author} {\bibfnamefont {L.~F.}\ \bibnamefont {Cugliandolo}},
  \bibinfo {author} {\bibfnamefont {D.}~\bibnamefont {Dom\'{\i}nguez}},\ and\
  \bibinfo {author} {\bibfnamefont {N.}~\bibnamefont {Gr\o{}nbech-Jensen}},\
  }\bibfield  {title} {\bibinfo {title} {Effective temperature in driven vortex
  lattices with random pinning},\ }\href
  {https://doi.org/10.1103/PhysRevLett.89.227001} {\bibfield  {journal}
  {\bibinfo  {journal} {Phys. Rev. Lett.}\ }\textbf {\bibinfo {volume} {89}},\
  \bibinfo {pages} {227001} (\bibinfo {year} {2002})}\BibitemShut {NoStop}%
\bibitem [{\citenamefont {Kolton}(2006)}]{kolton2006}%
  \BibitemOpen
  \bibfield  {author} {\bibinfo {author} {\bibfnamefont {A.~B.}\ \bibnamefont
  {Kolton}},\ }\bibfield  {title} {\bibinfo {title} {Pinning induced
  fluctuations on driven vortices},\ }\href
  {https://doi.org/https://doi.org/10.1016/j.physc.2005.12.026} {\bibfield
  {journal} {\bibinfo  {journal} {Physica C}\ }\textbf {\bibinfo {volume}
  {437-438}},\ \bibinfo {pages} {153} (\bibinfo {year} {2006})},\ \bibinfo
  {note} {proceedings of the Fourth International Conference on Vortex Matter
  in Nanostructured Superconductors VORTEX IV}\BibitemShut {NoStop}%
\bibitem [{\citenamefont {Blatter}\ \emph {et~al.}(1994)\citenamefont
  {Blatter}, \citenamefont {Feigel'man}, \citenamefont {Geshkenbein},
  \citenamefont {Larkin},\ and\ \citenamefont {Vinokur}}]{Blatter1994}%
  \BibitemOpen
  \bibfield  {author} {\bibinfo {author} {\bibfnamefont {G.}~\bibnamefont
  {Blatter}}, \bibinfo {author} {\bibfnamefont {M.~V.}\ \bibnamefont
  {Feigel'man}}, \bibinfo {author} {\bibfnamefont {V.~B.}\ \bibnamefont
  {Geshkenbein}}, \bibinfo {author} {\bibfnamefont {A.~I.}\ \bibnamefont
  {Larkin}},\ and\ \bibinfo {author} {\bibfnamefont {V.~M.}\ \bibnamefont
  {Vinokur}},\ }\bibfield  {title} {\bibinfo {title} {Vortices in
  high-temperature superconductors},\ }\href
  {https://doi.org/10.1103/RevModPhys.66.1125} {\bibfield  {journal} {\bibinfo
  {journal} {Rev. Mod. Phys.}\ }\textbf {\bibinfo {volume} {66}},\ \bibinfo
  {pages} {1125} (\bibinfo {year} {1994})}\BibitemShut {NoStop}%
\bibitem [{\citenamefont {Tinkham}(2004)}]{Tinkham2004}%
  \BibitemOpen
  \bibfield  {author} {\bibinfo {author} {\bibfnamefont {M.}~\bibnamefont
  {Tinkham}},\ }\href {https://store.doverpublications.com/0486435032.html}
  {\emph {\bibinfo {title} {Introduction to Superconductivity}}},\ \bibinfo
  {edition} {2nd}\ ed.\ (\bibinfo  {publisher} {Dover Publications},\ \bibinfo
  {year} {2004})\BibitemShut {NoStop}%
\bibitem [{\citenamefont {{Le Doussal}}\ and\ \citenamefont
  {Wiese}(2007)}]{LeDoussalWiese2007}%
  \BibitemOpen
  \bibfield  {author} {\bibinfo {author} {\bibfnamefont {P.}~\bibnamefont {{Le
  Doussal}}}\ and\ \bibinfo {author} {\bibfnamefont {K.}~\bibnamefont
  {Wiese}},\ }\bibfield  {title} {\bibinfo {title} {How to measure {Functional
  RG} fixed-point functions for dynamics and at depinning},\ }\href
  {https://doi.org/10.1209/0295-5075/77/66001} {\bibfield  {journal} {\bibinfo
  {journal} {EPL}\ }\textbf {\bibinfo {volume} {77}},\ \bibinfo {pages} {66001}
  (\bibinfo {year} {2007})},\ \Eprint {https://arxiv.org/abs/cond-mat/0610525}
  {cond-mat/0610525} \BibitemShut {NoStop}%
\bibitem [{\citenamefont {Doussal}\ \emph {et~al.}(2009)\citenamefont
  {Doussal}, \citenamefont {Wiese}, \citenamefont {Moulinet},\ and\
  \citenamefont {Rolley}}]{LeDoussalWieseMoulinetRolley2009}%
  \BibitemOpen
  \bibfield  {author} {\bibinfo {author} {\bibfnamefont {P.~L.}\ \bibnamefont
  {Doussal}}, \bibinfo {author} {\bibfnamefont {K.}~\bibnamefont {Wiese}},
  \bibinfo {author} {\bibfnamefont {S.}~\bibnamefont {Moulinet}},\ and\
  \bibinfo {author} {\bibfnamefont {E.}~\bibnamefont {Rolley}},\ }\bibfield
  {title} {\bibinfo {title} {Height fluctuations of a contact line: {A} direct
  measurement of the renormalized disorder correlator},\ }\href
  {https://doi.org/10.1209/0295-5075/87/56001} {\bibfield  {journal} {\bibinfo
  {journal} {EPL}\ }\textbf {\bibinfo {volume} {87}},\ \bibinfo {pages} {56001}
  (\bibinfo {year} {2009})},\ \Eprint {https://arxiv.org/abs/arXiv:0904.4156}
  {arXiv:0904.4156} \BibitemShut {NoStop}%
\bibitem [{\citenamefont {Wiese}\ \emph {et~al.}(2020)\citenamefont {Wiese},
  \citenamefont {Bercy}, \citenamefont {Melkonyan},\ and\ \citenamefont
  {Bizebard}}]{WieseBercyMelkonyanBizebard2019}%
  \BibitemOpen
  \bibfield  {author} {\bibinfo {author} {\bibfnamefont {K.}~\bibnamefont
  {Wiese}}, \bibinfo {author} {\bibfnamefont {M.}~\bibnamefont {Bercy}},
  \bibinfo {author} {\bibfnamefont {L.}~\bibnamefont {Melkonyan}},\ and\
  \bibinfo {author} {\bibfnamefont {T.}~\bibnamefont {Bizebard}},\ }\bibfield
  {title} {\bibinfo {title} {Universal force correlations in an {RNA-DNA}
  unzipping experiment},\ }\href
  {https://doi.org/10.1103/PhysRevResearch.2.043385} {\bibfield  {journal}
  {\bibinfo  {journal} {Phys. Rev. Research}\ }\textbf {\bibinfo {volume}
  {2}},\ \bibinfo {pages} {043385} (\bibinfo {year} {2020})},\ \Eprint
  {https://arxiv.org/abs/arXiv:1909.01319} {arXiv:1909.01319} \BibitemShut
  {NoStop}%
\bibitem [{\citenamefont {ter Burg}\ and\ \citenamefont
  {Wiese}(2021)}]{terBurgWiese2020}%
  \BibitemOpen
  \bibfield  {author} {\bibinfo {author} {\bibfnamefont {C.}~\bibnamefont {ter
  Burg}}\ and\ \bibinfo {author} {\bibfnamefont {K.}~\bibnamefont {Wiese}},\
  }\bibfield  {title} {\bibinfo {title} {Mean-field theories for depinning and
  their experimental signatures},\ }\href
  {https://doi.org/10.1103/PhysRevE.103.052114} {\bibfield  {journal} {\bibinfo
   {journal} {Phys. Rev. E}\ }\textbf {\bibinfo {volume} {103}},\ \bibinfo
  {pages} {052114} (\bibinfo {year} {2021})},\ \Eprint
  {https://arxiv.org/abs/arXiv:2010.16372} {arXiv:2010.16372} \BibitemShut
  {NoStop}%
\bibitem [{\citenamefont {Doussal}\ and\ \citenamefont
  {Wiese}(2009)}]{LeDoussalWiese2009}%
  \BibitemOpen
  \bibfield  {author} {\bibinfo {author} {\bibfnamefont {P.~L.}\ \bibnamefont
  {Doussal}}\ and\ \bibinfo {author} {\bibfnamefont {K.}~\bibnamefont
  {Wiese}},\ }\bibfield  {title} {\bibinfo {title} {Driven particle in a random
  landscape: disorder correlator, avalanche distribution and extreme value
  statistics of records},\ }\href {https://doi.org/10.1103/PhysRevE.79.051105}
  {\bibfield  {journal} {\bibinfo  {journal} {Phys. Rev. E}\ }\textbf {\bibinfo
  {volume} {79}},\ \bibinfo {pages} {051105} (\bibinfo {year} {2009})},\
  \Eprint {https://arxiv.org/abs/arXiv:0808.3217} {arXiv:0808.3217}
  \BibitemShut {NoStop}%
\bibitem [{\citenamefont {Doussal}\ \emph {et~al.}(2004)\citenamefont
  {Doussal}, \citenamefont {Wiese},\ and\ \citenamefont
  {Chauve}}]{LeDoussalWieseChauve2003}%
  \BibitemOpen
  \bibfield  {author} {\bibinfo {author} {\bibfnamefont {P.~L.}\ \bibnamefont
  {Doussal}}, \bibinfo {author} {\bibfnamefont {K.}~\bibnamefont {Wiese}},\
  and\ \bibinfo {author} {\bibfnamefont {P.}~\bibnamefont {Chauve}},\
  }\bibfield  {title} {\bibinfo {title} {Functional renormalization group and
  the field theory of disordered elastic systems},\ }\href
  {https://doi.org/10.1103/PhysRevE.69.026112} {\bibfield  {journal} {\bibinfo
  {journal} {Phys. Rev. E}\ }\textbf {\bibinfo {volume} {69}},\ \bibinfo
  {pages} {026112} (\bibinfo {year} {2004})},\ \Eprint
  {https://arxiv.org/abs/cond-mat/0304614} {cond-mat/0304614} \BibitemShut
  {NoStop}%
\bibitem [{Note2()}]{Note2}%
  \BibitemOpen
  \bibinfo {note} {We learned from L.~Ponson that the perpendicular roughness
  for fracture in $d=1$ is $\zeta _\perp \simeq 0$, consistent with the
  ``thermal'' exponent for LR elasticity $\zeta _\perp = (1-d)/2$.}\BibitemShut
  {Stop}%
\bibitem [{\citenamefont {Grassi}\ \emph {et~al.}(2018)\citenamefont {Grassi},
  \citenamefont {Kolton}, \citenamefont {Jeudy}, \citenamefont {Mougin},
  \citenamefont {Bustingorry},\ and\ \citenamefont
  {Curiale}}]{GrassiKoltonJeudyMouginBustingorryCuriale2018}%
  \BibitemOpen
  \bibfield  {author} {\bibinfo {author} {\bibfnamefont {M.}~\bibnamefont
  {Grassi}}, \bibinfo {author} {\bibfnamefont {A.~B.}\ \bibnamefont {Kolton}},
  \bibinfo {author} {\bibfnamefont {V.}~\bibnamefont {Jeudy}}, \bibinfo
  {author} {\bibfnamefont {A.}~\bibnamefont {Mougin}}, \bibinfo {author}
  {\bibfnamefont {S.}~\bibnamefont {Bustingorry}},\ and\ \bibinfo {author}
  {\bibfnamefont {J.}~\bibnamefont {Curiale}},\ }\bibfield  {title} {\bibinfo
  {title} {Intermittent collective dynamics of domain walls in the creep
  regime},\ }\href {https://doi.org/10.1103/PhysRevB.98.224201} {\bibfield
  {journal} {\bibinfo  {journal} {Phys. Rev. B}\ }\textbf {\bibinfo {volume}
  {98}},\ \bibinfo {pages} {224201} (\bibinfo {year} {2018})}\BibitemShut
  {NoStop}%
\bibitem [{\citenamefont {Rosso}\ \emph {et~al.}(2009)\citenamefont {Rosso},
  \citenamefont {{Le Doussal}},\ and\ \citenamefont
  {Wiese}}]{RossoLeDoussalWiese2009}%
  \BibitemOpen
  \bibfield  {author} {\bibinfo {author} {\bibfnamefont {A.}~\bibnamefont
  {Rosso}}, \bibinfo {author} {\bibfnamefont {P.}~\bibnamefont {{Le
  Doussal}}},\ and\ \bibinfo {author} {\bibfnamefont {K.~J.}\ \bibnamefont
  {Wiese}},\ }\bibfield  {title} {\bibinfo {title} {Avalanche-size distribution
  at the depinning transition: A numerical test of the theory},\ }\href
  {https://doi.org/10.1103/PhysRevB.80.144204} {\bibfield  {journal} {\bibinfo
  {journal} {Phys. Rev. B}\ }\textbf {\bibinfo {volume} {80}},\ \bibinfo
  {pages} {144204} (\bibinfo {year} {2009})}\BibitemShut {NoStop}%
\bibitem [{\citenamefont {{Le~Doussal}}\ \emph {et~al.}(2009)\citenamefont
  {{Le~Doussal}}, \citenamefont {Middleton},\ and\ \citenamefont
  {Wiese}}]{LeDoussalMiddletonWiese2008}%
  \BibitemOpen
  \bibfield  {author} {\bibinfo {author} {\bibfnamefont {P.}~\bibnamefont
  {{Le~Doussal}}}, \bibinfo {author} {\bibfnamefont {A.}~\bibnamefont
  {Middleton}},\ and\ \bibinfo {author} {\bibfnamefont {K.}~\bibnamefont
  {Wiese}},\ }\bibfield  {title} {\bibinfo {title} {Statistics of static
  avalanches in a random pinning landscape},\ }\href
  {https://doi.org/10.1103/PhysRevE.79.050101} {\bibfield  {journal} {\bibinfo
  {journal} {Phys. Rev. E}\ }\textbf {\bibinfo {volume} {79}},\ \bibinfo
  {pages} {050101 (R)} (\bibinfo {year} {2009})},\ \Eprint
  {https://arxiv.org/abs/arXiv:0803.1142} {arXiv:0803.1142} \BibitemShut
  {NoStop}%
\bibitem [{\citenamefont {Bolech}\ and\ \citenamefont
  {Rosso}(2004)}]{BolechRosso2004}%
  \BibitemOpen
  \bibfield  {author} {\bibinfo {author} {\bibfnamefont {C.}~\bibnamefont
  {Bolech}}\ and\ \bibinfo {author} {\bibfnamefont {A.}~\bibnamefont {Rosso}},\
  }\bibfield  {title} {\bibinfo {title} {Universal statistics of the critical
  depinning force of elastic systems in random media},\ }\href
  {https://doi.org/10.1103/PhysRevLett.93.125701} {\bibfield  {journal}
  {\bibinfo  {journal} {Phys. Rev. Lett.}\ }\textbf {\bibinfo {volume} {93}},\
  \bibinfo {pages} {125701} (\bibinfo {year} {2004})},\ \Eprint
  {https://arxiv.org/abs/cond-mat/0403023} {cond-mat/0403023} \BibitemShut
  {NoStop}%
\bibitem [{\citenamefont {Fedorenko}\ \emph
  {et~al.}(2006{\natexlab{b}})\citenamefont {Fedorenko}, \citenamefont
  {{Le~Doussal}},\ and\ \citenamefont {Wiese}}]{FedorenkoLeDoussalWiese2006}%
  \BibitemOpen
  \bibfield  {author} {\bibinfo {author} {\bibfnamefont {A.}~\bibnamefont
  {Fedorenko}}, \bibinfo {author} {\bibfnamefont {P.}~\bibnamefont
  {{Le~Doussal}}},\ and\ \bibinfo {author} {\bibfnamefont {K.}~\bibnamefont
  {Wiese}},\ }\bibfield  {title} {\bibinfo {title} {Universal distribution of
  threshold forces at the depinning transition},\ }\href
  {https://doi.org/10.1103/PhysRevE.74.041110} {\bibfield  {journal} {\bibinfo
  {journal} {Phys. Rev. E}\ }\textbf {\bibinfo {volume} {74}},\ \bibinfo
  {pages} {041110} (\bibinfo {year} {2006}{\natexlab{b}})},\ \Eprint
  {https://arxiv.org/abs/cond-mat/0607229} {cond-mat/0607229} \BibitemShut
  {NoStop}%
\bibitem [{\citenamefont {Kolton}\ \emph {et~al.}(2009)\citenamefont {Kolton},
  \citenamefont {Rosso}, \citenamefont {Giamarchi},\ and\ \citenamefont
  {Krauth}}]{KoltonRossoGiamarchiKrauth2009b}%
  \BibitemOpen
  \bibfield  {author} {\bibinfo {author} {\bibfnamefont {A.~B.}\ \bibnamefont
  {Kolton}}, \bibinfo {author} {\bibfnamefont {A.}~\bibnamefont {Rosso}},
  \bibinfo {author} {\bibfnamefont {T.}~\bibnamefont {Giamarchi}},\ and\
  \bibinfo {author} {\bibfnamefont {W.}~\bibnamefont {Krauth}},\ }\bibfield
  {title} {\bibinfo {title} {Creep dynamics of elastic manifolds via exact
  transition pathways},\ }\href {https://doi.org/10.1103/PhysRevB.79.184207}
  {\bibfield  {journal} {\bibinfo  {journal} {Phys. Rev. B}\ }\textbf {\bibinfo
  {volume} {79}},\ \bibinfo {pages} {184207} (\bibinfo {year}
  {2009})}\BibitemShut {NoStop}%
\bibitem [{\citenamefont {Cugliandolo}(2011)}]{Cugliandolo2011}%
  \BibitemOpen
  \bibfield  {author} {\bibinfo {author} {\bibfnamefont {L.~F.}\ \bibnamefont
  {Cugliandolo}},\ }\bibfield  {title} {\bibinfo {title} {The effective
  temperature},\ }\href@noop {} {\bibfield  {journal} {\bibinfo  {journal}
  {Journal of Physics A: Mathematical and Theoretical}\ }\textbf {\bibinfo
  {volume} {44}},\ \bibinfo {pages} {483001} (\bibinfo {year}
  {2011})}\BibitemShut {NoStop}%
\bibitem [{\citenamefont {Le~Doussal}\ \emph {et~al.}(2012)\citenamefont
  {Le~Doussal}, \citenamefont {Petkovi\ifmmode~\acute{c}\else \'{c}\fi{}},\
  and\ \citenamefont {Wiese}}]{LeDoussalPetkovicWiese2012}%
  \BibitemOpen
  \bibfield  {author} {\bibinfo {author} {\bibfnamefont {P.}~\bibnamefont
  {Le~Doussal}}, \bibinfo {author} {\bibfnamefont {A.}~\bibnamefont
  {Petkovi\ifmmode~\acute{c}\else \'{c}\fi{}}},\ and\ \bibinfo {author}
  {\bibfnamefont {K.}~\bibnamefont {Wiese}},\ }\bibfield  {title} {\bibinfo
  {title} {Distribution of velocities and acceleration for a particle in
  {Brownian} correlated disorder: Inertial case},\ }\href
  {https://doi.org/10.1103/PhysRevE.85.061116} {\bibfield  {journal} {\bibinfo
  {journal} {Phys. Rev. E}\ }\textbf {\bibinfo {volume} {85}},\ \bibinfo
  {pages} {061116} (\bibinfo {year} {2012})},\ \Eprint
  {https://arxiv.org/abs/arXiv:1203.5620} {arXiv:1203.5620} \BibitemShut
  {NoStop}%
\bibitem [{\citenamefont {Sparfel}\ and\ \citenamefont
  {Wiese}(2021)}]{SparfelWiese2021}%
  \BibitemOpen
  \bibfield  {author} {\bibinfo {author} {\bibfnamefont {J.}~\bibnamefont
  {Sparfel}}\ and\ \bibinfo {author} {\bibfnamefont {K.}~\bibnamefont
  {Wiese}},\ }\bibfield  {title} {\bibinfo {title} {Skewness at depinning, and
  conformal invariance},\ }\href@noop {} {\bibfield  {journal} {\bibinfo
  {journal} {unpublished}\ } (\bibinfo {year} {2021})}\BibitemShut {NoStop}%
\bibitem [{\citenamefont {Ramanathan}\ \emph {et~al.}(1997)\citenamefont
  {Ramanathan}, \citenamefont {Ertas},\ and\ \citenamefont
  {Fisher}}]{RamanathanErtasFisher1997}%
  \BibitemOpen
  \bibfield  {author} {\bibinfo {author} {\bibfnamefont {S.}~\bibnamefont
  {Ramanathan}}, \bibinfo {author} {\bibfnamefont {D.}~\bibnamefont {Ertas}},\
  and\ \bibinfo {author} {\bibfnamefont {D.}~\bibnamefont {Fisher}},\
  }\bibfield  {title} {\bibinfo {title} {Quasistatic crack propagation in
  heterogeneous media},\ }\href {https://doi.org/10.1103/PhysRevLett.79.873}
  {\bibfield  {journal} {\bibinfo  {journal} {Phys. Rev. Lett.}\ }\textbf
  {\bibinfo {volume} {79}},\ \bibinfo {pages} {873} (\bibinfo {year}
  {1997})}\BibitemShut {NoStop}%
\bibitem [{\citenamefont {Dalmas}\ \emph {et~al.}(2008)\citenamefont {Dalmas},
  \citenamefont {Lelarge},\ and\ \citenamefont
  {Vandembroucq}}]{DalmasLelargeVandembroucq2008}%
  \BibitemOpen
  \bibfield  {author} {\bibinfo {author} {\bibfnamefont {D.}~\bibnamefont
  {Dalmas}}, \bibinfo {author} {\bibfnamefont {A.}~\bibnamefont {Lelarge}},\
  and\ \bibinfo {author} {\bibfnamefont {D.}~\bibnamefont {Vandembroucq}},\
  }\bibfield  {title} {\bibinfo {title} {Crack propagation through
  phase-separated glasses: Effect of the characteristic size of disorder},\
  }\href {https://doi.org/10.1103/PhysRevLett.101.255501} {\bibfield  {journal}
  {\bibinfo  {journal} {Phys. Rev. Lett.}\ }\textbf {\bibinfo {volume} {101}},\
  \bibinfo {pages} {255501} (\bibinfo {year} {2008})}\BibitemShut {NoStop}%
\end{thebibliography}%
\end{document}